\documentclass[11pt,a4paper]{article}
\pdfoutput=1
\usepackage{jheppub}
\usepackage[english]{babel}
\usepackage[babel]{csquotes}
\usepackage{amsfonts}
\usepackage{array, tabularx}
\usepackage{mathtools}
\usepackage{amsthm}
\usepackage{url}
\usepackage{multicol, multirow}
\usepackage{emptypage}
\usepackage{verbatim}
\usepackage{dsfont}
\usepackage{float}
\usepackage{hyperref}
\usepackage{amsmath}
\usepackage{amssymb,tikz}
\usepackage{verbatim}
\usepackage{xspace}
\usepackage{tensor}
\usepackage{youngtab}
\usepackage{colortbl}

\usepackage{mathrsfs}   

\usepackage[font=small,labelfont={sf,bf}]{caption}

\makeatother

\usepackage{bm}
\newcommand{\mtrx}[1]{{\bm{\mathsf{#1}}}}
\newcommand{\vc}[1]{\boldsymbol{#1}}
\newcommand{\eps}{{\varepsilon}}

\newcommand{\zb}{{\bar{z}}}
\newcommand{\1}{{\mathds{1}}}
\renewcommand{\O}{{\mathcal O}}
\newcommand{\Z}{{\mathbb Z}}
\newcommand{\G}{{\mathcal G}}
\newcommand{\de}{{\partial}}

\allowdisplaybreaks[2]

\title{The tricritical Ising CFT and conformal bootstrap}

\author[\phi,\phi^2,\phi^3]{Johan Henriksson}
\affiliation[\phi]{Dipartimento di Fisica ``E.\ Fermi'', Universit\`a di Pisa \emph{and} INFN, sezione di Pisa, Largo Bruno Pontecorvo 3, 56127 Pisa, Italy}
\affiliation[\phi^2]{Universit\'e Paris--Saclay, CEA, Institut de Physique Th\'eorique, 91191, Gif-sur-Yvette, France}
\affiliation[\phi^3]{Theoretical Physics Department, CERN, 1211, Geneva, Switzerland}

\emailAdd{johan.henriksson@cern.ch}

\abstract{
The tricritical Ising CFT is the IR fixed-point of $\lambda\phi^6$ theory. It can be seen as a one-parameter family of CFTs connecting between an $\varepsilon$-expansion near the upper critical dimension $3$ and the exactly solved minimal model in $d=2$. We review what is known about the tricritical Ising CFT, and study it with the numerical conformal bootstrap for various dimensions. Using a mixed system with three external operators $\{\phi\sim\sigma,\,\phi^2\sim \epsilon,\,\phi^3\sim\sigma'\}$, we find three-dimensional ``bootstrap islands'' in $d=2.75$ and $d=2.5$ dimensions consistent with interpolations between the perturbative estimates and the 2d exact values. In $d=2$ and $d=2.25$ the setup is not strong enough to isolate the theory. This paper also contains a survey of the perturbative spectrum and a review of results from the literature.}

\begin{document}

	\makeatletter
	\let\old@fpheader\@fpheader
	\renewcommand{\@fpheader}{  \vspace*{-0.1cm} \hfill CERN-TH-2025-028}
	\makeatother

\maketitle

\noindent \textbf{Note added.} \textsl{Since the publication of this paper, references \cite{Adzhemyan:2026whu,Jack:2026npe} have found mistakes in \cite{Hager:2002uq} from which some expressions in this paper are quoted. The updates concerns six-loop anomalous dimensions for the operators $\phi^4$ and $\phi^6$, and we give the correct expressions from \cite{Adzhemyan:2026whu,Jack:2026npe} in appendix~\ref{app:renormalisation}. No equations in the main text have been updated. Note that the bootstrap study of this paper is not sensitive to these precise form of these expressions; it depends only on the imposed gaps, which are chosen by hand (see table~\ref{tab:operatorsAndGaps}).}

\section{Introduction}

It is well-known that the scalar theory with interaction $\mathcal L=\frac12(\de\phi)^2+\frac12m^2\phi^2+\frac1{24}\lambda\phi^4$ has a non-trivial infrared (IR) fixed-point of the renormalisation group (RG), which is attained by tuning the renormalised mass parameter $m^2$ to zero. This fixed-point is described by the Ising conformal field theory (CFT) and exists in spacetime dimensions $d$ below the upper critical dimension $4$. In this paper we study instead the action
\begin{equation}
\label{eq:actionIntro}
S=\int d^dx\left(\frac12\de_\mu\phi\de^\mu\phi +\frac{u_2}2\phi^2+\frac{u_4}{4!}\phi^4+\frac{\lambda}{6!}\phi^6\right).
\end{equation}
By tuning both $u_2$ and $u_4$, this theory flows to an IR fixed-point which describes tricritical behaviour and is denoted the \emph{tricritical Ising CFT}. The interaction $\phi^6$ becomes marginal in the upper critical dimension $d=3$, leading to a perturbative expansion in $\eps=3-d$. 
The Ising and tricritical Ising CFTs are the first instances in a sequence of multicritical fixed-points which are dominated by the interaction $\lambda\phi^{2k}$ and have upper critical dimensions $d_c(k)=\frac{2k}{k-1}$. They are believed to be connected to the diagonal minimal models $\mathcal M_{k+2,k+1}$ \cite{Zamolodchikov:1986db}, which are a sequence of exactly solved CFTs in two dimensions \cite{Belavin:1984vu,Friedan:1983xq,Cappelli:1987xt}. Although the picture of a sequence of 2d CFTs connecting to perturbative expansions near the upper critical dimensions is natural from field theory and has been supported by non-perturbative RG studies in fractional spacetime dimensions \cite{Codello:2012sc,Codello:2014yfa,Hellwig:2015woa}, it remains conjectural.

\begin{figure}
\centering
\includegraphics{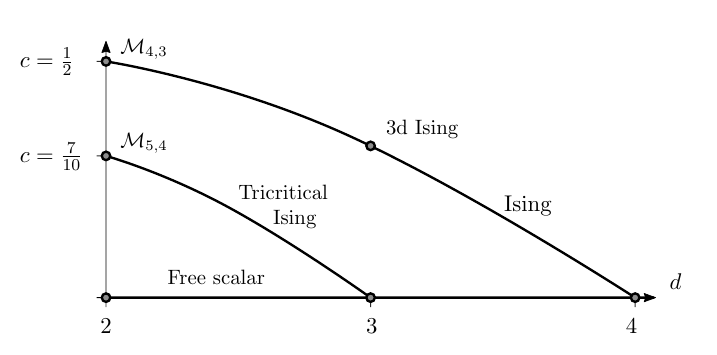}
\caption{Continuous families of CFTs discussed in this work. The circles represent unitary theories in integer dimensions. The vertical axis is a proxy for interaction strength, and $c$ is the 2d central charge.}\label{fig:operahouse}
\end{figure}

Here we will consider the tricritical Ising CFT from the perspective of the modern conformal bootstrap \cite{Rattazzi:2008pe}. This programme has led to great success for a variety of theories, see \cite{Poland:2018epd,Rychkov:2023wsd} for reviews. To a large extent, the development of the conformal bootstrap has been guided by the following sequence of results for the Ising CFT:
\begin{itemize}
\item Single correlator bootstrap produced an exclusion plot in the space spanned by the lowest two scaling dimensions, with a kink at values corresponding to the Ising CFT in $d=2$ \cite{Rychkov:2009ij} and $d=3$ \cite{El-Showk:2012cjh} dimensions.
\item Using mixed-correlator bootstrap, the kink was converted to an isolated island in parameter space \cite{Kos:2014bka}. With improved algorithms and numerical strength, the island has been shrunk dramatically to give high-precision data for the 3d Ising CFT \cite{Kos:2016ysd,Chang:2024whx}.
\item An estimate for the low-lying spectrum can be generated using the extremal functional method \cite{ElShowk:2012hu,El-Showk:2014dwa}. This has produced a large set of data for low-lying operators in the Ising CFT \cite{Simmons-Duffin:2016wlq}. 
\item Finally, studies for fractional spacetime dimensions $d$ have corroborated the existence of the Ising CFT as a family of CFTs parametrised by $d$ \cite{El-Showk:2013nia,Cappelli:2018vir,Henriksson:2022gpa,Bonanno:2022ztf}.\footnote{There are subtleties with this interpretation, related to evanescent operators \cite{Hogervorst:2015akt,Binder:2019zqc}, see section~\ref{sec:continuationInD} for comments.}
\end{itemize}
In this paper we initiate an attempt to repeat this bootstrap success for the tricritical Ising CFT, and pave the way for future work on related theories, in particular tricritical $O(n)$ models (see discussion in section~\ref{sec:tricriticalON}). Aware that the 2d theory is exactly solved, our primary goal is to isolate the theory to islands for intermediate, fractional, values of $d$.

\begin{figure}
\centering
\includegraphics[width=0.9\textwidth]{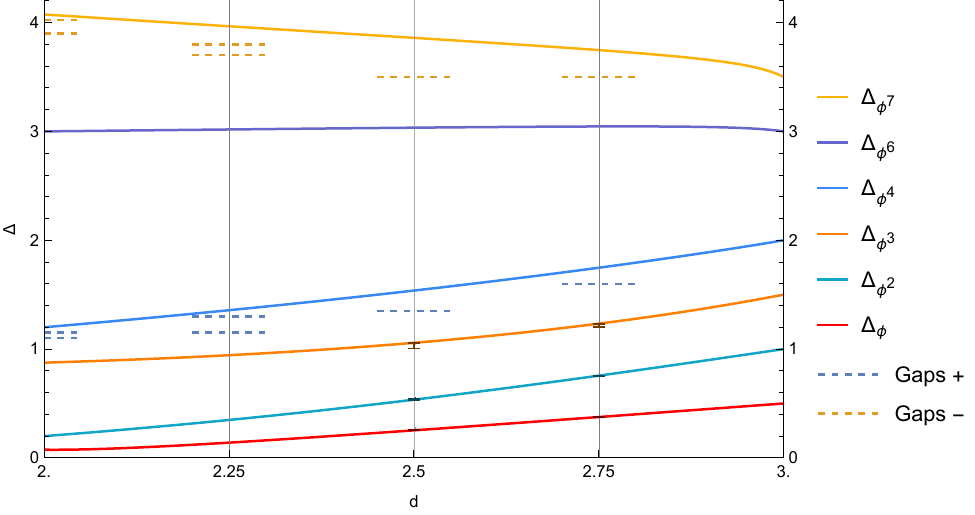}
\caption{Figure describing our setup and results. The error-bars in $d=2.5$ and $d=2.75$ represent our three-dimensional islands \eqref{eq:island275}--\eqref{eq:island25}. Solid lines are Pad\'e approximants \eqref{eq:Pades} and the dashed lines are the spectral gaps we assume above the external operators in the $\Z_2$-even (blue tones) and $\Z_2$-odd (orange tones) sectors. The islands are almost invisible at this scale, see figure~\ref{fig:islandsLog} for a log plot of anomalous dimensions. In $d=2$ and $d=2.25$ we do not find islands with the current precision.
}\label{fig:islands}
\end{figure}

\begin{figure}
\centering
\includegraphics[width=0.92\textwidth]{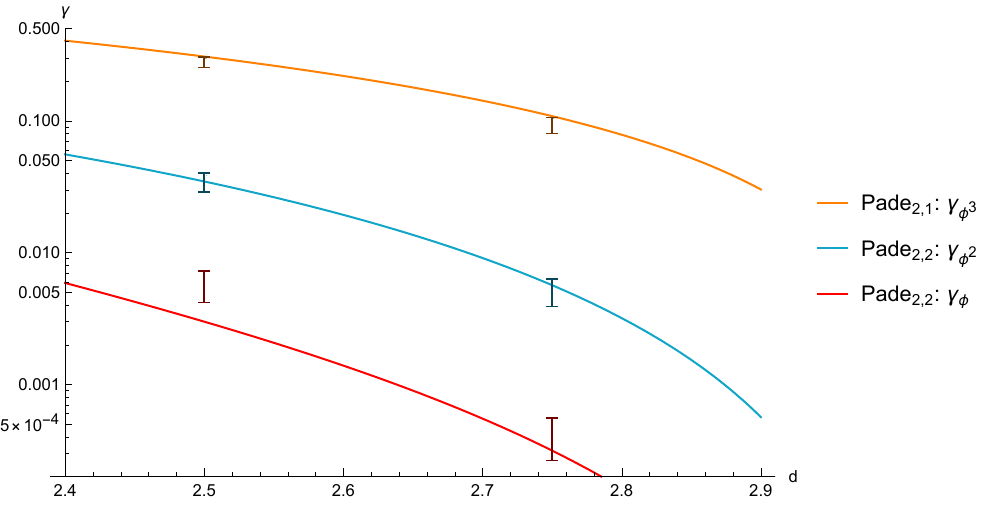}
\caption{Logarithmic plot of our islands \eqref{eq:island275}--\eqref{eq:island25} in terms of anomalous dimensions $\gamma$. The three intervals for a fixed $d$ represent three-dimensional islands (rigorous intervals) and the solid lines are Pad\'e approximants \eqref{eq:Pades}.}\label{fig:islandsLog}
\end{figure}

\begin{figure}
\centering
\includegraphics[width=0.72\textwidth]{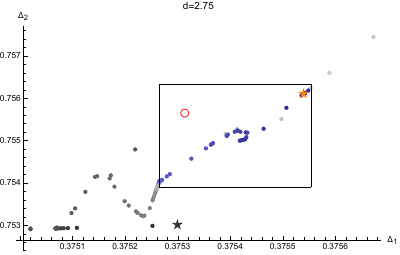}
\caption{Navigator run in $d=2.75$ projected onto the subspace $(\Delta_\phi,\Delta_{\phi^2})$. Gray is positive navigator values (ruled-out points), purple is negative (allowed points). We display plots including the $\Delta_{\phi^3}$ axis in figure~\ref{fig:nav-275-more}. The rectangle represents our rigorous bound, found using additional navigator searches along the different axes (appendix~\ref{app:NavigatorIslands}), and the red circle the Pad\'e approximant \eqref{eq:Pades}. Although the starting point (dark star) is farther away from the allowed points than the Pad\'e approximant, the navigator search still finds the island.}\label{fig:nav-275}
\end{figure}

Our main results follow from numerical conformal bootstrap of the system of all four-point correlators involving the three lowest-lying scalar operators in the theory: $\{\phi,\phi^2,\phi^3\}$, in 2d denoted $\{\sigma,\epsilon,\sigma'\}$. We design a set of gap assumptions to single out this theory from other theories or solutions to crossing -- in particular we impose a large gap in the $\Z_2$-odd sector after $\phi^3$, motivated by the fact that $\phi^5$ is missing in the spectrum and the next $\Z_2$-odd operator, perturbatively, is $\phi^7$.\footnote{Moreover, by including $\phi^2$ and $\phi^3$ as external operators, our system should be sensitive to the presence or not of the operator $\phi^5$, knowing that the corresponding free-theory OPE coefficient $\lambda^2_{\phi^2\phi^3\phi^5}=10$ is large.} With this setup, we find bootstrap islands in $d=2.75$ and $d=2.5$ dimensions, with scaling dimensions rigorously confined to the intervals
\begin{align}
\label{eq:island275}
d&=2.75: &    \Delta_{\phi} &=0.375405(145),&    \Delta_{\phi^2} &=0.75511(121),&    \Delta_{\phi^3} &= 1.2181(128),
\\
\label{eq:island25}
d&=2.5: &    \Delta_{\phi} &=0.25574(155),&    \Delta_{\phi^2} &=0.5344(57),&    \Delta_{\phi^3} &= 1.029(25),
\end{align}
shown also in figures~\ref{fig:islands} and \ref{fig:islandsLog}. These intervals represent the maximal extent along the three coordinate axes of our bootstrap islands. We have not mapped out the precise shape of these islands, instead we used the recent ``navigator'' approach \cite{Reehorst:2021ykw} to work out their delimitations. The navigator was also used for locating the islands in parameter space, as illustrated in figure~\ref{fig:nav-275}.

In $d=2$ and $d=2.25$, our navigator search did not terminate but drifted away, indicating that with the assumptions used and at the present derivative-order ($\Lambda=19$), the numerics is not strong enough to isolate the theory to an island, but rather to a peninsula. This happened even when starting from the exact values in $d=2$: $\Delta_{\phi}=\frac3{40}$, $\Delta_{\phi^2}=\frac15$, $\Delta_{\phi^3}=\frac78$ and using somewhat stringent gap assumptions. 
In order to convert these peninsulas to actual islands in $d=2$ and $d=2.25$, we would need to either increase the derivative-order, or consider scanning over more observables/parameters. We discuss the latter in section~\ref{sec:sevenParams}.

In the body of the paper, we review the tricritical Ising CFT, discuss strategies and general lessons for the bootstrap, which will pertain to future studies, and present our results. 
In section~\ref{sec:triIsing} we outline the experimental picture, before gathering known results from the literature in the $3-\eps$ expansion and in $d=2$. We also present a small collection of new order-$\eps$ anomalous dimensions computed using the one-loop dilatation operator. 
In section~\ref{sec:bootstrapApproach}, we give an instrumental view of the bootstrap, emphasising the functionality of recent computational frameworks (specifically we use \texttt{Simpleboot} by Ning Su \cite{simpleboot}) which automatise many technical parts of the implementation, allowing the user to focus on the physically meaningful aspects. In section~\ref{sec:results} we explain in detail our setup and present all the results. Finally, in section~\ref{sec:disc} we give an outlook covering several proposed directions.
With some minimal prerequisites, the sections \ref{sec:triIsing}, \ref{sec:bootstrapApproach}, \ref{sec:results} and \ref{sec:disc} of this paper can be read independently.

\section{The tricritical Ising CFT}
\label{sec:triIsing}

In this section, we review various facts about the tricritical Ising CFT. We take a non-perturbative perspective on the theory, viewing a CFT as a set of conformal primary operators with consistent $n$-point correlators. The experimentally measurable critical exponents are related to the scaling dimensions of certain low-lying operators in the spectrum. 
Within this non-perturbative perspective, the spectrum can be organised by scaling dimension $\Delta$, spin $\ell$, and global-symmetry representation $R$ of the primary operators. Thus, a presentation of the spectrum would be to give for each $(R,\ell)$, a list of all operators with these quantum numbers ordered by increasing scaling dimension. 

Apart from the spectrum, the data characterising a CFT also contains the OPE coefficients $\lambda_{ijk}$. Together the spectrum and the set of OPE coefficients completely fix two- and three-point functions of the CFT. Higher-point functions can be determined iteratively by the OPE; for the four-point function it leads to the famous conformal block decomposition 
\begin{equation}
\G_{ijkl}(u,v)=\sum_{\O}\lambda_{ij\O}\lambda_{kl\O}g_{\tau_\O,\ell_\O}(u,v),
\end{equation} 
where $g_{\tau,\ell}(u,v)$ are the conformal blocks (depending implicitly also on the combinations $\Delta_i-\Delta_j$ and $\Delta_k-\Delta_l$), $u,v$ are the conformal cross-ratios, and $\tau=\Delta-\ell$ is the twist. 

For the theory studied here, we have access to a perturbative expansion in $d=3-\eps$ dimensions, which allows us to connect the non-perturbative perspective with a perturbative construction of operators. With this in mind, we can switch between the non-perturbative notation (first singlet scalar, second singlet scalar etc.) and a perturbative notation ($\phi^2$, $\phi^4$, etc.).\footnote{When perturbative extrapolations cross, operators are expected to repel non-perturbatively \cite{Korchemsky:2015cyx} meaning that we lose control of perturbative naming conventions. By inspecting perturbative estimates, we note that the first few levels to not appear to cross (see figures~\ref{fig:spec-E-0-2} etc. below), and we can use perturbative description for these low-lying operators. In the Ising CFT, such avoided level crossing involves the third and the fourth $\Z_2$-even scalars \cite{Henriksson:2022gpa}, and in $\mathcal N=4$ SYM at large enough $N$ the first two scalar singlets are expected to repel \cite{Korchemsky:2015cyx,Beem:2016wfs,Bissi:2020jve,Chester:2023ehi}.}

Near the upper critical dimension, operator dimensions have been computed perturbatively, for instance
\begin{align}
\Delta_\phi&=\frac12-\frac\eps2+\frac{\eps^2}{1000}+\left(\frac{2279}{375000}+\frac{27 \pi ^2}{40000}\right)\eps^3+O(\eps^4),
\label{eq:deltaPhiN1}
\\
\Delta_{\phi^2}&=1-\eps+\frac{4\eps^2}{125}+\left(\frac{3866}{46875}+\frac{171 \pi ^2}{10000}\right)\eps^3+O(\eps^4),
\label{eq:deltaPhi2N1}
\end{align}
which are results of a six-loop computation \cite{Hager1999,Hager:2002uq} reviewed below. In two dimensions, the theory is exactly solvable \cite{Belavin:1984vu}, and scaling dimensions take exact values,
\begin{align}
\nonumber
\Delta_\phi|_{2d}&=\frac3{40}=0.075\,, & \Delta_{\phi^2}|_{2d}&=\frac15=0.2\,,&   \Delta_{\phi^3}|_{2d}&=\frac78=0.875\,,\\
   \Delta_{\phi^4}|_{2d}&=\frac65=1.2\,,&  \Delta_{\phi^6}|_{2d}&=3\,.
\end{align}
It is expected that resummation of the results from the $3-\eps$-expansion should reproduce these values in $d=2$ with some accuracy, however with the orders available this has not been successful.

\subsection{Tricritical physics}

The tricritical Ising CFT is physically relevant in the context of systems with $\Z_2$ symmetry and two relevant symmetry-preserving parameters. In the example below, these parameters are temperature and a chemical potential. Considering a two-dimensional phase diagram in these parameters, the parameter space can be divided into different phases distinguished by lines of phase transitions. These transitions are either first-order or continuous, and the continuous transitions are generically described by the (normal) Ising CFT. The tricritical Ising CFT appears when several lines of phase transitions meet.

The canonical example is the Blume--Capel model, also known as the Ising model with vacancies \cite{Blume:1966zz,Capel1966}. Here we follow \cite{Cardy:1996xt}, and define the Hamiltonian of this lattice model as 
\begin{equation}
\label{eq:BCmodel}
\mathcal H=-J\sum_{\langle i,j\rangle} s_is_j+\Delta\sum_i s_i^2-h\sum_i s_i,
\end{equation}
where $\langle i,j\rangle$ denotes the sum over nearest-neighbour lattice sites. Setting $J=1$, the chemical potential $\Delta$ and the temperature $T=1/\beta$ can be taken as the two relevant parameters. Here we consider $h=0$, however including $h\neq0$, gives an interesting three-dimensional phase diagram, see e.g. \cite{Kaufman1981,Cardy:1996xt}. 

\begin{figure}
\centering
\includegraphics{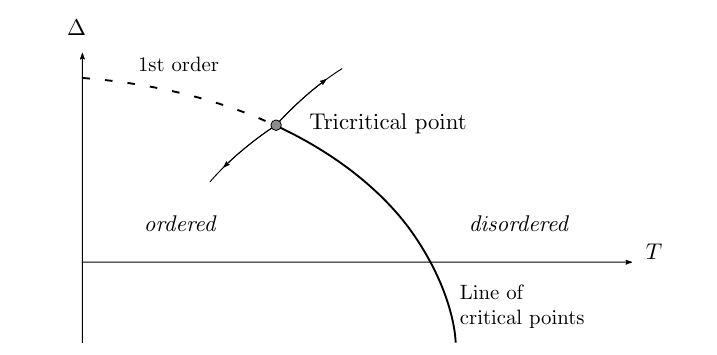}
\caption{Schematic phase diagram of the Blume--Capel model \eqref{eq:BCmodel}. Along the dashed line, the phase transition is first-order; along the solid line, the phase transition is second-order and described by the Ising CFT. The first- and second-order transitions meet at the tricritical point. Figure adapted from \cite{Cardy:1996xt}.}\label{fig:phase-diagram}
\end{figure}

Figure~\ref{fig:phase-diagram} shows the expected phase diagram for the Blume--Capel model \eqref{eq:BCmodel} with $h=0$, displaying an ordered phase and a disordered phase separated by a line of phase transitions. At very low temperature, the transition is first-order,\footnote{At $T=0$, this is the discontinuity fixed-point, which is first-order.} while at higher temperatures the transition is second order and described by the Ising CFT.\footnote{The fact that the second-order transitions are described by the Ising universality class is quoted as ``well established'' in \cite{Moueddene2024}, with references to \cite{Fytas2012,Zierenberg2017}.} At some point the two different transitions meet, and the transition there is tricritical, described by the tricritical Ising CFT in $d<3$.
At this point, tricritical exponents are defined by the following semi-conventional names: 
\begin{equation}
y_h=d-\Delta_\phi=\frac{d+2-\eta}2,\qquad y_t=d-\Delta_{\phi^2}=\nu^{-1}, \qquad y_g=d-\Delta_{\phi^4}\,.
\end{equation}

\begin{figure}
\centering
\includegraphics{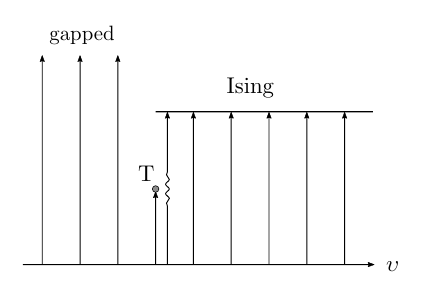}
\caption{RG flow on the phase boundary, parametrised by $\upsilon$. $\mathrm T$ represents the tricritical Ising transition. The wavy line near the point $\mathrm T$ represents a slowdown in the RG flow near the tricritical point, before ultimately flowing towards the Ising (or gapped on the left).}\label{fig:flow-on-line}
\end{figure}

The RG flow on different points along the phase transition line is shown in figure~\ref{fig:flow-on-line}, parametrised by some variable $\upsilon$. To the left of the tricritical point where the transition is first-order, the theory flows to the gapped phase (trivial CFT). To the right, it flows to the Ising CFT. Exactly at the tricritical point, it flows to the tricritical Ising CFT, which stresses that two parameters need to be tuned to reach this point (one more relevant parameter to reach the critical line, another less relevant parameter to reach the specific point of the critical line). In the vicinity of the tricritical point, the RG flow is slow and one can expect the tricritical exponents to describe the systems over a range of scales before finally transitioning to Ising/gapped. 

Other models that also have tricritical fixed-points are the spin-1 Ising model (also known as Blume--Emert--Griffiths model \cite{Blume:1971zza}), directly related to the above, and spin-$\frac12$ Ising metamagnets \cite{Nienhuis1976,Landau1981}. Here metamagnets refer to models with sizeable interactions of both nearest-neighbour and next-to-nearest-neighbour type near the tricritical point.

While the tricritical Ising CFT does not exist in $d=3$, tricritical phenomena can still occur, but are then described by the Gaussian fixed point (free theory). Some references mention logarithmic corrections but their status does not appear to be definite \cite{Maciolek2004}. 
Examples of real-world systems with tricritical fixed-points are $^3\mathrm{He}$--$^4\mathrm{He}$ mixtures
\cite{Blume:1971zza,FarahmandBafi2015} (see also \cite{Maciolek2004}), four-component fluid mixtures near room temperature \cite{Radyshevskaya1962,Lang1975}, and physical metamagnets \cite{Fisher1975,Fisher1975b} (e.g.~$\mathrm{FeCl_2}$ \cite{Shang1980}
$\mathrm{Dy_3Al_5O_{12}}$ \cite{Giordano1975}).

In two dimensions, no experimental determinations of classical tricritical behaviour exists to the best of our knowledge. Quantum-tricritical behaviour (referring to $1+1$ dimensions) was discussed in \cite{Ejima:2016nkr,Maffi:2023tkn}. Instead significant effort has gone into studying the lattice models discussed above, especially the Blume--Capel model \eqref{eq:BCmodel}. Contrary to the usual Ising model (recovered as $\Delta\to-\infty$), the Blume--Capel lattice model has not been exactly solved. 
Simulating it on a lattice, the picture of figure~\ref{fig:phase-diagram} has been confirmed and the tricritical temperature and chemical potential have been estimated around 
\begin{equation}
(T_c,\Delta_c)\approx (0.608,\, 1.966),
\end{equation}
for $J=1$. 
See \cite{Kwak2015} for a collection of estimates and references. 
Tuning to the tricritical point, the (tri)critical exponents can then be estimated and tested against the exact values.\footnote{Some of them were conjectured before the exact solution of the 2d tricritical Ising CFT, see \cite{Nijs1979,Nienhuis:1979mb,Pearson:1980wt,Nienhuis1982}. Note that the tricritical exponents do not satisfy all of the usual scaling laws of a critical theory.} Methods to simulate the Blume--Capel model include Monte Carlo RG
\cite{Landau1981,Landau1986}, transfer matrix
\cite{Beale1986,Xavier1998}, and the Wang--Landau method
\cite{Silva2006,Kwak2015}; see also \cite{Graham2006} for some formal aspects.

\subsection{Review of perturbative results}

In this section we summarise the set of perturbative results available in the literature, presenting a picture of the spectrum valid in $d-3-\eps$ dimensions. 
In the limit $\eps\to0$, the spectrum of the tricritical model agrees with that of a free scalar, up to a few exceptions implied by the equations of motion. The scaling dimensions and OPE coefficients are then perturbations from the free-theory values, given in an (asymptotic) series in $\eps$:
\begin{equation}
\Delta_\O = \Delta_\O^{\mathrm{free}}+\gamma_\O(\eps), \qquad \lambda_{ijk}= \lambda_{ijk}^{\mathrm{free}}+\lambda_{ijk}^{(1)}\eps+\ldots.
\end{equation}
The anomalous dimensions $\gamma_\O(\eps)=\gamma_\O^{(1)}\eps+\gamma_\O^{(2)}\eps^2+\ldots$ can be computed by renormalisation of local operators using Feynman diagrammatic computations. OPE coefficients can, in principle, be computed by constructing perturbatively the precise form of the operators and then computing three-point functions, however this is rarely attempted and results for OPE coefficients are sporadic. 

In table~\ref{tab:lowLyingSchematic}, we give an overview of the spectrum of the model. Conformal primary operators are constructed from scalar field $\phi$ and partial derivatives $\de_\mu$, which must be combined in specific ways to satisfy the primary condition. The free-theory dimension of an operator with $k$ fields and $l$ derivatives is $\Delta_\O^{\mathrm{free}}=k\frac{d-2}2+l$. It is convenient to organise the spectrum by twist $\tau=\Delta-\ell$ and spin $\ell$, as presented in table~\ref{tab:lowLyingSchematic}. At twist $\frac12$ (we refer to the values of the twist at $d=3$), there is a single primary operator which is the scalar $\phi$. At higher twists, there are primary operators of unbounded spin, and generically also parity-odd operators.\footnote{These are not exchanged in the bootstrap system considered here, and we do not discuss them in the main text. The first parity-odd operator has spin 4 and dimension $\frac{13}2$, see table~\ref{tab:3dspecOddP} in appendix~\ref{eq:characterDecomp}.}

Among the scalar operators we have pure powers of the field, $\phi^k$, and from dimension $6$ and onwards also operators constructed using contracted derivatives. The operator $\phi^5$ is missing from the spectrum, since the operator at this level is the equation-of-motion operator $\de^2\phi- \lambda\phi^5$, which is redundant.
Moreover, in table~\ref{tab:lowLyingSchematic} we also indicated the expected identification in the 2d tricritical Ising, following \cite{Zamolodchikov:1986db} (see also (7.120) of \cite{DiFrancesco:1997nk}).

For spinning operators, there are some general patterns at low-lying twists, as indicated in table~\ref{tab:lowLyingSchematic}: At $\tau=1$, there is a single operator $\mathcal J_\ell$ at each even spin. These are conserved in the free theory, but for $\ell>2$ acquire anomalous dimensions \eqref{eq:twist-1-ops} at order $\eps^2$ in the interacting theory. Also at $\tau=\frac32$, the leading anomalous dimension is at order $\eps^2$, but in this case they have not been computed. Starting from $\tau=2$, spinning operators generically acquire $O(\eps)$ anomalous dimensions.

\begin{table}\caption{Structure of the perturbative spectrum of conformal primary operators, organised by twist $\tau=\Delta-\ell$ (in $d=3$) and spin $\ell$. }
\label{tab:lowLyingSchematic}
\centering
\small
\def\arraystretch{1.5}
\begin{tabular}{|l|p{0.4\textwidth}|p{0.4\textwidth}|}
\hline
\textbf{Twist} & \textbf{Scalar operators} & \textbf{Spinning operators}
\\\hline
$1/2$ & 
$\phi\to\sigma(\frac3{40})$. Fundamental field/order parameter. $\gamma_\phi$ known to $O(\eps^3)$ \eqref{eq:deltaPhiN1}.
&
\emph{No primary operators}
\\\hline
$1$ & 
$\phi^2\to\epsilon(\frac15)$. Mass/energy operator. $\gamma_{\phi^2}$ known to $O(\eps^3)$ \eqref{eq:deltaPhi2N1}.
&
$\mathcal J_\ell = \phi\de^\ell\phi$ at even $\ell$. $\mathcal J_2^{\mu\nu}=T^{\mu\nu}$. $\gamma_\ell$ known to $O(\eps^2)$ \eqref{eq:twist-1-ops}. 
\\\hline
$3/2$ & 
$\phi^3\to\sigma'(\frac78)$. Subleading order parameter. $\gamma_{\phi^3}$ known to $O(\eps^2)$ \eqref{eq:gammaphi3}. 
&
$\de^\ell\phi^3$. $\Sigma_i d_iq^i=\frac1{(1-q^2)(1-q^3)}$. $\gamma=O(\eps^2)$ unknown.
\\\hline
$2$ & 
$\phi^4\to\epsilon'(\frac65)$. Subleading energy operator. $\gamma_{\phi^4}$ known to $O(\eps^3)$ \eqref{eq:gammaphi4}.
&
$\de^\ell\phi^4$. $\Sigma_i d_iq^i=\frac1{(1-q^2)(1-q^3)(1-q^4)}$. One operator per spin has non-zero $\gamma$ at $O(\eps)$ \eqref{eq:twist-2-ops}.
\\\hline
$5/2$ & 
\emph{No primary operator (due to equation of motion)}
&
\multirow{2}{*}{\parbox{0.36\textwidth}{\vspace{8pt}\emph{Generic structure. $\gamma$ at $O(\eps)$ can be determined using one-loop dilatation operator, sec.~\ref{sec:OneLoopDil}.}}}
\\\cline{1-2}
$3$ & 
$\phi^6\to\eps''(3)$. Irrelevant deformation. $\gamma_{\phi^6}$ known to $O(\eps^3)$ \eqref{eq:gammaphi4}.
&
\\\cline{1-2}
$>3$ & 
$\phi^k$ operators \eqref{eq:gammaphiK}. First primary with derivative $\square^2\phi^4$ at $\Delta=6$ \eqref{eq:gammabox2phiK}. 
&
\\\hline
\end{tabular}
\end{table}

In tables~\ref{tab:lowlyingE} and \ref{tab:lowlyingO} we list all individual operators with $\Delta\leqslant 6$ alongside their anomalous dimensions and OPE coefficients with our external operators $\phi$, $\phi^2$ and $\phi^3$. The results compiled in these tables are a combination of results extracted from the literature, and some leading-order computations performed in section~\ref{sec:OneLoopDil} below.

\subsubsection{Six-loop renormalisation}

In work by Hager and Sch\"afer \cite{Hager1999,Hager:2002uq}, the theory defined by \eqref{eq:actionIntro} was renormalised in dimensional regularisation in $\eps=3-d$. This completely mimics standard renormalisation of $\phi^4$ theories in $\eps=4-d$ \cite{Kleinert:2001hn}, although the Feynman integrals that need to be evaluated differ.\footnote{For an incomplete list of diagrammatic work in $d=3-\eps$ dimensions, see \cite{Minahan:2009wg,Gracey:2016tuh,Badel:2019khk,Jack:2020wvs}.}
While \cite{Hager1999,Hager:2002uq} gave results for the $O(n)$ symmetric generalisation of our theory, we present here the case $n=1$.
The renormalisation of the action \eqref{eq:actionIntro} at six loops gives the following anomalous dimensions at order $\eps^3$:\footnote{\textbf{Note added.} \textsl{Since the publication of this paper, references \cite{Adzhemyan:2026whu,Jack:2026npe} have found mistakes in \cite{Hager:2002uq} on which the expressions below are based. We give the correct expressions from \cite{Adzhemyan:2026whu,Jack:2026npe} in appendix~\ref{app:renormalisation}, whereas no equations in the main text have been updated.}}
\begin{align}
\label{eq:gammaphi1}
\gamma_\phi &=0.001\eps^2+0.012739 \eps^3+O(\eps^4),
\\
\gamma_{\phi^2}&=0.032 \eps^2+0.25124\eps^3+O(\eps^4),
\label{eq:gammaphi2}
\\
\gamma_{\phi^4} &=0.8\eps+2.29397 \eps^2-28.05898 \eps^3+O(\eps^4),
\label{eq:gammaphi4}
\\
\gamma_{\phi^6}&=
4 \eps-13.40598 \eps^2+337.6492 \eps^3+O(\eps^4),
\label{eq:gammaphi6}
\end{align}
where the total scaling dimension is given by $
\Delta_{\phi^k}=k\left(\frac12-\frac\eps2\right)+\gamma_{\phi^k}$. 
The exact expressions behind the numerical values are given in equations~\eqref{eq:gamma1expl}--\eqref{eq:gamma6expl} in the appendix, displaying the dependence on the following numbers: $\pi^2$, $\zeta_3$, $\pi^4$, $\beta_2$, $\beta_4$, $\ln 2$.

Unfortunately, the computation in \cite{Hager1999,Hager:2002uq} does not include any interaction $\frac{u_3}{3!}\phi^3$, so we do not have access to any six-loop result for this operator. 
However, in \cite{ODwyer:2007brp}, the anomalous dimension $\phi^k$ was given on closed form to order $\eps^2$:
\begin{align}
\label{eq:gammaphiK}
\gamma_{\phi^k}=\frac{k(k-1) (k-2) }{30}\eps
 -k\! \left(\tfrac{131 k^4-1010 k^3+1999 k^2-1117 k-18}{15000}+\tfrac{9(k-5) (k-2) (k-1)}{1600}\pi^2 \right)\!\eps^2+O(\eps^3),
\end{align}
where $k\neq5$.\footnote{
It would be desirable to compute this anomalous dimension to the next order. One ingredient in such a derivation would be matching with the predictions from \cite{Antipin:2024ekk} for the highest $k$ power. The coefficients $\frac1{30}$ and $\frac{-131}{15000}$ of the highest $k$ power at each order agree perfectly with \cite{Antipin:2024ekk}. By the extension of \cite{Antipin:2024ekk}, the expression entering \eqref{eq:gammaphiK} at order $\eps^3$ would be $k$ times a degree-$6$ polynomial in $k$ with the coefficient of $k^6$ fixed to $\frac{983}{225000} $ (I thank the authors of \cite{Antipin:2024ekk} for sharing this value).} This evaluates to
\begin{equation}
\label{eq:gammaphi3}
\gamma_{\phi^3}=
0.2\eps+ 
 1.073598\eps^2+O(\eps^3).
\end{equation}
for $k=3$. 

\subsubsection{Pad\'e approximants}

Instead of a directly using the truncated series above, due to their asymptotic nature one typically resorts to resummation methods \cite{LeGuillou:1979ixc}.
Here, with access only to a few orders in the expansion, we make use of the simplest method, namely the Pad\'e approximant, which requires a minimal amount of theory and choices. 
A Pad\'e approximant is an expression
\begin{equation}
\text{Pad\'e}_{m,n}(\eps)=\frac{a_0+a_1\eps+\ldots a_m\eps^m}{1+b_1\eps+b_2\eps^2+\ldots + b_n\eps^n}.
\end{equation}
where the $m+n+1$ constants are fixed by matching with a series expansion to order $\eps^{m+n}$. A slight modification, used here, is to construct ``Pad\'e approximants tied to 2d,'' which means that we use a series expansion to one order less, $O(\eps^{m+n-1})$, and impose that $\text{Pad\'e}_{m,n}(1)$ agrees with the known value at $\eps=1$. 

Applying this procedure for instance to $\Delta_\phi$, we find
\begin{equation}
\label{eq:PadePhi}
\text{Pad\'e}_{2,2}[\Delta_\phi](\eps)=\frac{\frac{1}{2}-\frac{8579056-34425 \pi ^2}{8898000}\eps+\frac{418477-2835 \pi ^2}{889800}\eps^2}{1-\frac{4130056-34425 \pi ^2 }{4449000}\eps+\frac{15272+2025 \pi ^2}{1483000}\eps^2}=\frac{0.5-0.925972 \eps+0.438859 \eps^2}{1-0.851943\eps+0.0237747 \eps^2}.
\end{equation}
Indeed its small-$\eps$ expansion matches \eqref{eq:deltaPhiN1} and it satisfies $\text{Pad\'e}_{2,2}[\Delta_\phi](1)=\frac3{40}$. In the same way, we construct Pad\'e approximants for $\phi^k$, $k\leqslant 7$ using the $O(\eps^3)$ results for $k=1,2,4,6$ and $O(\eps^2)$ from \eqref{eq:gammaphiK} for $k=3,7$. This gives a list of the following Pad\'e approximants, all tied to 2d:
\begin{equation}
\label{eq:Pades}
\text{Pad\'e}_{2,2}[\Delta_\phi], \quad 
\text{Pad\'e}_{2,2}[\Delta_{\phi^2}], \quad 
\text{Pad\'e}_{2,1}[\Delta_{\phi^3}], \quad 
\text{Pad\'e}_{2,2}[\Delta_{\phi^4}], \quad 
\text{Pad\'e}_{1,2}[\Delta_{\phi^6}], \quad 
\text{Pad\'e}_{1,2}[\Delta_{\phi^7}].
\end{equation}
In constructing these, we have made choices for the parameters $m$ and $n$, which are normally chosen to be roughly equal. Sometimes a Pad\'e approximant has spurious poles, for instance the approximant $\text{Pad\'e}_{2,2}[\Delta_{\phi^6}](\eps)$ has a pole at $0.120745335$. In this case we discarded the approximant and chose one of a lower order.

\begin{table}\caption{Even operators with engineering dimension $\leqslant 6$, with leading anomalous dimensions, and OPE coefficients computed from Wick contractions.
In parentheses we show the leading $\eps$ correction where results are available \cite{Codello:2017qek,Codello:2017hhh}.
}
\label{tab:lowlyingE}
\centering
\small
\def\arraystretch{1.15}
\begin{tabular}{|lll|c|r|r|r|r|}
\hline
$\ell$ & $\Delta$ &  $\O$  & $\gamma^{(1)}$ &$\lambda_{\phi\phi\O}^2$ & $\lambda_{\phi^2\phi^2\O}^2$ & $\lambda_{\phi^3\phi^3\O}^2$ & $\lambda_{\phi\phi^3\O}^2$ 
\\\hline
$0$ &$1$ & $\phi^2$                & $0 $       & $2 $       & $8 $       & $ 18$ & $ 3 $     
\\
$0$ &$2$ & $\phi^4$               & $ \frac45$     & $(\frac{3\eps^2}{200}) 
$       & $6 $       & $ 54$ & $4 $   
\\
$0$ &$3$ & $\phi^6$               & $4 $    & $ (\frac{\eps^2}{500})
$       & $0 $       & $20 $ & $(\frac{27 \eps^2}{40}) $   
\\
$0$ &$4$ & $\phi^8$              & $\frac{56}5 $      & $(\frac{7 \eps^4}{16000})$       & $ 0$     & $0 $ & $ 0$     
\\
$0$ &$5$ & $\phi^{10}$          & $24 $       & $(\frac{63 \eps ^4}{25000000})
$       & $ 0$     & $0 $ & $ 0$      
\\
$0$ &$6$ & $\square^2\phi^4$          & $0 $         & $0 $       & $ \frac1{50}$    &  $ \frac9{50}$& $0 $    
\\
$0$ &$6$ & $\phi^{12}$            & $ 44$       & $0 $      & $0 $      & $0 $ & $0 $     
\\\hline
$2$ &$3$ & $T_{\mu\nu}$            & $ 0$        & $\frac3{32} $     & $\frac38 $       & $\frac{27}{32} $ & ---
\\
$2$ &$4$ & $\de^2\phi^4$            & $\frac8{25} $       & $0 $       & $\frac12 $       & $\frac92 $ & $ \frac3{16}$    
\\
$2$ &$5$ & $\de^2\phi^6$             & $\frac{68}{25} $      & $0 $       & $ 0$      & $\frac{81}{32} $ & $ 0$     
\\
$2$ &$6$ & $\square\de^2\phi^4$             & $ 0$      & $0 $      & $ \frac{3}{196} $       & $ \frac{27}{196}$ & $0$    
\\
$2$ &$6$ & $\de^2\phi^8$        & $ \frac{44}5$           & $ 0$      & $ 0$      & $ 0$ & $0 $      
\\\hline
$3$ &$5$ & $\de^3\phi^4$         & $\frac4{35} $          & ---     & ---      & --- & $ \frac5{256}$    
\\\hline
$4$ &$5$ & $\mathcal J^{(4)}$           & $0 $         & $\frac{35}{8192} $      & $ \frac{35}{2048}$       &$\frac{315}{8192} $ & $ 0$   
\\
$4$ &$6$ & $(\de^4\phi^4)_a$             & $ 0$       & $ 0$      & $\frac{9}{1120} $       & $\frac{81}{1120}$ & $ 0$     
\\
$4$ &$6$ & $(\de^4\phi^4)_b$          & $\frac{4}{15} $         & $0 $       & $\frac{1}{35}$       & $\frac{9}{35}$ & $\frac{105}{8192} $     
\\\hline
\end{tabular}
\end{table}

\subsubsection{Further results from the literature}

Here we collect references and additional results for the tricritical Ising CFT in $d=3-\eps$ dimensions. Apart from \cite{Hager1999,Hager:2002uq}, classical references for diagrammatic computations are \cite{Stephen1973,Stephen1975,Lewis:1978zz,Boyanovsky:1979qf,McKeon:1992cs}. For the tricritical $n$-vector model (tricritical $O(n)$ CFT), classical references are \cite{Pisarski:1982vz,Pisarski:1983gn}. 
More recent works on these theories are \cite{ODwyer:2007brp,Basu:2015gpa,Codello:2017qek,Codello:2017epp,Codello:2017hhh,Sakhi:2021uir}. 

The leading anomalous dimensions of the twist-1 operators (broken currents) $\mathcal J_\ell= \phi\de^\ell\phi$ were computed in \cite{Gliozzi:2017gzh} using multiplet-recombination methods (\cite{Rychkov:2015naa})
\begin{equation}\label{eq:twist-1-ops}
\Delta_{\mathcal J_\ell}=1-\eps+\ell+\frac{(\ell^2-4)\eps^2}{125(4\ell^2-1)}+O(\eps^3).
\end{equation}
Together with the results \eqref{eq:gammaphi1}--\eqref{eq:gammaphi4} and \eqref{eq:gammaphiK} above, this represents the complete set of literature values of anomalous dimensions as known to the author. 

\begin{table}\caption{Odd operators with engineering dimension $< 6$, with leading anomalous dimensions, and OPE coefficients computed from Wick contractions and/or by decomposing the free-theory correlators.
We insert, in parentheses, the order $\eps$ correction where available from \cite{Codello:2017hhh}.
}
\label{tab:lowlyingO}
\centering
\small
\def\arraystretch{1.15}
\begin{tabular}{|lll|c|r|r|}
\hline
$\ell$ & $\Delta$ &  $\O$  & $\gamma^{(1)}$ &$\lambda_{\phi\phi^2\O}^2$ & $\lambda_{\phi^2\phi^3\O}^2$
\\\hline
$0$ &$1/2$ & $\phi$                & $0 $       & $ 2$            & $3 $    
\\
$0$ &$3/2$ & $\phi^3$                & $\frac15 $       & $ 3 $            & $ 18 $    
\\
$0$ &$7/2$ & $\phi^7$                & $7 $       & $0 $            & $ (\frac{1889 \eps ^2}{20})$
\\
$0$ &$9/2$ & $\phi^9$                & $\frac{84}5 $       & $0 $            & $ 0$    
\\
$0$ &$11/2$ & $\phi^{11}$                & $33 $       & $0 $            & $0 $    
\\\hline
$2$ &$7/2$ & $\de^2\phi^3$            & $0 $ & $\frac3{20} $     & $\frac9{10} $    
\\
$2$ &$9/2$ & $\de^2\phi^5$            & $\frac{29}{25} $ & $0 $     & $\frac{27}{28} $    
\\
$2$ &$11/2$ & $\de^2\phi^7$            & $\frac{26}5 $ & $ 0$     & $0 $    
\\\hline
$3$ &$9/2$ & $\de^3\phi^3$         & $0 $          & $ \frac5{616}$        & $\frac{15}{308} $    
\\
$3$ &$11/2$ & $\de^3\phi^5$         & $\frac57 $          & $ 0$         & $\frac5{312} $    
\\\hline
$4$ &$11/2$ & $\de^4\phi^3$         & $0 $          & $\frac7{832} $         & $\frac{21}{416} $    
\\\hline
\end{tabular}
\end{table}

OPE coefficients in the free theory ($\eps=0$) can be computed using Wick contractions. For scalar operators $\phi^k$, a combinatorial exercise gives
\begin{equation}
\label{eq:purePhiPowerOPE}
\lambda_{\phi^i\phi^j\phi^k}=\frac{\sqrt{i! j! k!}}{\left(\frac{1}{2} (i+j-k)\right)! \left(\frac{1}{2} (i-j+k)\right)! \left(\frac{1}{2} (-i+j+k)\right)!}.
\end{equation}
For more complicated operators, the Wick contraction method requires knowing the precise form of the operators, and we delay this discussion to section~\ref{sec:OneLoopDil} below. 
For corrections to the OPE coefficients beyond the free theory, only sporadic results are available. We reproduce the following results from \cite{Codello:2017qek} 
\begin{align}
\nonumber
\lambda_{\phi\phi\phi^4}^2&=\frac{3 \eps^2}{200}, & \lambda_{\phi\phi\phi^6}^2&=\frac{ \eps^2}{500}, & \lambda_{\phi\phi^3\phi^6}^2&=\frac{27 \eps^2}{40}
\\\label{eq:CodelloA}
\lambda_{\phi\phi\phi^8}^2&=\frac{7 \eps^4}{16000}, &\lambda_{\phi\phi\phi^{10}}^2&=\frac{63 \eps ^4}{25000000},
\end{align}
and from \cite{Codello:2017hhh}
\begin{align}\label{eq:CodelloB}
\lambda_{\phi\phi^3\phi^4}^2&=4-\frac{24 \eps}{5},& \lambda_{\phi^2\phi^4\phi^4}^2&= 32-\frac{12}{5} \left(32+3 \pi ^2\right) \eps, & \lambda_{\phi^2\phi^2\phi^6}^2&= \frac{256 \eps^2}{125}, &  \lambda_{\phi^2\phi^3\phi^7}^2&=\frac{189\eps^2}{20}.
\end{align}
The last of these expressions forms part of a tower of OPE coefficients $\lambda_{\phi^2\phi^k\phi^{k+4}}$, $k=3,4,\ldots$ given in \cite{Codello:2017hhh}. In writing \eqref{eq:CodelloA}--\eqref{eq:CodelloB} we converted the results of \cite{Codello:2017qek,Codello:2017hhh} to the CFT conventions with unit-normalised position-space two-point functions. 
In section~6.3 of \cite{Henriksson:2020jwk}, the OPE coefficients of the broken currents $\phi\de^\ell\phi$ in the $\phi\times\phi$ OPE were computed to order $\eps^2$, we print these in \eqref{eq:OPEJell} in appendix~\ref{app:large}. The case $\ell=2$ gives the following result for the central charge 
\begin{equation}
\frac{C_T}{C_{T,\mathrm{free}}}=1-\frac{4\eps^2}{1875}+O(\eps^3),
\end{equation}
where $C_{T,\mathrm{free}}=\frac d{d-1}$ in our conventions. 

An alternative way to obtain some OPE coefficients in the free theory is to perform the conformal block decomposition of free-theory correlators, which can be computed using Wick contractions for the external field. The computation is standard, where the most difficult step is the conformal blocks in three dimensions for unequal external operators. These blocks can for instance be computed using the subcollinear expansion explained in appendix~A of \cite{Bertucci:2022ptt}, or by the radial expansion \cite{Kos:2013tga,Hogervorst:2013sma}. 
As a simple example, we find\footnote{The case $\G_{\phi\phi\phi\phi}(u,v)$ is less interesting and simply gives the free-theory OPE coefficients \eqref{eq:FreeTheoryOPE}.}
\begin{align}
\nonumber
\G_{\phi^2\phi^2\phi^2\phi^2}(u,v)&=1+u+\frac{u}{v}+4 \sqrt{u}+\frac{4 u}{\sqrt{v}}+\frac{4 \sqrt{u}}{\sqrt{v}}=1+8g_{1,0}+\frac38g_{1,2}+\frac{35}{2048}g_{1,4}
\\&\quad +\frac{231}{262144}g_{1,6}+\ldots+6g_{2,0}+\frac12 g_{2,2}+\frac{41}{1120}g_{2,4}+\ldots ,
\label{eq:CBdec}
\end{align}
where $g_{\tau,\ell}$ represent the conformal blocks of operators with dimension $\Delta=\tau+\ell$ and spin $\ell$. For non-degenerate operators, the OPE coefficients $\lambda_{\phi^2\phi^2\O}^2$ can now be read off; compare \eqref{eq:CBdec} with table~\ref{tab:lowlyingE}. A case with degeneracy is $(\tau,\ell)=(2,4)$, where we can check that the individual OPE coefficients for $\O_a$ and $\O_b$ computed in \eqref{eq:OPEab1}--\eqref{eq:OPEab2} below sum up to $\frac{41}{1120}$.

\subsection{New perturbative results from one-loop dilatation operator}
\label{sec:OneLoopDil}

In this section we are concerned with anomalous dimensions of arbitrary composite operators. In principle these can be computed systematically to any loop order using diagrams, however we are not aware of any such systematic computation in $3-\eps$ dimensions. Leaving multiloop results to future work, we use here a convenient method to derive leading-order results,\footnote{We refer to this as one-loop results since the computation is leading order, however diagrammatically they would be the result of a two-loop computation.} which makes use of conformal perturbation theory \cite{Cardy:1996xt}, section~5.2, see also \cite{Komargodski:2016auf,Amoretti:2017aze}. 

Consider a UV CFT perturbed by an operator $\alpha \O_{\mathrm{int}}$ such that there is an IR fixed-point at $\alpha=\alpha_*\ll1$. Then the leading correction to the anomalous dimension matrix $\tensor\Gamma{_i^j}$ is proportional to the OPE coefficient $\tensor\lambda{_{i\O_{\mathrm{int}}}^j}$, where the index is raised with the two-point function. Here the UV CFT is the free theory, the IR CFT the tricritical Ising CFT, the interaction is
 $\O_{\mathrm{int}}=\phi^6$ and we have $\alpha\sim\lambda \sim \eps$. 
Taking combinatorial and kinematic factors into account, we find the formula,
\begin{equation}
\label{eq:gammaIJmat}
\tensor\Gamma{_i^j} = \frac{\eps}{\sqrt{500}}\tensor\lambda{_{i\,\phi^6}^j}. 
\end{equation}
As a direct example, for the operator $\phi^k$, this formula together with \eqref{eq:purePhiPowerOPE} gives the expression $\gamma_{\phi^k}=\frac{k(k-1)(k-2)\eps}{30}$, in agreement with the leading term in \eqref{eq:gammaphiK}.

For more complicated operators, using \eqref{eq:gammaIJmat} requires a few steps, similar to section~4.2 of \cite{Henriksson:2022rnm}: 1) write a basis of operators with a given scaling dimension and spin, 2) compute the set of OPE coefficients $\tensor\lambda{_{i\,\phi^6}^j}$ involving these operators, 3) diagonalise the anomalous dimension matrix, and 4) identify primary operators among the eigenvectors. 
For the second step rather than using direct Wick contractions, it is convenient to use the implementation from \cite{Hogervorst:2015akt,Hogervorst:2015tka}, which directly gives $\tensor\lambda{_{i\,\O_\mathrm{int}}^j}$ with the $j$ index raised, implying that the basis elements chosen in the first step do not need to be unit-normalised. 

We give one example, dimension-$6$ operators of spin 4 with four fields. We take as basis,
\begin{align}
\nonumber
\O_1&=\phi^3(\alpha.\de)^4\phi\,,& 
\O_2&=\phi^2(\alpha.\de)\phi(\alpha.\de)^3\phi\,, & 
\O_3&=\phi[(\alpha.\de)\phi]^2(\alpha.\de)^2\phi\,, \\
\O_4&=\phi^2[(\alpha.\de)^2\phi]^2 ,& 
\O_5&=[(\alpha.\de)\phi]^4 , 
\end{align}
where $\alpha.\de=\alpha^\mu \de_\mu$ for an ancillary null vector $\alpha$, $\alpha^2=0$. Evaluating the OPE coefficients using the method of \cite{Hogervorst:2015akt,Hogervorst:2015tka}, we find the anomalous dimension matrix
\begin{equation}
\label{eq:gammamat}
\Gamma=\frac1{525}{\small \begin{pmatrix}
210 & 840 & 1260 & 630 & 0 \\
 10 & 230 & 390 & 60 & 90 \\
 1 & 26 & 162 & 27 & 36 \\
 6 & 48 & 324 & 162 & 0 \\
 0 & 12 & 72 & 0 & 24
\end{pmatrix}},
\end{equation}
suppressing $\eps$. 
The eigenvalues are $\left\{0,\frac{4}{35},\frac{4}{15},\frac{8}{25},\frac{4}{5}\right\}$, and include both primaries and descendants. Either by considering the primary condition (operator annihilated by special conformal transformation), or by identifying the eigenvalues $\left\{\frac{4}{5},\frac{8}{25},\frac{4}{35}\right\}$ as anomalous dimensions of primaries at lower spin ($\ell=0,2,3$), we find two new primary operators (left eigenvectors of \eqref{eq:gammamat}):
\begin{align}
\label{eq:spin4op1}
\gamma&=0: && \O_a\propto -6 \O_3+\O_4+9 \O_5 ,\\
\label{eq:spin4op2}
\gamma&=\frac4{15}: && \O_b\propto -\O_1+28 \O_2-12 \O_3-33 \O_4+18 \O_5.
\end{align}

Performing this computation at a number of spins, we discover a structure in the $O(\eps)$ spectrum of $\phi^4$-type operators:\footnote{This is similar to $\phi^3$ type operators in $\lambda\phi^4$ theory \cite{Kehrein:1992fn,Kehrein:1994ff}. Generalising the pattern to the general $\phi^{2m}$ theory we expect: At $k<m$ fields, $\gamma^{(1)}=0$, with $k=m$ fields, $\gamma^{(1)}=0$ except at spin $\ell=0$, with $k=m+1$ fields, $\gamma^{(1)}=0$ except at for a single operator at each spin $\ell\neq1$, and for $k\geqslant m+2$, operators generically acquire $O(\eps)$ anomalous dimensions.} Exactly one operator at each spin has a non-vanishing anomalous dimension at order $\eps$, which takes the following values:
\begin{equation}
\label{eq:twist-2-ops}
\gamma^{(1)}_{\neq0}=\frac15+\frac{3(-1)^\ell}{10(\ell+\frac12)}, \qquad \ell=0,2,3,4,5,\ldots.
\end{equation}
The first instances are $\frac45,\frac{8}{25},\frac{4}{35},\frac{4}{15},\frac{8}{55},\frac{16}{65},\frac{4}{25},\frac{4}{17},\frac{16}{95}$. In addition to those, we find operators with $\gamma^{(1)}=0$ existing at spins $4,6,7,\ldots$. The total number of operators at spin $\ell$ is given by the generating function shown in table~\ref{tab:lowLyingSchematic}. 
From twist $\frac52$ with five fields and onwards, one arrives at a generic picture, where several operators have non-vanishing $\gamma^{(1)}$, similar to the picture from twist $4$ onwards in $\phi^4$ theory. 

By executing the computations above, we also discover a collection of towers of operators with fixed number of derivatives but increasing number of fields, directly analogous to \cite{Kehrein:1994ff}.\footnote{\textbf{Note added}: \textsl{Some towers of this type have also been discussed in \cite{Gliozzi:2017gzh}, see equation (4.9) there.}} 
For these towers, the anomalous dimensions are given by expressions in closed form in $k$, valid for some $k\geqslant k_0$, where $k_0\leqslant l$ and $l$ is the number of partial derivatives used to construct the operators. We report the following towers:
 \begin{itemize}
\item The operators $\phi^k$, with $\gamma$ given by \eqref{eq:gammaphiK}. 
\item Operators of the form $\de^2\phi^k$:
\begin{equation}
\gamma^{(1)}=\frac{(k-3) (k-2) (5 k+4)}{150},\qquad k=2,3,4,5,6,\ldots.
\end{equation}
The first few cases are $0,0,\frac{8}{25},\frac{29}{25},\frac{68}{25},\frac{26}5,\frac{44}{5},\frac{343}{25}$. 
\item Operators of the form $\de^3\phi^k$:
\begin{equation}
\gamma^{(1)}=\frac{(k-3) (7 k^2-12 k-40)}{210},\qquad k=3,4,6,\ldots.
\end{equation}
The first few cases are $0,\frac{4}{35},\frac{5}{7},\frac{146}{35},\frac{52}{7},\frac{419}{35}$. The case $k=5$ ($\gamma=2$) is excluded from the primary spectrum due to equations of motion. 
\item Operators of the form $\square^2\phi^k$:
\begin{equation}
\label{eq:gammabox2phiK}
\gamma^{(1)}=\frac{(k-4) (k-2) (5 k+3)}{150},\qquad k=4,5,6,\ldots.
\end{equation}
The first few cases are $0,\frac{14}{25},\frac{44}{25},\frac{19}{5},\frac{172}{25},\frac{56}{5}$
\item Two operators of the form $\de^4\phi^k$, with anomalous dimensions
\begin{equation}
\label{eq:spin4Gen}
\gamma^{(1)}= \frac{35 k^3-182 k^2+32 k+684\mp \sqrt{49 k^4+812 k^3+37228 k^2-287184 k+508176}}{1050}.
\end{equation}
These exist for $k=2,3,4,5,\ldots$ (upper sign), and for $k=4,5,\ldots$ (lower sign) respectively.
\end{itemize}
For all of these towers, the leading large-$k$ behaviour is always $k^3/30$.

The diagonalisation of the anomalous dimension matrix gives not only the eigenvalues, but also the leading form of the eigenvectors. They can then be used to compute unit-normalised OPE coefficients by Wick contractions. 
For instance, for the operators \eqref{eq:spin4op1}--\eqref{eq:spin4op2}, we find
\begin{align}
\label{eq:OPEab1}
\lambda_{\phi^2\phi^2\O_a}&=\frac{3}{\sqrt{1120}}\,, & \lambda_{\phi^3\phi^3\O_a}&=\frac9{\sqrt{1120}}\,, &  \lambda_{\phi\phi^3\O_a}&=0\,,
\\
\label{eq:OPEab2}
\lambda_{\phi^2\phi^2\O_b}&=\frac{1}{\sqrt{35}}\,, & \lambda_{\phi^3\phi^3\O_b}&=\frac{3}{\sqrt{35}}\,, &  \lambda_{\phi\phi^3\O_b}&=\frac{\sqrt{210}}{128}\,,
\end{align}
as reported in table~\ref{tab:lowlyingE}.\footnote{Note that $\lambda_{\phi\phi^3\O_a}=0$ where $\gamma^{(1)}_{\O_a}=0$. It is interesting to compare with $\phi^3$ type operators for Ising. In \cite{Bertucci:2022ptt} it was found that only the operators (one per spin) with non-zero one-loop anomalous dimensions have non-zero tree-level OPE coefficient in $\phi\times \phi^2$. This suggests a general vanishing of tree-level OPE coefficients for $\phi^{k+1}$ type operators with $\gamma^{(1)}=0$, in the OPE $\phi\times \phi^{k}$ in $\phi^{2k}$ theory.}

\subsection{Two-dimensional minimal model}

We now review the case $d=2$, where the tricritical CFT becomes the two-dimensional minimal model $\mathcal M_{5,4}$. The material here is standard and can for instance be found in~\cite{DiFrancesco:1997nk}; important original references are \cite{Zamolodchikov:1986db,Lassig:1990xy}. 
The 2d tricritical Ising model has various interesting properties, such as Kramers--Wanniers duality and integrable RG flows, see \cite{Lassig:1990xy,Delfino:1995zk,Frohlich:2006ch} and \cite{Zamolodchikov:1987ti,Zamolodchikov:1991vx,Zamolodchikov:1991vh,Fendley:1993xa} for some references. Here we focus on aspects related to the spectrum, OPE, and four-point correlators.

In a two-dimensional CFT, the equivalent of conformal primaries are called quasi-primaries, and belong to Virasoro multiplets. In a minimal model, there is a finite number of multiplets, each consisting of a Virasoro primary and an infinite number of Virasoro descendants. 
The Virasoro primaries are organised in a Kac table, e.g. table~\ref{tab:minimalModel}.
In this table, we have identified three sets of names for the Virasoro primaries: Kac labels $\phi_{p,q}$; conventional names $\mathbb I$ (identity), $\sigma$ ($\Z_2$-odd) and $\epsilon$ ($\Z_2$-even), where primes denote subleading operators in the respective global symmetry irrep, and the identification with the $3-\eps$ expansion through the names $\phi^k$.

\begin{table}\caption{Kac table showing the Virasoro primaries in the minimal model. Based on (7.120) of \cite{DiFrancesco:1997nk}.}
\label{tab:minimalModel}
\centering
\small
\def\arraystretch{1.5}
\begin{tabular}{||c|c|c|c||}
\hline\hline
     $ \phi_{3,1} =\epsilon''(3)\sim \phi^6$          &     $ \phi_{3,2}=\epsilon'(\frac65) \sim \phi^4$         &           $ \phi_{3,3}=\epsilon(\frac15)  \sim \phi^2$  &    $ \phi_{3,4}=\mathbb I(0)  \sim \1$  
\\\hline
$\phi_{2,1}= \sigma'(\frac78)  \sim \phi^3$     & $ \phi_{2,2}=\sigma(\frac3{40})\sim\phi$                   &       $ \phi_{2,3}=\sigma(\frac3{40})\sim\phi$            &   $\phi_{2,4}= \sigma'(\frac78)  \sim \phi^3$   
\\\hline
  $\phi_{1,1}=\mathbb I(0) \sim \1$         &       $ \phi_{1,2}=\epsilon(\frac15)  \sim \phi^2$        &      $ \phi_{1,3}=\epsilon'(\frac65) \sim \phi^4$            &          $ \phi_{1,4} =\epsilon''(3)\sim \phi^6$  
\\\hline\hline
\end{tabular}
\end{table}

Among the Virasoro primaries, there are restrictions on which operators have non-zero OPE coefficients, determined by fusion rules: 
\begin{align}
\sigma\times\sigma& =\mathbb I+\epsilon+\epsilon'+\epsilon'',
&
\sigma\times \epsilon &=\sigma+\sigma',
&
\epsilon\times\epsilon&=\mathbb I+\epsilon',
 \quad \text{etc.}
\end{align}
These lead to additional conditions beyond those imposed by $\Z_2$ symmetry, for instance, $
\lambda_{\epsilon\epsilon\epsilon}=0
$, which are not expected to be valid at generic $d>2$ (for instance, in the free 3d theory we have $\lambda^2_{\phi^2\phi^2\phi^2}=8$).
The complete set of fusion rules with OPE coefficients is
\begin{align}
\sigma\times\sigma& =\mathbb I+A\epsilon+\frac A6\epsilon'+\frac1{56}\epsilon'',
&
\sigma\times \epsilon &=A\sigma+\frac12\sigma',
\nonumber
&
\sigma'\times\sigma&=\frac12\epsilon+\frac34\epsilon',
\\
\sigma'\times\epsilon &=\frac12 \sigma\,,
\nonumber
&
\sigma'\times\sigma'&=\mathbb I+\frac78\epsilon'',
&
\epsilon\times\epsilon&=\mathbb I +\frac{2A}3\epsilon',
\nonumber
\\
\sigma\times \epsilon' &= \frac A6\sigma+\frac34\sigma',
&
\epsilon\times \epsilon' &= \frac{2A}3\epsilon+\frac37\epsilon'',
\nonumber
&
\sigma'\times \epsilon' &=\frac34 \sigma\,,
\\
\epsilon'\times\epsilon' &= \mathbb I+\frac{2A}3\epsilon',
\nonumber
&
\sigma\times \epsilon'' &= \frac1{56}\sigma\,,
&
\epsilon\times \epsilon'' &= \frac37\epsilon',
\nonumber
\\
\sigma'\times \epsilon'' &=\frac78 \sigma',
&
\epsilon'\times\epsilon'' &=\frac37\epsilon\,,
&
\epsilon''\times\epsilon''&=\mathbb I,
\label{eq:FusionRules}
\end{align}
where 
\begin{equation}
\label{eq:Aconst}
A=\sqrt{\frac{\Gamma(4/5)\Gamma(2/5)^3}{\Gamma(1/5)\Gamma(3/5)^3}}=0.915453111397\ldots.
\end{equation}
We (re)computed the OPE coefficients using formula (A.5) from \cite{Poghossian:2013fda}, taken from \cite{Poghossian:1989plv} and valid for the choice of Kac labels compatible with (A.3) in the same paper, however original references are \cite{Dotsenko:1984ad,Dotsenko:1985hi}.

The torus partition function takes the well-known form of a diagonal modular invariant of a minimal model CFT:
\begin{equation}
\label{eq:pfmn}
Z_{m,n}=\sum_{r,s} \chi_{r,s}(\tau)\overline{\chi_{r,s}(\tau)}.
\end{equation}
where $\chi_{r,s}$ are the characters of the minimal models. 
For the tri-critical minimal model (and periodic boundary conditions), the diagonal is the only modular invariant, and the sum is over the states of table~\ref{tab:minimalModel}.
For higher minimal models, there are additional modular invariants. 
In appendix~\ref{eq:characterDecomp}, we give more details in the partition function \eqref{eq:pfmn} and discuss the decomposition in quasiprimaries. 

Using Virasoro conformal blocks, one can write down four-point correlation functions of Virasoro primaries. For instance the four-point function of $\sigma\sim\phi\sim\phi_{2,2}$ is given by
\begin{align}
\label{eq:Gsigma2d}
\G_{\sigma\sigma\sigma\sigma}^{2d}(z,\bar z)&= \left|\mathscr F_{\1}(z)\right|^2+\lambda_{\sigma\sigma\epsilon}^2\left|\mathscr F_{\phi_{3,3}}(z)\right|^2+
\lambda_{\sigma\sigma\epsilon'}^2\left|\mathscr F_{\phi_{1,3}}(z)\right|^2+\lambda_{\sigma\sigma\epsilon''}^2\left|\mathscr F_{\phi_{3,1}}(z)\right|^2,
\end{align}
where $\mathscr F_{\phi_{p,q}}(z)$ 
are Virasoro conformal blocks for four $\phi_{2,2}$ operators, see appendix~\ref{app:fourpointSigma}. The OPE coefficients entering are
\begin{equation}
\lambda_{\sigma\sigma\epsilon}^2=A^2, \qquad \lambda_{\sigma\sigma\epsilon'}^2=\frac{A^2}{36}, \qquad \lambda_{\sigma\sigma\epsilon''}^2=\frac1{3136},
\end{equation} as read off from \eqref{eq:FusionRules}. 
Other correlators are constructed in similar ways; for instance an explicit expression for $\G_{\epsilon\epsilon\epsilon\epsilon}(z,\bar z)$ was given in \cite{Friedan:1984rv,Behan:2017rca} and likewise for $\G_{\sigma'\sigma'\sigma'\sigma'}(z,\bar z)$ in \cite{Maloney:2016kee}.

Finally we note that there is a close relation between the tricritical Ising CFT and the first supersymmetric minimal model \cite{Friedan:1983xq,Friedan:1984rv,Qiu:1986if}, where the supersymmetric model is obtained by a Jordan--Wigner transform, see e.g.~\cite{Hsieh:2020uwb} for a recent treatment. Concrete realisations of this model were discussed in \cite{Grover:2013rc,Rahmani:2015qpa}.

\subsection[Continuation in $d$]{Continuation in $\boldsymbol d$}
\label{sec:continuationInD}

Finally, we would like to comment on the fact that a fundamental assumption in this paper is that there is a way to continue the tricritical Ising CFT across spacetime dimensions in a consistent way. We do not know of any rigorous treatment of this, but it seems likely that such continuation can be made well-defined. Here we outline a scenario for how this could be formalised, capturing also continuation in group parameters. 

We postulate the existence of a family of consistent but non-unitary theories which exist across continuous values of $d$, $\mathcal{CFT}_d$. They contain evanescent operators (operators which vanish identically at some integer dimensions due to trace relations \cite{Buras:1989xd,Dugan:1990df,Hogervorst:2015akt}), and may be defined in the sense of Deligne categories following \cite{Binder:2019zqc}. At some integer dimensions $d=D\in \Z$, we postulate then that there are two theories live on top of each other: One is the specification to $d=D$ of the continuous family: $\mathcal{CFT}_D$. The other is the unitary theory $\texttt{CFT}_D$. The latter is a subsector of the former:
\begin{equation}
\texttt{CFT}_D\subset \mathcal{CFT}_D\,,
\end{equation}
in the sense that whenever the same observable exist in both theories, they agree. Since the OPE is closed within operators in $\texttt{CFT}_D$, it means that as $d\to D$, a subset of operators in $\mathcal{CFT}_d$ forms a closed subsector under the OPE, isomorphic to $\texttt{CFT}_d$. 

The same idea would hold for continuation in $n$ where $n$ is a group-theory parameter. The work \cite{Binder:2019zqc} lists for which symmetry groups this is possible.\footnote{This list contains $O(n)$ but not $SO(n)$. This indicates that only parity-respecting theories (invariant under $O(d)$ and not just $SO(d)$) may be continued in $d$.} We mention some examples where such continuation has been discussed in detail: 
\begin{itemize}
\item 2d $O(n)$ models \cite{Grans-Samuelsson:2021uor,Jacobsen:2022nxs}. $\mathcal{CFT}_{n}$ is defined as a loop gas model for $n\in [-2,2]$. For $n=1$, the theory $\mathcal{CFT}_{1}$ is still a loop model and has observables like the fractal dimension $d_f = \Delta_T$, where $\Delta_T$ is the dimension of the leading operator in the (continued) rank-2 traceless-symmetric representation. On the other hand $\texttt{CFT}_{1}$ is the 2d Ising CFT, in which $d_f$ is not an observable. The limit $n\to0$ is also interesting, since there $\mathcal{CFT}_n$ becomes a logarithmic CFT \cite{Cardy:1999zp,Cardy:2013rqg,Movahed:2004nr,Hogervorst:2016itc}. 
\item One can also consider $d$-dimensional critical $O(n)$ loop models, and simultaneously continue in both $n$ and $d$ \cite{Shimada:2015gda}. The 3d critical $O(n)$ loop model are amendable for simulations, including the case at non-integer values of $n$ \cite{Liu:2012ca}. 
\item Similar to 2d $O(n)$ models are $q$-state Potts models, where the group is the symmetric group $S_q$ which admits a continuation in $q$ \cite{Nivesvivat:2022qvw,Jacobsen:2022nxs}. 
\end{itemize}
We remark that the limit $d\to2$ is expected to be rather special with respect to the outlined story, since the conformal algebra needs is enhanced to the Virasoro algebra. This imposes an infinite number of constraints that can only be achieved by a drastic reorganisation of the spectrum, in particular at large spin. For Ising, the situation was studied numerically in  \cite{Cappelli:2018vir}, and analytically in \cite{Li:2021uki}, however more work is needed to clarify the picture.

For the bootstrap, there is also the complication that the theory is non-unitary away from integer dimensions, as pointed out in \cite{Hogervorst:2015akt}. However, several bootstrap studies have been successfully completed in non-integer dimensions: \cite{El-Showk:2013nia,Cappelli:2018vir,Henriksson:2022gpa,Bonanno:2022ztf} for Ising, and \cite{Sirois:2022vth,Chester:2022hzt,Reehorst:2024vyq,Nakayama:2024iiw} for other theories. In these works, the non-unitarity does not seem to cause any problem, most likely because the unitarity-violating operators are high up in the spectrum and/or couple weakly to the external operators.

\section{The bootstrap approach}
\label{sec:bootstrapApproach}

In this section we give an overview of the numerical conformal bootstrap approach. The content is relatively standard by now, as reviewed in \cite{Poland:2018epd,Chester:2019wfx,PerimeterCourse,Rychkov:2023wsd}.
The purpose of this section is to complement these reviews by emphasising the various options available when designing bootstrap algorithms with a specific problem in mind.

\subsection{Standard formulation}

Consider a set $\mathscr E$ of external operators. 
The starting point of the bootstrap is a set of crossing equations
\begin{equation}
\label{eq:theCrossingEqs}
\left\{
0=\sum_{\mathcal O\in i\times j\,\cap\, k\times l} \lambda_{ij\O}\lambda_{kl\O} g^{ijkl}_{\Delta_\O,\ell_\O}(u,v) -
\sum_{\mathcal O\in i\times l\,\cap\, k\times j} \lambda_{il\O}\lambda_{kj\O} \mathbf C_{ijkl}(u,v) g^{ilkj}_{\Delta_\O,\ell_\O}(v,u) 
\right\}_{i,j,k,l\in \mathscr E}\!,
\end{equation}
where $\mathbf C_{ijkl}(u,v)$ is a standard kinematic factor. 

The main approach, canonised after the development of mixed-correlator bootstrap \cite{Kos:2014bka,Kos:2016ysd}, is to rewrite this set of crossing equations on the form
\begin{equation}
\label{eq:masterBootstrapEquation}
\vec V_{\text{OPE}}+\sum_Q\sum_{\O\in Q} \vc{\lambda}_Q^T\, \vec {\mtrx V}_{Q}\,\vc \lambda_Q=0\,.
\end{equation}
We refer to this as the bootstrap equation. 
In this expression, $Q$ denotes the quantum numbers of the exchanged states, which will be spin and global symmetry representations, $\vc\lambda_Q=(\lambda_{ij\O})_{(ij)}$ is a vector of OPE coefficients for the pairs $(ij)$ of external operators whose OPE contains operators with quantum numbers $Q$. 
$\vec{\mtrx V}_Q$ is a vector of matrices which contains (in a rather sparse way) the sums/differences of conformal blocks, and $\vec V_{\text{OPE}}$ is a specific contraction involving the external operators and will be discussed below. 

To illustrate this, consider the mixed two-operator system used to find the 3d Ising island in \cite{Kos:2014bka}. In this case, $\mathscr E=\{\phi,\phi^2\}$, we have $Q\leftrightarrow (R,\ell)$, where $R=\pm$ is the $\Z_2$ representation. For $R=+$, only even spins contribute. The bootstrap equation \eqref{eq:masterBootstrapEquation} becomes\footnote{Here we have assumed a truncation in spin defined by including all spins up to $\ell_{\mathrm{max}}$ and no subsequent ones. This can be replaced by any finite set of spins $\mathscr L_Q$.}
\begin{equation}
\label{eq:mixedCorrelatorKos}
\sum_{\ell=0,2,\ldots}^{\ell_{\mathrm{max}}}\sum_{\mathcal O\in (+,\ell)}\left(\lambda_{\phi\phi\O},\lambda_{\phi^2\phi^2\O} \right)\vec{\mtrx V}_{(+,\ell)}\begin{pmatrix}
\lambda_{\phi\phi\O}\\\lambda_{\phi^2\phi^2\O}\end{pmatrix}
+
\sum_{\ell=0,1,\ldots}^{\ell_{\mathrm{max}}}\sum_{\mathcal O\in (-,\ell)}\lambda_{\phi\phi^2\O}^2 \vec V_{(-,\ell)}
=0\,.
\end{equation}
This is exactly equation (3.11) of \cite{Kos:2014bka}, and explicit expressions for the vectors $\vec{\mtrx V}_{(+,\ell)}$,  $\vec V_{(-,\ell)}$ can be found there. 
The system used for the main computations of the present paper is rather big and is shown in appendix~\ref{app:technicalDetails}.

The elements in the bootstrap equation~\eqref{eq:masterBootstrapEquation} depend on the scaling dimensions $\Delta$ of the operators $\O\in Q$, which we will assume lie in some set of intervals $\Delta\in \mathscr D_Q$ . The main idea is then to act on the bootstrap equation \eqref{eq:masterBootstrapEquation} with linear functionals. If we can find a functional that is positive on all intervals, equation \eqref{eq:masterBootstrapEquation} cannot be satisfies and the theory with spectrum $\bigcup_Q\mathscr D_Q$ is ruled out. 

The minimal choice for $\mathscr D_Q$ is that of unitarity
\begin{equation}
\label{eq:UBregions}
\mathscr D_Q^{\text{unitarity}} = [\Delta_{\text{u.b.}},\infty)\,, \qquad \Delta_{\text{u.b.}}=\begin{cases}
\frac{d-2}2\,,& \ell=0\,,
\\
d-2+\ell\, ,& \ell>0\,.
\end{cases}
\end{equation}
To use a different lower limit $\Delta_*$ of this interval is called imposing a ``gap assumption,'' $\mathscr D_Q=[\Delta_*,\infty)$, however it is also possible to use a collection of disjoints points or intervals inside $ [\Delta_{\text{u.b.}},\infty)$. The scalar singlet sector should also allow for the identity operator $\Delta=0$, however it is often separated and used as a normalisation (see below).

The linear functionals that we will use are denoted $\vec \alpha$, where each component is a set of functionals defined by evaluating derivatives $\de_z^m\de_\zb^n$ at the crossing symmetric point $z=\bar z=\frac12 $, with $m+n \leqslant \Lambda$. The goal is to find for functionals that are positive on all $\mathscr D_Q$; this search can be performed by a semidefinite problem (SDP) solver such as \texttt{SDPB} \cite{Simmons-Duffin:2015qma,Landry:2019qug}, now standard in conformal bootstrap.  
If we study a problem depending on some parameters $p\in \mathscr P$, finding a positive functional then rigorously rules out the point $p$ in parameter space. By arguments of continuity, in this way regions of parameter space can be ruled out. 
Positivity is interpreted as positive-semidefiniteness of the matrices $\vec \alpha\cdot [\vec {\mtrx V}_Q]$ for all $Q$. Then it follows that for any unknown (but real) vector $\vc \lambda_Q$ of OPE coeffients, $\vc \lambda_Q^T\vec \alpha\cdot [\vec{\mtrx V}_Q]\vc \lambda_Q\geqslant 0$.

We now return to $\vec V_{\text{OPE}}$, sometimes called OPE angle block, needed to impose the non-degeneracy of the external operators in the spectrum \cite{Kos:2016ysd}. It is given by
\begin{equation}
\vec V_{\text{OPE}} = \vc \lambda_{\text{ext}} ^T\vec {\mtrx V}_{\text{ext}}\vc \lambda_{\text{ext}}\,,
\end{equation}
where $\vc \lambda_{\text{ext}}$ is a vector consisting of all allowed OPE coefficients involving only external operators. For instance, in the mixed correlator system \eqref{eq:mixedCorrelatorKos}, this vector would read $\vc \lambda_{\text{ext}}=(\lambda_{\phi\phi\phi^2},\lambda_{\phi^2\phi^2\phi^2})^T$. The OPE angle block can be treated in different ways: 1) agnostic to external OPE coefficients, in which we impose positive-semidefiniteness $\vec \alpha\cdot[\vec {\mtrx V}_{\text{ext}}]\succcurlyeq0$, 2) fix a complete set of ratios of OPE coefficients, in which case we impose the scalar-valued inequality $\vec \alpha\cdot[\vc \lambda_{\text{ext}} ^T\vec {\mtrx V}_{\text{ext}}\vc \lambda_{\text{ext}}]\geqslant 0$ where $\vc \lambda_{\text{ext}}$ is a fixed vector, or 3) partial fixing of ratios of OPE coeffients, in which case we impose $\vec \alpha\cdot[\vec {\mtrx V}'_{\text{ext}}]\succcurlyeq0$ for a smaller matrix $\vec {\mtrx V}'_{\text{ext}}=\mtrx B^T\vec {\mtrx V}_{\text{ext}}\mtrx B$.\footnote{\label{foot:Bmat}For instance, with three OPE coefficients $\vc\lambda=(\lambda_1,\lambda_2,\lambda_3)^T$, choosing matrix $\mtrx B=\left(\begin{smallmatrix}1&r&0\\0&0&1\end{smallmatrix}\right)^T$ implements the fixed ratio $\lambda_2/\lambda_1=r$ while keeping $\lambda_1$ and $\lambda_3$ free.}

The bootstrap solver also requires specifying a normalisation for the functional. It is common to separate out the contribution from the identity operator, to use as a normalisation in the algorithm. It also allows the choice of an objective to maximise. The semidefinite problem is then
\begin{quote}
Find $\alpha$ such that $\vec\alpha \cdot \vec V_{\mathrm{norm}}=1$, and $\vec \alpha \cdot [\vec{ \mtrx V}_Q] \succcurlyeq0$, for all $Q$ and for all $\Delta\in \mathscr D_Q$, and (optionally) maximises $\vec\alpha\cdot \vec V_{\mathrm{obj.}}$. 
\end{quote}
Any parameter that enters the bootstrap equation linearly can be chosen as normalisation and objective. For instance, it is possible to use the contribution from an individual operator at a fixed scaling dimension.

\subsection{Using the bootstrap}

While in principle it is possible to code every step from first principles, here we assume the availability of certain ``frameworks'' that take care of various routine steps.
To use the conformal bootstrap, the logical workflow then looks as following:
\begin{equation}
 \begin{matrix}
& & \underset{\displaystyle \downarrow}{\text{Settings}} & & &
\\
 \raisebox{2pt}{\text{Parameters}} & \longrightarrow & \raisebox{-2pt}{\text{local}} & \longrightarrow &   \raisebox{-2pt}{\text{cluster}} & \longrightarrow &   \raisebox{-2pt}{\text{SDP}}
\\
 \raisebox{-2pt}{\text{Algorithm}} & \longrightarrow & \raisebox{2pt}{\text{ framework }} & \longleftarrow &   \raisebox{2pt}{\text{ framework }} & \longleftarrow &   \raisebox{2pt}{\text{solver}}
\\
& & \overset{\displaystyle \downarrow}{\text{Output}} & & &
\end{matrix}
\end{equation}
Each entry in this chart will be discussed below. The user needs to specify the settings (external operators, crossing equations etc), the parameter space $\mathscr P$, and a scanning algorithm. This is done within a local framework. This local framework communicates with a separate framework running on a computational cluster. It performs resource-consuming steps of the program, and in particular runs the SDP solver. Finally, the output is loaded back into the local framework, and can be studied by the user.

\paragraph{Frameworks}
Executing the bootstrap algorithm requires several necessary blocks of code, such as the computation of (derivatives of) conformal blocks, repackaging the components of the linear functionals, scanning algorithms, job submission on the cluster etc. A framework for the conformal bootstrap is a predefined code that automatises some or all of these ingredients. 

Since the formulation of the systematic approach described above, several frameworks have been developed by various authors. 
Recent frameworks are
 \texttt{Simpleboot} \cite{simpleboot}, \texttt{Autoboot} \cite{Go:2019lke,Go:2020ahx}, and the Hyperion framework (presented in \cite{PerimeterCourse}), while earlier frameworks are found in \cite{Paulos:2014vya,Ohtsuki:2016wim,Behan:2016dtz}.
Here we use \texttt{Simpleboot}. It has the advantage of being easy to use with an intuitive Mathematica interface, and great flexibility. 
A disadvantage is that it uses Mathematica on the cluster, which requires the use of central Mathematica licences. 

\paragraph{Settings}

The settings are a set of bootstrap equations, including the definitions of the sectors $Q$, global symmetry and spins from a set of spins $\mathscr L_Q$. The bootstrap equations can be generated by hand for simple systems, or for several symmetry groups automatically generated by \texttt{Autoboot} \cite{Go:2019lke,Go:2020ahx} and loaded into \texttt{Simpleboot}. 
The settings also specify the derivative order $\Lambda$ as well as parameters used for the SDP solver.

\paragraph{Parameters}

The parameters are properties of the bootstrap setup that can be varied during the execution of a scanning algorithm. These include
\begin{itemize}
\item Scaling dimensions $\Delta_i$ of external operators. 
\item Gap assumptions in certain sectors, or more generally positivity intervals $\mathscr D_Q$ for all sectors. 
\item Parameters entering the OPE angle block $\vec V_{\text{OPE}}$. 
\end{itemize}
There is no canonical distinction between settings and parameters. For instance \texttt{Simpleboot} allows the dimension $d$ to be implemented as a parameter while other frameworks keep $d$ fixed. 
A \emph{linear parameter} is a parameter that enters the bootstrap equation \eqref{eq:masterBootstrapEquation} linearly, and its multiplier can be used as normalisation or objective.

For a given bootstrap algorithm, the parameters are divided into fixed parameters and scanning parameters $\mathscr P=\mathscr P_0 \times \mathscr P_{\text{scan}}$, where a $N$-parameter scan refers to $N=\dim \mathscr P_{\text{scan}}$. 
\paragraph{Algorithms} 
We now review various bootstrap algorithms, all available in \texttt{Simpleboot}. More details and a summary of the developments of these methods and their applications can be found in \cite{Rychkov:2023wsd}. 
\begin{itemize}
\item ``Oracle'' mode, which is the simplest setup where one tests if a point $p$ in parameter space is ruled out: 
\begin{equation}
p\in \mathscr P \Rightarrow \begin{cases}
\text{find $\vec\alpha$} & \Rightarrow \ \text{ruled out},
\\
\text{no $\vec\alpha$} & \Rightarrow\ \text{maybe}.
\end{cases}
\end{equation}
By testing a grid of points, or performing a more sophisticated Delaunay triangulation, this divides the parameter space in allowed and rule-out regions, up to some resolution. 
\item Oracle mode with OPE scan uses the cutting-surface algorithm which implements in an efficient way to use the constraints from the OPE ratios in $\vec V_{\text{OPE}}$ \cite{Chester:2019ifh}. 
\item Optimisation mode. Given two linear parameters $L_1$ and $L_2$, the terms proportional to $L_1$ and $L_2$ can be separated from the rest of \eqref{eq:masterBootstrapEquation}, and it is possible to optimize over ratios $L_1/L_2$. A common choice is $C_T$ minimisation \cite{El-Showk:2014dwa}, which maximises the contribution of the stress-tensor normalised against the contribution from the identity operator, using that the squared OPE coefficients with the stress-tensor are proportional to the inverse central charge. 
\item Oracle or optimisation mode with fixed contributions in the crossing equation. This is achieved by adding vectors $\vec V$ corresponding to specific operators to the normalisation while removing the corresponding points from $\mathscr D_Q$. 
\item The ``Navigator'' \cite{Reehorst:2021ykw} allows a search towards allowed regions in parameter space, see below.
\item The ``Skydive'' algorithm \cite{Liu:2023elz} uses the idea of the navigator but allows to move to the next point in parameter space without fully solving the semidefinite problem at the previous points, thus more rapidly approaching the allowed region.\footnote{Running the skydive algorithm requires choosing the value of several internal parameters. We found it difficult to use the skydive algorithm for the system studied in this paper.}  
\end{itemize}

\paragraph{SDP solver} To solve the semidefinite program, \texttt{Simpleboot} uses the purpose-build solver \texttt{SDPB} \cite{Simmons-Duffin:2015qma,Landry:2019qug}. 
It requires a set of polynomial matrices and specifying a normalisation, and it has an optional choice of specifying an objective.

\subsection{Remarks}
\label{sec:remarksOnBootstrap}

\paragraph{Intuition}
In order to get intuition for why bootstrap bounds are at all possible, the original paper \cite{Rattazzi:2008pe} gives some pedagogical examples. To illustrate the workings of the semidefinite problem solver \texttt{SDPB}, consider the test example in \cite{SDPBmanual}, or, for a more physical example, the minimisation of the ratio of Wilson coefficients $g_3/g_2$ discussed in \cite{Caron-Huot:2020cmc}. Both of these examples admit analytic solutions.

\paragraph{Hot-starting}

A way to speed up the semidefinite problem is to use ``hot-starting'' \cite{Go:2019lke}. In this case, the solver uses a check-point from a nearby point in parameter space in order to cut a substantial number of internal steps in \texttt{SDPB}. This is implemented in \texttt{Simpleboot}.

\paragraph{Extremal functional method}

If a bootstrap bound has been found using an objective (so not in the ``oracle'' mode), the resulting \emph{extremal functional} carries meaningful information. To saturate crossing, the extremal functional is expected to have zeros, as a function of $\Delta$, at the positions of operators in the physical spectrum. 
This idea, called the extremal functional method, was first developed in \cite{El-Showk:2012cjh,El-Showk:2014dwa} and has been used in many subsequent works. With high-precision islands, it gives a rather good picture of the low-lying spectrum \cite{Simmons-Duffin:2016wlq,Liu:2020tpf,Chang:2024whx}.

\paragraph{Navigator approach}

In \cite{Reehorst:2021ykw}, a new algorithm was proposed for the bootstrap denoted the \emph{Navigator}, which implements a quasi-Newton search for finding allowed points. Specifically, to use a navigator, one constructs a function that is positive when there is no bootstrap solution, and negative where there is a solution (at a given derivative-order). In \cite{Reehorst:2021ykw}, two such functions were proposed: A physically motivated option is to use the ``GFF navigator,'' in which a generalised-free-field solution to crossing is added to the system with a coefficient that constitutes the navigator function. A second option is an ``automatic'' navigator, denoted the ``sigma navigator'' in \cite{Reehorst:2021ykw}. 

Having constructed the navigator function, reference \cite{Reehorst:2021ykw} proposes a modified Broyden--Fletcher--Goldfarb--Shanno (BFGS) algorithm which is used to ``navigate'' to the minimum of the navigator function, and in this way find allowed points (if the navigator value at the minimum is negative). To determine whether this allowed point is inside an island, one can then maximize/minimize in all scan parameters while keeping the navigator negative. In this paper, we use both approaches with the ``sigma navigator.''

\section{Bootstrap implementation and results}
\label{sec:results}

In this paper, we use the conformal bootstrap with the goal of isolating a specific theory, rather than deriving a set of universal bounds.
The bootstrap has no Lagrangian input, so we need a strategy -- a set of crossing equations to study supplemented by suitable gap assumptions -- which has the potential to single out our target theory. 

In the first subsection, we describe this strategy. We then give results from a preliminary study with a two-correlator system, before moving to our main study in sections~\ref{sec:navigatorRes}--\ref{sec:runoff} where we perform three-operator navigator runs. In section~\ref{sec:extremalSpec} we look at the extremal spectrum, in section~\ref{sec:previousRes} we compare with previous results, and in section~\ref{sec:ResDisc} we discuss possible improvements for future studies.
For all runs, we used \texttt{Simpleboot} \cite{simpleboot} with $\Lambda=19$. 

\subsection{Strategy}

The tricritical Ising CFT in $2<d<3$ dimensions has not been bootstrapped before, so we cannot simply build on improving previous results.\footnote{
Reference \cite{Gowdigere:2018lxz} performed a bootstrap study based on the assumption that $\lambda_{\phi^2\phi^2\phi^2}=0$. They found an interesting coincidence between two bounds at a value for $\Delta_\phi$ compatible with the 2d tricritical Ising theory. This signature continued to exist for $d>2$, but for $d>2$ it is not clear how to interpret the results given that $\lambda_{\phi^2\phi^2\phi^2}$ is generically expected to be nonzero there (compare with \eqref{eq:purePhiPowerOPE}).} We therefore need to develop a novel strategy based on perturbative estimates and experience from bootstrap studies of other systems. Hence, we need to identify a set of assumptions which are only satisfied by the target theory and no other ``competing'' theories or solutions to crossing, at least not in the vicinity of the target (whose precise location may be unknown). With sufficient numerical strength, this would lead to a small island, which may be accompanied by a larger allowed region far away. 

The challenge lies in finding an implementation that singles out the theory, but is also feasible from the constraints of numerical resources.
Aware of the computational cost of adding additional external operators, we make first an attempt with a two-operator system. As anticipated, this does not lead to the isolation to islands, so we then proceed to the three-operator system advertised earlier in this paper.

The main rationale for our setup comes from comparing with how the Ising CFT was isolated, namely by assuming that there is only one $\Z_2$-even and one $\Z_2$-odd scalar below $\Delta=d$. The success of this assumption can be credited to the relative sparseness of the Ising spectrum. We illustrate this sparseness by comparing with the $O(n)$ CFT, which could serve as a competing theory for the 3d Ising CFT but is ruled out by the assumptions. For instance, the $O(2)$ CFT seen as a theory with only $\Z_2$ symmetry $\vec\varphi\mapsto-\vec\varphi$ only has two $\Z_2$-even scalars with $\Delta<3$:  $\varphi^2_T\approx 1.24$ and $\varphi^2_S\approx 1.51$ \cite{Chester:2019ifh,Liu:2020tpf}. 
The same idea of sparseness led to the isolation of the $O(n)$ CFT \cite{Kos:2015mba}, which is more sparse than other more complicated $\phi^4$ theories. 
With these considerations in mind, we are in a good position, since the tricritical Ising CFT is (at least perturbatively) more sparse compared to $\phi^6$ theories with other global symmetry, so we anticipate that some suitable scalar gaps might be enough for our study. 

Next we consider which gaps to impose. A key assumption for isolating the Ising CFT was the gap in the $\Z_2$-odd sector, since $\Delta\geqslant d$ excludes any $\phi^4$ theory that has $\phi^3$ type operators. 
The Ising CFT has no such operator due to the equations of motion, which brings the perturbative gap of the theory to $\phi^5$ with $\Delta_{\phi^5}=5+\frac56\eps+\ldots$. 
For the tricritical Ising CFT, the corresponding equation-of-motion effect is that the operator $\phi^5$ is absent with respect to the free theory and with respect to $\phi^6$ theories with other global symmetries. This means that we can impose a large gap in the $\Z_2$-odd scalar spectrum between the operators $\phi^3$ with $\Delta=\frac32-\frac{13}{10}\eps$ and $\phi^7$ with $\Delta=\frac72+\frac72\eps$. While for Ising, the value $d$ for the gap was chosen for aesthetical reasons (it corresponds to marginality), we will simply select gap assumptions compatible with perturbative estimates through our Pad\'e approximants~\eqref{eq:Pades}. 
\begin{figure}
\centering
\includegraphics[width=0.75\textwidth]{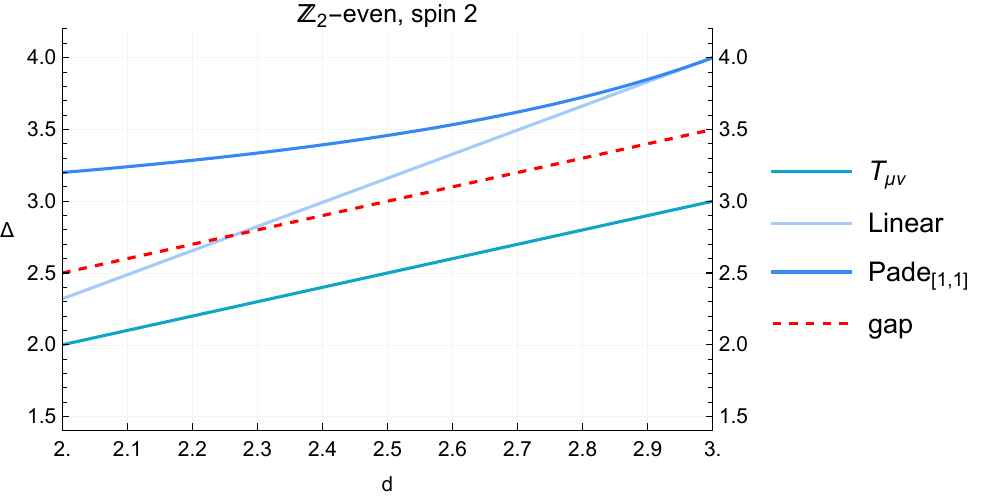}
\caption{The first two $\Z_2$-even operators at spin $2$. We impose the gap assumption $\Delta_{T'}\geqslant d+0.5$ (dashed red). The linear extrapolation of the one-loop dimension of the next operator seems to violate this, however assuming that it connects to the second spin-2 operator in $d=2$ gives a Pad\'e$_{[1,1]}$ approximant that is safely above our assumption.}\label{fig:gaps-spin-2}
\end{figure}

In the preliminary study, we impose this gap in using the $\{\phi,\phi^2\}$ system, but find that it is not sufficient to isolate the theory. This is unsurprising given that the free-theory OPE coefficient $\lambda^2_{\phi\phi^2\phi^5}=0$, meaning that the $\{\phi,\phi^2\}$ system might not be sensitive enough to detect to the presence or not of $\phi^5$. Including $\phi^3$ as external would solve this, since $\lambda^2_{\phi^2\phi^3\phi^5}=10$ in the free theory (see \eqref{eq:purePhiPowerOPE}), hence we expect that the full $\{\phi,\phi^2,\phi^3\}$ system should be sensitive to gap assumptions excluding $\phi^5$. 
We discuss the precise values for the gap assumptions below. 

Finally, following other bootstrap works we also impose a gap in the spin-2 sector after the stress-tensor. We choose this gap to be $\frac12$, compatible with the perturbative spectrum if we assume that the subleading spin-2 operator continues in a regular way to the subleading spin-2 operator in $d=2$, see figure~\ref{fig:gaps-spin-2}.

\subsection{Preliminary study from two-operator system}

We now report the results from the preliminary study, in which we considered the $\{\phi,\phi^2\}$ system, i.e. using one $\Z_2$-odd and one $\Z_2$-even external scalar. This is the system \eqref{eq:mixedCorrelatorKos} used in multiple studies for the Ising CFT, however, we now imposed completely different gap assumptions. 
We focused on $d=2.75$ and assumed
\begin{align}
\Delta_e&>1.6\,,\\
\Delta_o&\in[1.1,1.6]\cup[3.5,\infty)\,,
\end{align}
where $\Delta_e$ and $\Delta_o$ represent any subleading operators in the $\Z_2$-even and $\Z_2$-odd representations. The scan parameters were $(\Delta_{\phi},\Delta_{\phi^2})$. The result of this study is a rather wide peninsula, starting close to (but not including) the free-theory point $(0.375,\,0.75)$ and extending towards the upper-right. We do not reproduce here.

\begin{figure}
\centering
\includegraphics[width=0.85\textwidth]{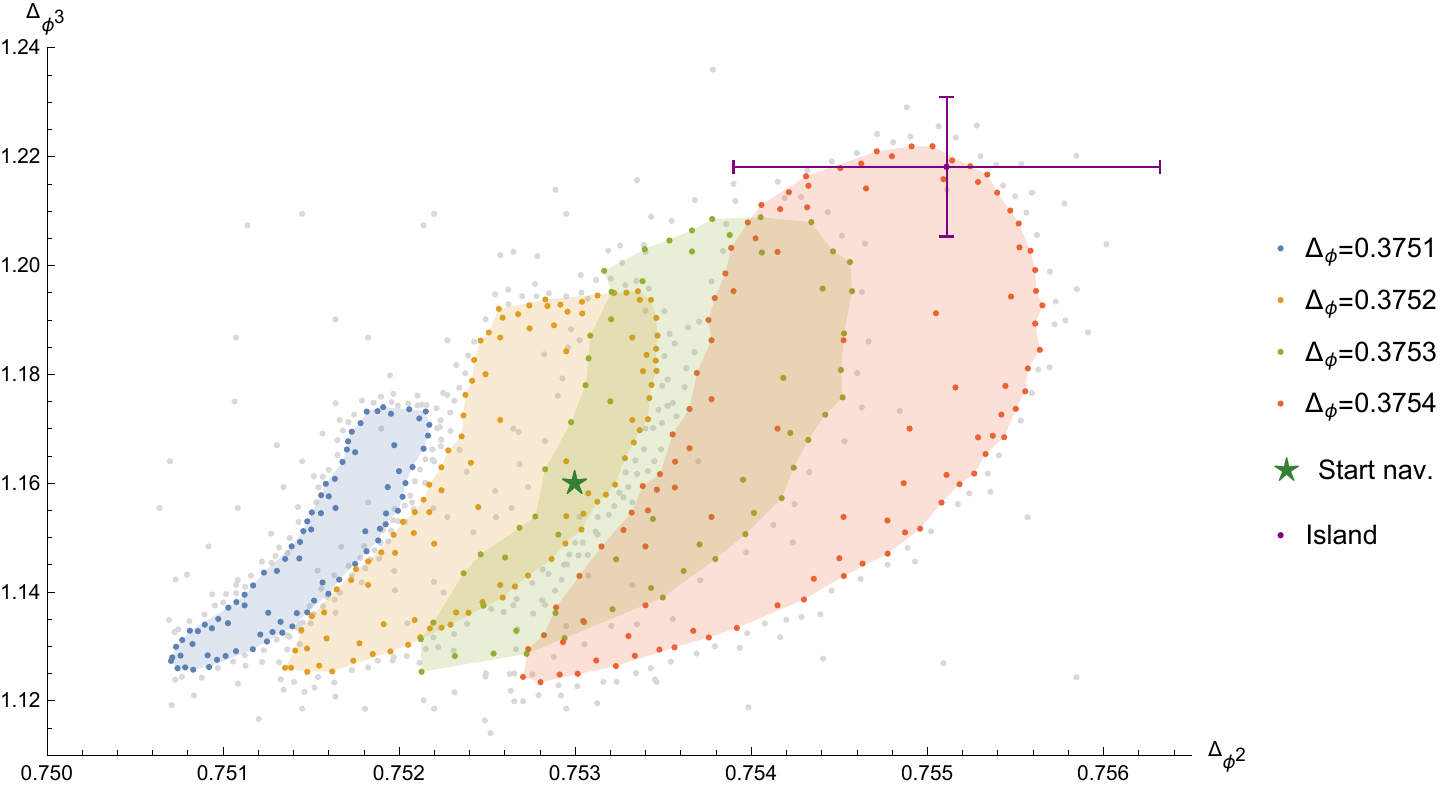}
\caption{Results of preliminary study. Allowed regions for $(\Delta_{\phi^2},\Delta_{\phi^3})$ with fixed $\Delta_\phi$ from the two-correlator system. For comparison, we also show in purple (with error bars) the island found later using the three-correlator system with $\Delta_\phi\in[0.37526, 0.37555]$, and the green star which is the start of the navigator run, \eqref{eq:navigatorStart}, with $\Delta_\phi=0.3753$.}\label{fig:prelRes}
\end{figure}

Next, still using the $\{\phi,\phi^2\}$ system, we added $\Delta_{\phi^3}$ as another scan parameter by replacing the interval $[1.1,1.6]$ with the exchange of an operator with fixed dimension $\Delta_{\phi^3}$. The assumptions were
\begin{align}
\Delta_e&>1.6\,,
\\
\Delta_{\phi^3}\!&\ \quad \text{isolated exchange},
\\
\Delta_o&>3.5\,,
\\
\label{eq:Tap}
\Delta_{T'}&>d+0.5=3.25\,,
\end{align} 
where $\Delta_{T'}$ denotes any operator after the stress-tensor in the $\Z_2$-even spin-2 sector.
We fixed $\Delta_\phi$ to a few different values and looked for allowed regions in the variables $(\Delta_{\phi^2},\Delta_{\phi^3})$ using a Delaunay scan. We finished the scans after a rough outline of the allowed region was observed, finding the following number of points: 
\begin{align}
\Delta_{\phi}&=0.3751:\ \text{\small 85 allowed, 115 ruled out,} & 
\Delta_{\phi}&=0.3752:\ \text{\small 103 allowed, 127 ruled out,} \\
\Delta_{\phi}&=0.3753:\ \text{\small 60 allowed, 76 ruled out,} & 
\Delta_{\phi}&=0.3754:\ \text{\small 104 allowed, 118 ruled out.} 
\end{align} 
The implementation of the Delaunay scan was done straightforwardly in \texttt{Simpleboot}.

The result of this search is displayed in figure~\ref{fig:prelRes}. Imagining the colour (value of $\Delta_\phi$) as a third direction, this picture represents a widening three-dimensional peninsula. Looking only at the extension in $\Delta_{\phi^2}$ at fixed $\Delta_\phi$, the size of this peninsula is similar in size to the one derived without including $\Delta_{\phi^3}$.

\subsection{Three-operator system and navigator minimisation}
\label{sec:navigatorRes}

This section represents our main numerical effort. We considered a system with three external operators: $\{\phi, \phi^2,\phi^3\}$ implemented as odd/even/odd scalars of an imposed global $\Z_2$ symmetry. Using \texttt{Autoboot} \cite{Go:2019lke,Go:2020ahx}, the crossing equations for this system were generated and imported to \texttt{Simpleboot}. There are 17 crossing equations (compared to 5 for $\{\phi, \phi^2\}$), four choices of quantum numbers ($\Z_2$-even/odd and $(-1)^\ell$ even/odd), and the vectors $\vec{\mtrx V}_Q$ of matrices of \eqref{eq:masterBootstrapEquation} have size $4\times 4$, $2\times2$, $2\times 2$ and $1\times 1$, see appendix~\ref{app:technicalDetails}. 

We used \texttt{Simpleboot} to perform a navigator scan following \cite{Reehorst:2021ykw}.
Specifically, we used the ``sigma navigator'' which is automatically constructed in \texttt{Simpleboot}. 
For the gap assumptions, we used the unitarity bounds \eqref{eq:UBregions}, except for the following cases:
\begin{align}
\mathscr D_{+,0}=[\Delta_e,\infty), \qquad 
\mathscr D_{-,0}=[\Delta_o,\infty), \qquad 
\mathscr D_{+,2}=[d+\tfrac12,\infty),
\end{align}
We also imposed the presence of the stress-tensor with OPE coefficients proportional to the external scaling dimensions $(\lambda_{\phi\phi T},\lambda_{\phi\phi^3 T},\lambda_{\phi^2\phi^2 T},\lambda_{\phi^3\phi^3 T})\propto (\Delta_\phi, 0,\Delta_{\phi^2}, \Delta_{\phi^3}) $. 

The gaps were chosen in a way compatible with Pad\'e approximants, based on the identification (see table~\ref{tab:minimalModel}) between the operators in the $\eps$-expansion and those in $d=2$. Allowing for some margin, we selected the gaps
\begin{align}
\label{eq:navigatorStart}
d&=2.75:\quad \Delta_e=1.6, \quad \Delta_o=3.5,
\\
d&=2.5:\quad \Delta_e=1.35, \quad \Delta_o=3.5,
\end{align}
see figure~\ref{fig:islands}. 
Finally, we used the following starting points for our three-parameter navigator runs: 
\begin{align}
\label{eq:navigatorStart}
d&=2.75:\quad p_{\mathrm{init.}}=(0.3753, 0.753, 1.16),
\\
d&=2.5:\quad  p_{\mathrm{init.}}= (0.25217, 0.5223,   0.984).
\end{align}
For $d=2.75$, this was chosen as a central point within figure~\ref{fig:prelRes}, and for $d=2.5$, we used Pad\'e approximants together with an allowed point in $d=2.75$ to select the initial point. 
Alternatively, the Pad\'e approximants \eqref{eq:Pades} alone could have been used directly as starting points.

\begin{figure}
\centering
\includegraphics[width=0.72\textwidth]{figs/nav-275-12-p.pdf}
\\
\includegraphics[width=0.48\textwidth]{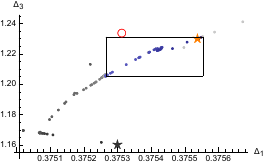}
\includegraphics[width=0.48\textwidth]{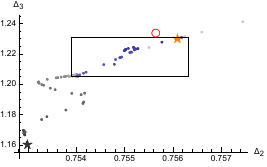}
\caption{Navigator run in $d=2.75$, giving a complementary view to figure~\ref{fig:nav-275}.}\label{fig:nav-275-more}
\end{figure}

\begin{figure}
\centering
\includegraphics[width=0.72\textwidth]{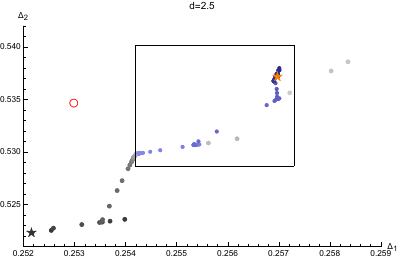}
\\
\includegraphics[width=0.48\textwidth]{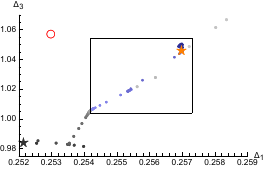}
\includegraphics[width=0.48\textwidth]{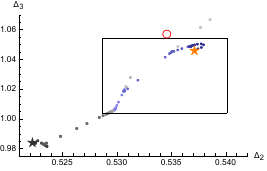}
\caption{Navigator run in $d=2.5$, using the same notation as figure~\ref{fig:nav-275}.}\label{fig:nav-25}
\end{figure}

\subsubsection*{Results} 

Our navigator runs terminated after $134$ and $113$ iterations in $d=2.75$ and $d=2.5$ dimensions respectively, arriving at the following final points
\begin{align}
d=2.75:&\qquad (0.3755406408,\, 0.7561071048,\, 1.230056059),
 \\
d=2.5:&\qquad (
0.2569800178, \,
0.5371399927, \,
1.0454992842).
\end{align}
We visualise the runs in figures~\ref{fig:nav-275-more} and~\ref{fig:nav-25} respectively. 
We also present a summary of the setup and results in table~\ref{tab:operatorsAndGaps}.

\begin{table}\caption{Some operators relevant for numerics, alongside our bounds and imposed gaps. The gaps in red did not lead to the isolation of the theory.}
\label{tab:operatorsAndGaps}
\centering
\small
\def\arraystretch{1.25}
\begin{tabular}{|clcllll|}
\hline
$\O$ & $\eps$-exp&  &  $d=2.75$ & $d=2.5$ & $d=2.25$ & $d=2$
\\\hline\hline
\multirow{2}*{$\phi$} &\multirow{2}*{$[\tfrac12-\tfrac\eps2]\!+\!\frac{\eps^2}{1000}$\!\! \!\!} & Pad\'e$_{2,2}$ 
& $0.37531$  & $0.25299$ & $0.13990$& $0.075$
\\
&& Bound & $[0.37526, 0.37555]$ & $[0.25419,0.25729]$ & NA & NA
\\\hline
\multirow{2}*{$\phi^2$} &\multirow{2}*{$[1-\eps]+\frac{4\eps^2}{125}$} & Pad\'e$_{2,2}$ 
& $0.75566$ & $0.53465$ & $0.34835$& $0.2$ 
\\
&& Bound & $[0.75390, 0.75632]$& $[0.52870,0.54013]$ &NA & NA
\\\hline

\multirow{2}*{$\phi^3$} &\multirow{2}*{$[\tfrac32-\tfrac{3\eps}2]+\frac\eps5$} & Pad\'e$_{2,1}$ 
&  $1.23347$ & $1.05722$ & $0.94354$ &  $0.875$
\\
&& Bound & $[1.20529,   1.23098]$ & $[1.00395,1.05415]$ &NA & NA
\\\hline
\multirow{2}*{$\phi^4$} &\multirow{2}*{$[2-2\eps]+\frac{4\eps}5$} & Pad\'e$_{2,2}$ 
& $1.74734$ & $1.53744$ & $1.35699$  &  $1.2$
\\
&& Gap & $1.6$ & $1.35$ &{\color{red} $1.15$}; {\color{red} $1.3$\!} &{\color{red} $1.1$}; {\color{red} $ 1.15$} 
\\\hline
$\phi^6$ & $[3-3\eps]+4\eps$ &  Pad\'e$_{1,2}$ & $3.04548$ & $3.03431$ &$ 3.01787$  & $3$
\\\hline
\multirow{2}*{$\phi^7$} &\multirow{2}*{$[\tfrac72-\tfrac{7\eps}2]+7\eps$}& Pad\'e$_{1,2}$ 
& $3.74737$ & $3.85975$ & $3.96631$  & $4.075$ 
\\
&& Gap & $3.5$ & $3.5$ & {\color{red} $3.7$}; {\color{red} $3.8$}& {\color{red} $3.9$}; {\color{red} $4.025$\!}
\\\hline
\end{tabular}
\end{table}

\subsubsection*{Navigator behaviour}

\begin{figure}
\centering
\includegraphics[width=0.48\textwidth]{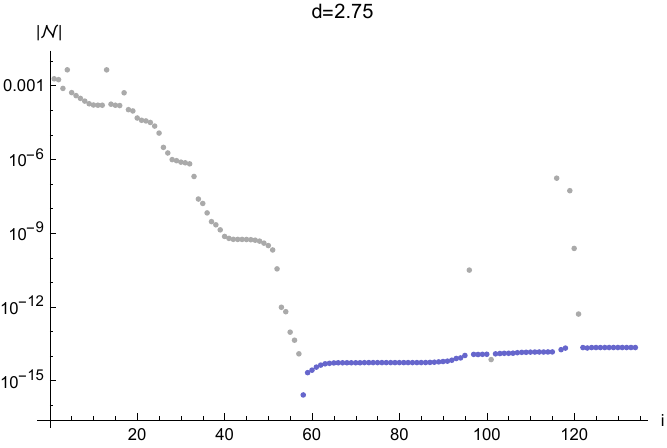}
\includegraphics[width=0.48\textwidth]{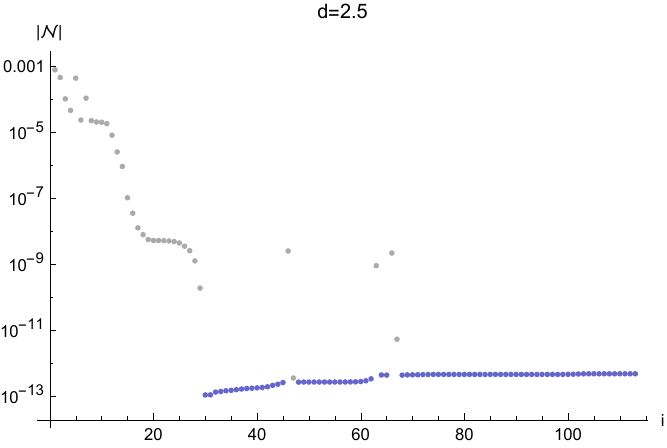}
\caption{Navigator value.}\label{fig:nav-i}
\end{figure}

\begin{figure}
\centering
\includegraphics[width=0.48\textwidth]{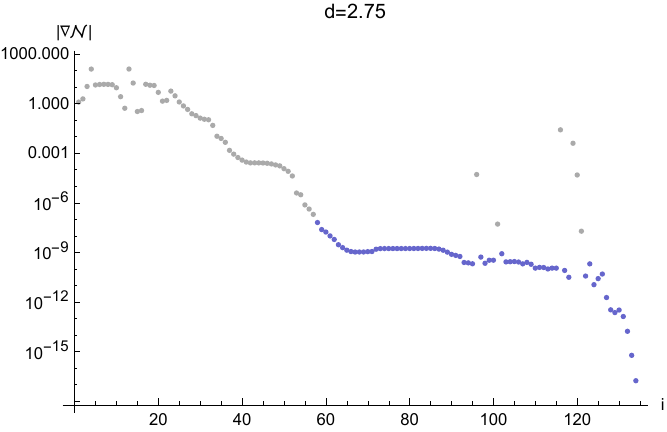}
\includegraphics[width=0.48\textwidth]{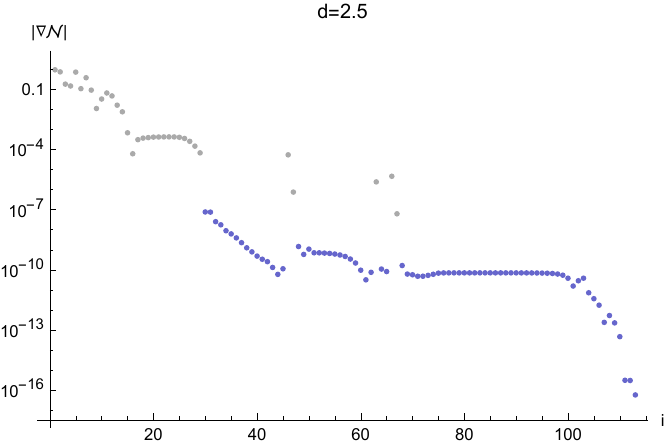}
\caption{Gradient of navigator.}\label{fig:grad-i}
\end{figure}

\begin{figure}
\centering
\includegraphics[width=0.48\textwidth]{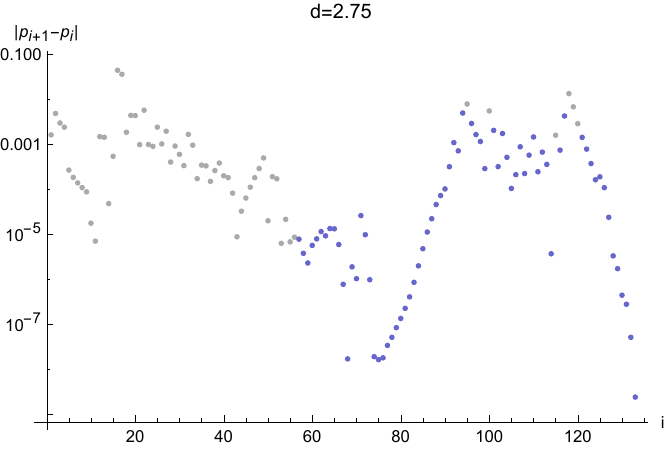}
\includegraphics[width=0.48\textwidth]{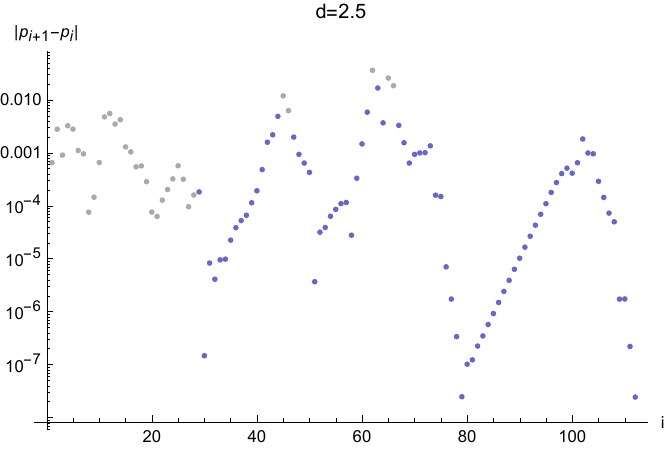}
\caption{Step lengths.}\label{fig:step-i}
\end{figure}

To give further details on the navigator run, we present three sets of plots showing properties of the navigator as a function of the iteration step $i$. Figure~\ref{fig:nav-i} shows the navigator value $|\mathcal N|$ as a function of $i$, figure~\ref{fig:grad-i} the gradient $|\nabla\mathcal N|$, and figure~\ref{fig:step-i} the step length.
In all of these plots, gray points are those with positive navigator value (disallowed), and purple with negative (allowed). 

We see that the runs in $d=2.75$ and $d=2.5$ dimensions are qualitatively similar. In both cases, the approach towards the allowed region and then towards the minimum inside the allowed region is not uniform -- the navigator has some plateaus with small step length, and then some phases with rather long step length. In the way out of such plateau, the step length is increased in an approximate geometric progression. 

\subsection{Extremising external scaling dimensions}

Next, after finding allowed points and ultimately the navigator minima, we started six searches per value of $d$, with the objective of maximising/minimising in each of the parameter directions. All of these runs terminated, meaning that we obtained closed intervals in $(\Delta_\phi,\Delta_{\phi^2}, \Delta_{\phi^3})$ containing the navigator minimum, thus the delimitations of a bootstrap island.
These intervals were reported in equations~\eqref{eq:island275}--\eqref{eq:island25} in the introduction, and are also given in table~\ref{tab:operatorsAndGaps} here.

In figures~\ref{fig:isl-275} and \ref{fig:isl-25} in appendix~\ref{app:NavigatorIslands} we show all allowed points in $d=2.75$ and $d=2.5$, from the minimise-navigator run, and the extremisation runs. Although we did not map out the precise shape of the island, these plots indicate that is has an elongated shape.

To get a suitable estimate on how small/large our islands are, we change variables and display them in the space of anomalous dimensions. For $\phi^k$, we define the anomalous dimension by $\gamma_{\phi^k}=\Delta_{\phi^k}-k\frac{d-2}2$. In these variables, our islands correspond to the intervals 
\begin{align}
d=2.75&:& \gamma_{\phi}& \in [0.00026,\, 0.00055],& \gamma_{\phi^2}& \in [0.00390\,, 0.00632],& \gamma_{\phi^3}& \in [0.08029, \,0.10598],
\nonumber
\\
d=2.5&:& \gamma_{\phi}& \in [0.00419, \,0.00729],& \gamma_{\phi^2}& \in [0.0287, \,0.04013],& \gamma_{\phi^3}& \in [0.25395,\,0.30415].
\end{align}
Thus we can observe that the islands we have found are rather large in this scale.

\subsection[Run-off in $d=2.25$ and $d=2$]{Run-off in $\boldsymbol{d=2.25}$ and $\boldsymbol{d=2}$}
\label{sec:runoff}

Based on the success in $d=2.75$ and $d=2.5$, we attempted a similar setup for $d=2$ and $d=2.25$ dimensions. We used the same algorithm and the same derivative-order as above, but adapted the gaps and starting points appropriately. Neither of these runs terminated at a navigator minimum -- at the beginning of the runs they looked similar to the successful runs in figures~\ref{fig:grad-i}--\ref{fig:step-i}, but after reaching allowed points, the navigator showed quick runoffs towards larger values of all parameters $(\Delta_\phi,\Delta_{\phi^2}, \Delta_{\phi^3})$. This indicates that there are no islands in these dimensions with the current setup, but instead ``peninsulas.''

The gaps we used were the following: 
\begin{align}
\label{eq:navigatorStart}
d=2.25&:\quad \Delta_e=1.15, \ \Delta_o=3.7\ \text{(attempt 1)}, \quad  \Delta_e=1.3, \ \Delta_o=3.8\ \text{(attempt 2)},  
\\
d=2&:\quad  \Delta_e=1.1, \ \Delta_o=3.9\ \text{(attempt 1)}, \quad  \Delta_e=1.15, \ \Delta_o=4.025\ \text{(attempt 2)},  
\end{align}
see also table~\ref{tab:operatorsAndGaps}. 
We also tried a few different starting points.

Although we view the lack of islands as a disappointment, we make a generic note peninsulas can still have some meaning as an indication of the existence of CFTs \cite{Chester:2016wrc,Chester:2022hzt}. If such peninsula is stable against varying some assumptions or increasing the derivative-order, it may be an indication that a theory might be close to saturating the bound. In that case, one could minimise one direction, say $\Delta_\phi$, and take that minimum as an approximate proxy for the theory.

\subsection{Extremal spectrum}
\label{sec:extremalSpec}

As a final step, we can extract an estimate of the spectrum using the extremal functional method \cite{ElShowk:2012hu,El-Showk:2014dwa}. We choose to do this at the final navigator point in $d=2.75$ and $d=2.5$ dimensions.

\begin{figure}
\centering
\includegraphics[height=0.48\textwidth]{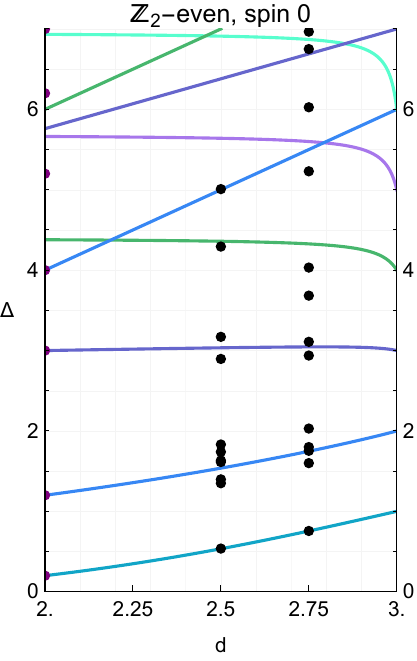}\qquad
\includegraphics[height=0.48\textwidth]{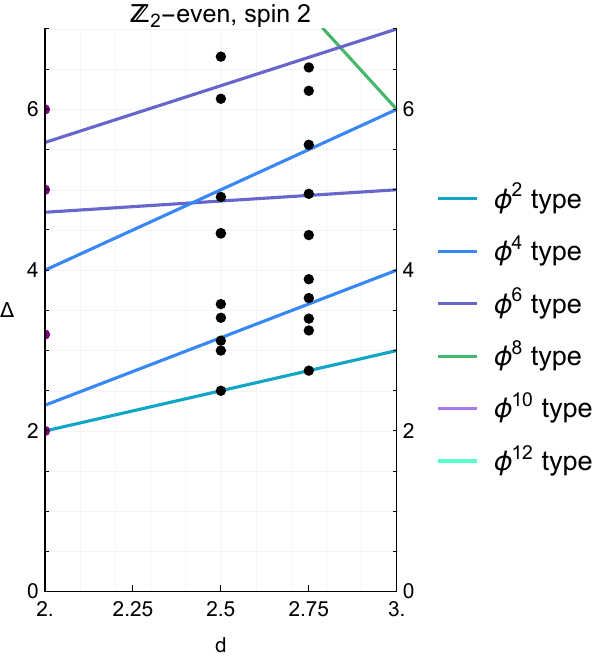}
\caption{Spectra for $\Z_2$-even operators of spin 0 and spin 2. For the first three scalar operators, we show the Pad\'e approximants \eqref{eq:Pades}. For the remaining scalar operators of form $\phi^{2k}$ we show $\text{Pad\'e}_{1,1}$ and for the rest we show $O(\eps)$ truncated results.}\label{fig:spec-E-0-2}
\end{figure}

In figure~\ref{fig:spec-E-0-2} we show $\Z_2$-even operators of spin 0 and 2.\footnote{We exclude spin 1 since there are not many low-lying such operators. Perturbatively, the lowest spin-1 operators have the schematic form $ \square^2\de\phi^k$ and anomalous dimension $\frac{35 k^3-207 k^2+64 k+600}{1050}$, for $k=5,6,\ldots$.} In this figure, we show the extremal spectra alongside perturbative estimates (our Pad\'e approximants \eqref{eq:Pades} for $\phi^{2k}$, $k=1,2,3$, Pad\'e$_{1,1}$ for $k>3$, and linear for other operators). More extremal spectra are shown in figures~\ref{fig:spec-E-3-4}, \ref{fig:spec-O-0-2} and \ref{fig:spec-O-3-4} in appendix~\ref{app:extremalSpectra}.

We can see that the extremal spectrum contains many points near our estimates, but the spread is large and there are many more number points than the expected number of operators.  
Compared to other studies, our extremal spectrum is disappointing and not very useful. For instance, in \cite{Henriksson:2022gpa} for the Ising CFT, the structure of many nearby zeroes was not observed. It should be noted that \cite{Henriksson:2022gpa} was performed with much higher numerical precision, $\Lambda=30$ instead of $\Lambda=19$, and the navigator minimum found there was inside much smaller islands than those found here.

\subsection{Comparison with previous results}
\label{sec:previousRes}
We would like to compare our results with the literature, however not much data is available at intermediate dimensions $d$. 
For $\Delta_\phi$ we can compare with figure~3 of \cite{Codello:2012sc}, and for $\Delta_{\phi^2}$ with figure~4 of \cite{Codello:2014yfa} (who also gave $\Delta_{\phi^4}$). Both these are results from the non-perturbative RG. Using the relations
$\Delta_\phi = \frac{d-2}2+\frac12\eta $, $\Delta_{\phi^2}=d-y_{1,3}$ and $\Delta_{\phi^4}=d-y_{2,3}$, 
the results in \cite{Codello:2012sc,Codello:2014yfa} translate to\footnote{Lacking access to the raw data, we attempted a manual reading of the plots of \cite{Codello:2012sc,Codello:2014yfa}. }
\begin{align}
d&=2.75: &    \Delta_{\phi} &= 0.385\,,&    \Delta_{\phi^2} &=0.754\,,&  \Delta_{\phi^4} &=  1.77\,,
\\
d&=2.5: &    \Delta_{\phi} &=0.292\,,&    \Delta_{\phi^2} &=0.524\, ,&    \Delta_{\phi^4} &=  1.62\,.
\end{align}
We can see that especially $\Delta_\phi$ is rather off from our results. 
We do not regard this disagreement as a sign of issues with our results, noting that the non-perturbative RG results deviate substantially also from the exact results in two dimensions: $\eta=0.31$ instead of $0.15$, $y_{1,3}=1.90$ instead of $1.80$ and $y_{2,3}=0.35$ instead of $0.8$.

\subsection{Comments and possible improvements}
\label{sec:ResDisc}

Let us discuss a few potential improvements of the current setup, however beyond the scope of the present article. Some technical issues are deferred to appendix~\ref{app:inpractice}. 

\subsubsection*{Navigator vs other implementations}

The navigator approach is very useful for large dimensionality of the search space and for finding small bootstrap islands. However, we observed that it required many iterations inside the allowed region to find the minimum. Moreover, the scans to minimise and maximise along the directions in parameter space also required many iterations. One option for mapping out the shape of the allowed region is to go back to a Delaunay scan (with surface-cutting algorithm for OPE ratios) after finding an allowed point, working in transformed variables where the island is less elongated. This was done e.g. in \cite{Chang:2024whx} which studied the Ising CFT with the mixed $\{\phi,\phi^2,T_{\mu\nu}\}$ system. 
A strategic choice of starting points may also reduce the number of iterations needed, see appendix~\ref{app:inpractice}. 

Another way to minimise the number of iterations (or rather the computational time at each iteration) is to use the Skydiving algorithm \cite{Liu:2023elz}. However, it was not available at the beginning of this project, and our attempts to use it were not successful, probably due to the need of fine-tuning some internal parameters. 

\subsubsection*{Preliminary study with more observables}
\label{sec:sevenParams}

Here we observed that the three-parameter scan $(\Delta_{\phi},\Delta_{\phi^2},\Delta_{\phi^3})$ was not enough to find an island in $d=2$ and $d=2.25$ dimensions. A future direction is to try more sophisticated setups. The most direct is to include dimensions of exchanged operators (e.g. $\phi^4$) and/or OPE ratios in the scan. The author has experimented with a few of these setups, including fixing one OPE ratio (see footnote~\ref{foot:Bmat}), and a seven-parameter scan including both $\Delta_{\phi^4}$ and all OPE ratios of external operators. For the latter, we expect an interpolation between the 3d values \eqref{eq:purePhiPowerOPE}
\begin{equation}
\label{eq:OPEratios3d}
\frac1{\lambda_{\phi\phi\phi^2}}\left(\lambda_{\phi\phi\phi^2},\lambda_{\phi^2\phi^2\phi^2},\lambda_{\phi\phi^3\phi^2},\lambda_{\phi^3\phi^3\phi^2}\right)=\{{1, 2, 1.22474488, 3}\},
\end{equation}
and the 2d exact values \eqref{eq:FusionRules}
\begin{equation}\label{eq:OPEratios2d}
\frac1{\lambda_{\sigma\sigma\epsilon}}\left(\lambda_{\sigma\sigma\epsilon},\lambda_{\epsilon\epsilon\epsilon},\lambda_{\sigma\sigma'\epsilon},\lambda_{\sigma'\sigma'\epsilon}\right)=\{1,0,0.546178,0\}.
\end{equation}
Although promising, to perform a complete study with this system requires computational resources beyond the scope of this study.\footnote{We found an allowed point in $d=2.75$ with $\big(\Delta_{\phi},\Delta_{\phi^2},\Delta_{\phi^3},\Delta_{\phi^4},\frac{\lambda_{\phi^2\phi^2\phi^2}}{\lambda_{\phi\phi\phi^2}},\frac{\lambda_{\phi\phi^3\phi^2}}{\lambda_{\phi\phi\phi^2}},\frac{\lambda_{\phi^3\phi^3\phi^2}}{\lambda_{\phi\phi\phi^2}}\big)=(0.37538791, 0.7550169, 1.2208268, 
  1.76368299, 1.6531125, 1.0989969, 
  1.981568)$.}

\subsubsection*{Obtaining a better extremal spectrum}

Here we discuss how the estimate of the spectrum might be improved.\footnote{For technical reasons, the extremal spectrum could not be read off until the final stages of this project. } 
First, we only used a single point for the extremal spectrum, chosen at the minimum of the ``sigma navigator'' that we used here. This minimum was found quite far from the centre of the island, in disagreement with a hope that the navigator minimum should represent ``the most allowed point'' and that the island should shrink in a rather uniform way. To get a better estimate of the spectrum, other methods could be employed. 
Perhaps the GFF navigator would produce a minimum more central in the island and therefore a more realistic spectrum. 
One could also use the sampling method of \cite{Simmons-Duffin:2016wlq} (also used in \cite{Liu:2020tpf,Atanasov:2022bpi}), which amounts to selecting a sample of allowed points and extract the extremal spectrum at each point in the sample. Then one performs a statistical analysis where the operator dimensions that are relatively stable across the sample are retained, and their variation is used as an estimate of the error.

\section{Discussion and outlook}
\label{sec:disc}

In this paper, we have studied the tricritical Ising CFT in the range $2\leqslant d<3$, and obtained bootstrap islands in $d=2.75$ and $d=2.5$ dimensions. We performed a targeted bootstrap study, meaning that we aimed specifically at isolating a target CFT rather than deriving general bounds. To this end, we reviewed the perturbative data in section~\ref{sec:triIsing}, and also supplemented the literature with a small survey of the low-lying spectrum using the one-loop dilatation operator (section~\ref{sec:OneLoopDil}). We also summarised the bootstrap method in section~\ref{sec:bootstrapApproach}, with the aim of explaining in general terms how recent bootstrap technology can be used to study a target theory.

The islands we found agree well with Pad\'e approximants connecting the $3-\eps$ expansion to the exact values in $d=2$, see figures~\ref{fig:islands}--\ref{fig:islandsLog}. This gives direct evidence in favour of the anticipated scenario that the minimal models $\mathcal M_{k+2,k+1}$ are continuously connected to $\phi^{2k}$ theory with upper critical dimension $d_c(k)=\frac{2k}{k-1}$, here for $k=3 $. We now discuss various open directions.

\subsection[How to find islands in $d=2$ and $d=2.25$?]{How to find islands in $\boldsymbol{d=2}$ and $\boldsymbol{d=2.25}$?}

We still failed to isolate the theory in $d=2$ dimensions, which means that to date the only 2d CFT to be isolated to a bootstrap island is the Ising CFT \cite{delaFuente:2019hbl}. This contrasts the wealth of exact results available for 2d CFTs \cite{DiFrancesco:1997nk}, and it would be interesting to understand better what the obstructions are. 
It is possible that a more judicious choice of assumptions would have isolated our target theory. For instance, one could demand that the stress-tensor is exchanged with the central charge fixed to $c=\frac7{10}$; the tricritical Ising is the only unitary CFT with that central charge. We did not attempt such judicious assumptions since ultimately the goal here was to develop and test methods that can apply to a wider set of theories, including those where we do not have exact results.

With this restriction taken into account, there are still some ideas that would be interesting to attempt. One is to experiment with different assumptions for spinning operators, where we only put a gap above the stress-tensor of $\frac12$ (c.f. equation \eqref{eq:Tap} and figure~\ref{fig:gaps-spin-2}). 
The paper \cite{Li:2017kck} nicely discussed how various assumptions in a single-correlator bootstrap system contributed to the isolation of several theories in 3d, including Ising, super-Ising and $O(3)$. 
One could also try with improved numerical strength (derivative-order $\Lambda$), and/or use improved numerical algorithms. 
The \texttt{SDPB} code has recently been updated to run faster, see \cite{Chang:2024whx}, which may facilitate increased precision. It is entirelt possible that purely increasing $\Lambda$, keeping the same setup as here, will in fact isolate the theory to islands in $d=2$ and $d=2.25$, just like some islands in the literature cannot be obtained with too low $\Lambda$, e.g. in \cite{Kousvos:2021rar}. 

\subsection{Perturbative computations for future bootstrap studies}

One motivation for this work was to execute a complete sequence of steps to approach a specific target theory with the bootstrap. 
Although we discussed tricritical theories here, some lessons from this work should also apply to other systems -- some practical and technical issues are discussed in section~\ref{sec:ResDisc} and appendix~\ref{app:inpractice}. An important target for the bootstrap is experimentally relevant $\phi^4$ theories \cite{Pelissetto:2000ek}, some of which have not yet been isolated to precision islands with the bootstrap. Previous bootstrap studies of these theories \cite{Stergiou:2018gjj,Kousvos:2018rhl,Stergiou:2019dcv,Kousvos:2019hgc,Henriksson:2020fqi,Kousvos:2021rar,Kousvos:2022ewl,Henriksson:2021lwn,Reehorst:2024vyq} have shown many suggestive features, which would be interesting to investigate with further studies. 

An important ingredient in our work is good perturbative estimates. 
Their role is to provide estimates for the target theory, and more generally also for known ``competing'' theories. Since the bootstrap is sensitive to many exchanged operators, not just those giving measurable critical exponents, we are interested in good estimates for a large portion of the spectrum, including spinning operators and operators in non-singlet global-symmetry representations. 

Here we used a combination of multiloop results for low-lying operators, and a systematic study of the one-loop dilatation operator. This programme can also be executed for complicated global symmetry, such as the cubic $\phi^4$ theory \cite{Bednyakov:2023lfj}. A collection of such data facilitates a flexible choice of gap assumptions, and at small $\eps$ the perturbative estimates can be taken to be rather reliable. Using a set of gap assumptions, one can perform a bootstrap study first at small $\eps$, and then increase towards physically interesting values of $d$. Such tracking strategy across $d$ was recently executed for $\phi^4$ theories with $O(m)\times O(n)$ symmetry \cite{Reehorst:2024vyq}. 

The current perturbative state of the art is multi-loop results for low-lying operators only, complemented with one-loop results for more larger parts of the spectrum. It would be desirable to go beyond this, with a study of multi-loop anomalous dimensions of arbitrary composite operators. Steps towards this are taken in \cite{LoopsBootstrap}, where composite operators are renormalised in $4-\eps$ dimensions using Feynman diagrams and techniques for operator bases from effective field theory \cite{Cao:2021cdt}. It would be nice to extend this computation to $3-\eps$ dimensions.  

To generate perturbative results for theories with various global symmetries, it is convenient to work with interactions with ``open field indices,'' like in \cite{Bednyakov:2021ojn} for $\phi^4$ theories. Translating this idea to our case implies working with the general Lagrangian density
\begin{equation}
\mathcal L\!=\!u_0+u_1^i\phi_i+\frac{u_2^{ij}\!}2\phi_i\phi_j+\frac{u_3^{ijk}\!\!}{3!}\phi_i\phi_j\phi_k+\frac{u_4^{ijkl}\!\!}{4!}\phi_i\phi_j\phi_k\phi_l+\frac{u_5^{ijklm}\!\!\!}{5!}\phi_i\phi_j\phi_k\phi_l\phi_m+\frac{u_6^{ijklmn}\!\!\!\!}{6!}\phi_i\phi_j\phi_k\phi_l\phi_m\phi_n
\end{equation}
and computing beta functions for all interactions $u_i$ in $3-\eps$ dimensions. It should be possible to perform this computation by recycling diagrammatic technology from \cite{Hager1999,Hager:2002uq}. By using group theory, results for many different theories with different global symmetry could then be extracted.

\subsection[Tricritical $O(n)$ models]{Tricritical $\boldsymbol{O(n)}$ models}
\label{sec:tricriticalON}

One of the motivations for this work is to lay the ground for future works on tricritical CFTs with global symmetry $O(n)$ \cite{Pisarski:1982vz,Bardeen:1983rv,Amit:1984ri,David:1984we,Gudmundsdottir:1984vyf,Gudmundsdottir:1984rr,David:1985zz,Hager1999,Hager:2002uq,Kaplan:2009kr,Yabunaka:2017uox,Osborn:2017ucf,Katsis:2018bvc,Goykhman:2020ffn}, since the literature on these theories contains some interesting results.  
It is known that the $3-\eps$ expansion for the tricritical $O(n)$ model does not have a well-behaved large-$n$ limit. Specifically, leading order perturbative data suggest that this expansion is only valid for $\varepsilon<\frac{36}{\pi^2n}\approx \frac{3.6}n$ .
In \cite{Yabunaka:2017uox}, it was proposed with motivation from non-perturbative RG that the curve $\eps\approx\frac{3.6}n$ ends at some finite $n\approx 19$. Above this value, the tricritical $O(n)$ CFT should annihilate with a new non-perturbative fixed-point. Below this value, \cite{Yabunaka:2017uox} predicts that the theory continues to exist, and by going around the point $(d,n)\approx (2.8,19)$ to the bottom-left and then towards large $n$ one reaches another new non-perturbative fixed-point. It would be very interesting to study these putative non-perturbative CFTs with the bootstrap, where a mixed system $\{\varphi,\varphi^2_S,\varphi^3_V\}$ inspired by this paper might be the right tool. See \cite{Yabunaka:2018mju,Defenu:2020cji,Yabunaka:2021fow} for further comments, and \cite{Hawashin:2024dpp,Komargodski:2024zmt} for work on other putative new CFTs of similar type. 

\subsection{Multicritical models}

Another direction is to bootstrap the higher multicritical theories. It would be nice to establish with the bootstrap the connection between theories with $\phi^{2k}$ interaction near their upper critical dimensions, and the minimal models $\mathcal M_{k+2 ,k+1}$. 
In appendix~\ref{app:generalMulti} we collect some results for these theories. 

Allowing for multicritical theories with global symmetry gives a plethora of opportunities. A collection of perturbative fixed-points of this class, based on discrete subgroups of $O(N)$, was discussed in \cite{Zinati:2019gct}, see also \cite{Codello:2020mnt,Kapoor:2021lrr,BenAliZinati:2021rqc}. It is not established what happens to these theories as $\eps$ is increased, and likewise which two-dimensional CFTs belong to families of CFTs extending above $d=2$ dimensions. 
A related question, for the tricritical model considered here, is whether its supersymmetric version can be bootstrapped and continued above two dimensions to make contact with the three-dimensional minimal $\mathcal N=1$ supersymmetric Ising CFT \cite{Bashkirov:2013vya,Iliesiu:2015qra,Rong:2018okz,Atanasov:2022bpi}, see also \cite{Fei:2016sgs}.

\section*{Acknowledgements}

I thank Ning Su and Stefanos Kousvos for many discussions and indispensable help throughout this project. 
I also thank Oleg Antipin, Ant\'onio Antunes, Jamall Bersini, Alessandro Codello, Ian Jack, Ying-Hsuan Lin, Hugh Osborn, Junchen Rong, Jasper Roosmale Nepveu, Slava Rychkov, Andreas Stergiou, Emilio Trevisani, Alessandro Vichi, Jakub Vo\v{s}mera and Omar Zanusso for useful conversations.
This work received funding from the European Research Council (ERC) under grant agreements 758903 and 853507.

\appendix

\section{Details on numerics and more results}

In this appendix we give some bulky expressions and give further details on the numerics. 

\subsection{Crossing equations and bootstrap settings}
\label{app:technicalDetails}

The bootstrap equation \eqref{eq:masterBootstrapEquation} in our system with one $\Z_2$-even and two $\Z_2$-odd operators was generated in \texttt{Autoboot} \cite{Go:2019lke,Go:2020ahx} and takes the form\footnote{
This particular setup could also be used to study other theories, in particular the Ising CFT where the operators would be  $\{\phi,\phi^2,\phi^5\}$. This would allow the operator $\phi^7$ would be exchanged with a significant OPE coefficient; this operator was not detected in \cite{Simmons-Duffin:2016wlq,Henriksson:2022gpa}.}
\begin{equation}
0=\vec V_{\text{OPE}}+\sum_{\O_{+,+}} \vc \lambda_{++}^T \vec{\mtrx V}_{+,+}\vc \lambda_{++}+\sum_{\O_{+,-}}\lambda_{13\O}^2V_{+,-}+\sum_{\O_{-,+}} \vc \lambda_{-+}^T \vec{\mtrx V}_{-,+}\vc \lambda_{-+}+\sum_{\O_{-,-}} \vc \lambda_{--}^T \vec{\mtrx V}_{-,-}\vc \lambda_{--},
\end{equation}
where we labelled the sectors by $\Z_2$ representation $\pm$ and spin parity $(-1)^\ell$. 
In this expression, the vectors of matrices are given by
\begin{align}
\nonumber
\vec{\mtrx V}_{+,+}&=\bigg\{\left(
\begin{smallmatrix}
 F_{1111} & 0 & 0 & 0 \\
 0 & 0 & 0 & 0 \\
 0 & 0 & 0 & 0 \\
 0 & 0 & 0 & 0 \\
\end{smallmatrix}
\right),\tfrac12\!\left(
\begin{smallmatrix}
 0 & {F_{1113}} & 0 & 0 \\
 {F_{1113}} & 0 & 0 & 0 \\
 0 & 0 & 0 & 0 \\
 0 & 0 & 0 & 0 \\
\end{smallmatrix}
\right),\tfrac12\!\left(
\begin{smallmatrix}
 0 & 0 & H_{1122} & 0 \\
 0 & 0 & 0 & 0 \\
 H_{1122} & 0 & 0 & 0 \\
 0 & 0 & 0 & 0 \\
\end{smallmatrix}
\right),\tfrac12\!\left(
\begin{smallmatrix}
 0 & 0 & F_{1122} & 0 \\
 0 & 0 & 0 & 0 \\
 F_{1122}& 0 & 0 & 0 \\
 0 & 0 & 0 & 0 \\
\end{smallmatrix}
\right),
\\&\qquad
\nonumber
\tfrac12\!\left(
\begin{smallmatrix}
 0 & 0 & 0 & H_{1133} \\
 0 & -2H_{1331} & 0 & 0 \\
 0 & 0 & 0 & 0 \\
 H_{1133} & 0 & 0 & 0 \\
\end{smallmatrix}
\right),\tfrac12\!\left(
\begin{smallmatrix}
 0 & 0 & 0 & F_{1133} \\
 0 &2 F_{1331} & 0 & 0 \\
 0 & 0 & 0 & 0 \\
 F_{1133} & 0 & 0 & 0 \\
\end{smallmatrix}
\right),\mathbf 0,\tfrac12\!
\left(
\begin{smallmatrix}
 0 & 0 & 0 & 0 \\
 0 & 0 &H_{1322} & 0 \\
 0 & H_{1322} & 0 & 0 \\
 0 & 0 & 0 & 0 \\
\end{smallmatrix}
\right),\\
\nonumber&\qquad\tfrac12\!\left(
\begin{smallmatrix}
 0 & 0 & 0 & 0 \\
 0 & 0 & F_{1322} & 0 \\
 0 & F_{1322} & 0 & 0 \\
 0 & 0 & 0 & 0 \\
\end{smallmatrix}
\right),
\mathbf 0,
\left(
\begin{smallmatrix}
 0 & 0 & 0 & 0 \\
 0 & F_{1313} & 0 & 0 \\
 0 & 0 & 0 & 0 \\
 0 & 0 & 0 & 0 \\
\end{smallmatrix}
\right),\tfrac12\!\left(
\begin{smallmatrix}
 0 & 0 & 0 & 0 \\
 0 & 0 & 0 & F_{1333}\\
 0 & 0 & 0 & 0 \\
 0 & {F_{1333}} & 0 & 0 \\
\end{smallmatrix}
\right),\left(
\begin{smallmatrix}
 0 & 0 & 0 & 0 \\
 0 & 0 & 0 & 0 \\
 0 & 0 & F_{2222} & 0 \\
 0 & 0 & 0 & 0 \\
\end{smallmatrix}
\right),\\
& \qquad\tfrac12\!\left(
\begin{smallmatrix}
 0 & 0 & 0 & 0 \\
 0 & 0 & 0 & 0 \\
 0 & 0 & 0 &{H_{2233}} \\
 0 & 0 & {H_{2233}} & 0 \\
\end{smallmatrix}
\right),\tfrac12\!\left(
\begin{smallmatrix}
 0 & 0 & 0 & 0 \\
 0 & 0 & 0 & 0 \\
 0 & 0 & 0 & F_{2233} \\
 0 & 0 & F_{2233} & 0 \\
\end{smallmatrix}
\right),\mathbf 0,\left(
\begin{smallmatrix}
 0 & 0 & 0 & 0 \\
 0 & 0 & 0 & 0 \\
 0 & 0 & 0 & 0 \\
 0 & 0 & 0 & F_{3333} \\
\end{smallmatrix}
\right)\bigg\},
\\
\vec V_{+,-}&=\left\{0,0,0,0,H_{1331},-F_{1331},0,0,0,0,F_{1313},0,0,0,0,0,0\right\},
\\
\nonumber
\vec{\mtrx V}_{-,+}&=\bigg\{\mathbf 0,\mathbf 0,\left(
\begin{smallmatrix}
 -H_{1221} & 0 \\
 0 & 0 \\
\end{smallmatrix}
\right),\left(
\begin{smallmatrix}
 F_{1221} & 0 \\
 0 & 0 \\
\end{smallmatrix}
\right),\mathbf 0,\mathbf 0,\left(
\begin{smallmatrix}
 F_{1212} & 0 \\
 0 & 0 \\
\end{smallmatrix}
\right),\tfrac12\left(
\begin{smallmatrix}
 0 & -{H_{1223}} \\
 -{H_{1223}} & 0 \\
\end{smallmatrix}
\right),\tfrac12\left(
\begin{smallmatrix}
 0 & {F_{1223}} \\
 {F_{1223}} & 0 \\
\end{smallmatrix}
\right),\\
&\qquad\tfrac12\left(
\begin{smallmatrix}
 0 & {F_{1232}} \\
 {F_{1232}} & 0 \\
\end{smallmatrix}
\right),\mathbf 0,\mathbf 0,\mathbf 0,\left(
\begin{smallmatrix}
 0 & 0 \\
 0 & -H_{2332} \\
\end{smallmatrix}
\right),\left(
\begin{smallmatrix}
 0 & 0 \\
 0 & F_{2332} \\
\end{smallmatrix}
\right),\left(
\begin{smallmatrix}
 0 & 0 \\
 0 & F_{2323} \\
\end{smallmatrix}
\right),\mathbf 0\bigg\},
\\
\nonumber
\vec{\mtrx V}_{-,-}&=\bigg\{\mathbf 0,\mathbf 0,\left(
\begin{smallmatrix}
 H_{1221} & 0 \\
 0 & 0 \\
\end{smallmatrix}
\right),\left(
\begin{smallmatrix}
- F_{1221} & 0 \\
 0 & 0 \\
\end{smallmatrix}
\right),\mathbf 0,\mathbf 0,\left(
\begin{smallmatrix}
 F_{1212} & 0 \\
 0 & 0 \\
\end{smallmatrix}
\right),\tfrac12\left(
\begin{smallmatrix}
 0 & -{H_{1223}} \\
 -{H_{1223}} & 0 \\
\end{smallmatrix}
\right),\tfrac12\left(
\begin{smallmatrix}
 0 & {F_{1223}} \\
 {F_{1223}} & 0 \\
\end{smallmatrix}
\right),\\
&\qquad\tfrac12\left(
\begin{smallmatrix}
 0 &- {F_{1232}} \\
- {F_{1232}} & 0 \\
\end{smallmatrix}
\right),\mathbf 0,\mathbf 0,\mathbf 0,\left(
\begin{smallmatrix}
 0 & 0 \\
 0 & H_{2332} \\
\end{smallmatrix}
\right),\left(
\begin{smallmatrix}
 0 & 0 \\
 0 & -F_{2332} \\
\end{smallmatrix}
\right),\left(
\begin{smallmatrix}
 0 & 0 \\
 0 & F_{2323} \\
\end{smallmatrix}
\right),\mathbf 0\bigg\},
\end{align}
where $F_{ijkl}$ ($H_{ijkl}$) is the difference (sum) of conformal blocks involving the operators $\phi^i,\phi^j,\phi^k,\phi^l$ (compare with \eqref{eq:theCrossingEqs}), and the vectors of OPE coefficients are given by $\vc\lambda_{++}=(\lambda_{11\O},\lambda_{13\O},\lambda_{22\O},\lambda_{33\O})^T$ and $\vc\lambda_{-+}=\vc\lambda_{--}=(\lambda_{12\O},\lambda_{23\O})^T$. 
Moreover, the OPE coefficient block is given by exchanging the operator $\phi^2$ in $Q=(+,+)$, the operators $\phi$ and $\phi^3$ in $Q=(-,+)$ and repackaging the resulting expression to the form
\begin{equation}
\vec V_{\text{OPE}}=\vc \lambda^T\vec{\mtrx V} \vc \lambda\,, \qquad \vc \lambda=(\lambda_{112},\lambda_{123},\lambda_{222},\lambda_{233})^T.
\end{equation}
This gives $\vec{\mtrx V}$ of the form
\begin{equation}
\bigg\{\!\!\left(
\begin{smallmatrix}
 F_{1111}(\Delta_2) & 0 & 0 & 0 \\
 0 & 0 & 0 & 0 \\
 0 & 0 & 0 & 0 \\
 0 & 0 & 0 & 0 \\
\end{smallmatrix}
\right)\!,\tfrac12\!\left(
\begin{smallmatrix}
 0 & {F_{1113}(\Delta_2)} & 0 & 0 \\
{F_{1113}(\Delta_2)} & 0 & 0 & 0 \\
 0 & 0 & 0 & 0 \\
 0 & 0 & 0 & 0 \\
\end{smallmatrix}
\right)\!,\tfrac12\!\left(
\begin{smallmatrix}
 -2H_{1221}(\Delta_1) & 0 & {H_{1122}(\Delta_2)} & 0 \\
 0 & -2H_{1221}(\Delta_3) & 0 & 0 \\
 {H_{1122}(\Delta_2)} & 0 & 0 & 0 \\
 0 & 0 & 0 & 0 \\
\end{smallmatrix}
\right)\!,\ldots\bigg\}. 
\end{equation}
We also separate out the contribution from the stress-tensor and contract it with the OPE coefficient vector $\vc \lambda\propto(\Delta_{\phi},0,\Delta_{\phi^2},\Delta_{\phi^3})^T$, as dictated by the conformal Ward identities.

In \texttt{Simpleboot}, we use $\Lambda=19$, $\kappa=80$, $r_N=20$, and the following spins:
\begin{equation}
\mathscr L=\{0, 1, 2, 3,\ldots, 33, 34,\, 51,52\}
\end{equation}
For the navigator \texttt{SDPB} runs, we used the settings:

\texttt{--saveMiddleCheckpointMuThreshold=1e-8}

\texttt{--maxIterations=1000 }

\texttt{--dualityGapThreshold=1e-25 }

\texttt{--primalErrorThreshold=1e-15 }

\texttt{--dualErrorThreshold=1e-15 }

\texttt{--precision=1200 }

\texttt{--initialMatrixScalePrimal=1e+20 }

\texttt{--initialMatrixScaleDual=1e+20 }

\texttt{--maxComplementarity=1e+70}

\subsection{Navigator islands}
\label{app:NavigatorIslands}
\begin{figure}
\centering
\includegraphics[width=0.48\textwidth]{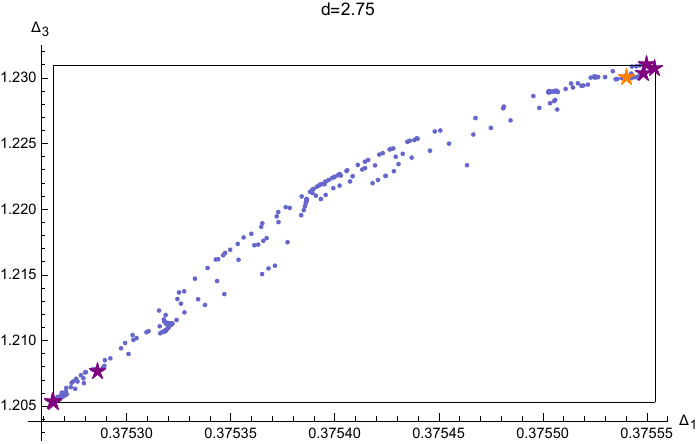}
\\
\includegraphics[width=0.48\textwidth]{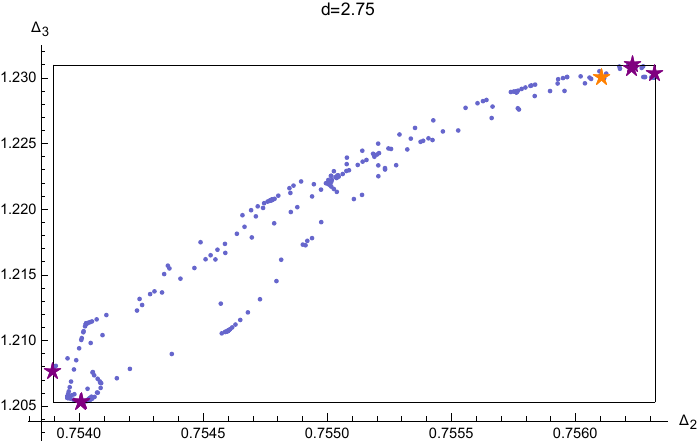}
\includegraphics[width=0.48\textwidth]{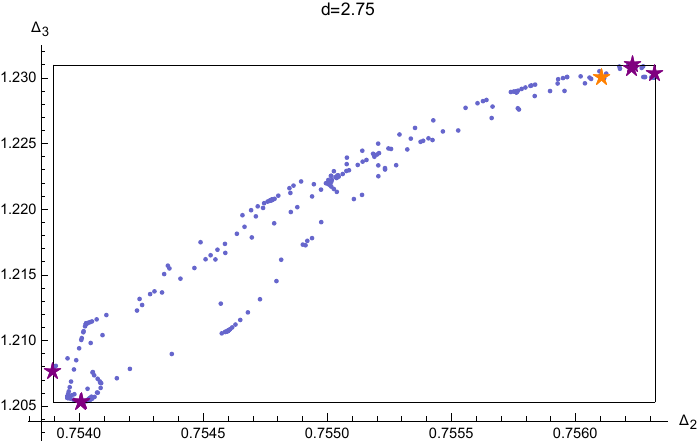}
\caption{All allowed navigator points in $d=2.75$. Stars denote the end-points of navigator runs extremising the directions.}\label{fig:isl-275}
\end{figure}

\begin{figure}
\centering
\includegraphics[width=0.48\textwidth]{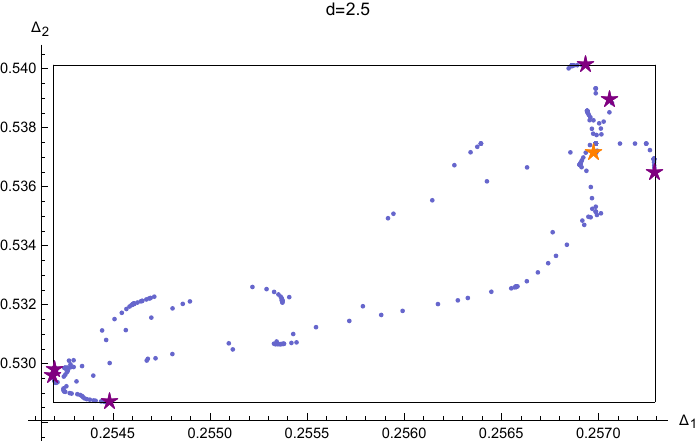}
\\
\includegraphics[width=0.48\textwidth]{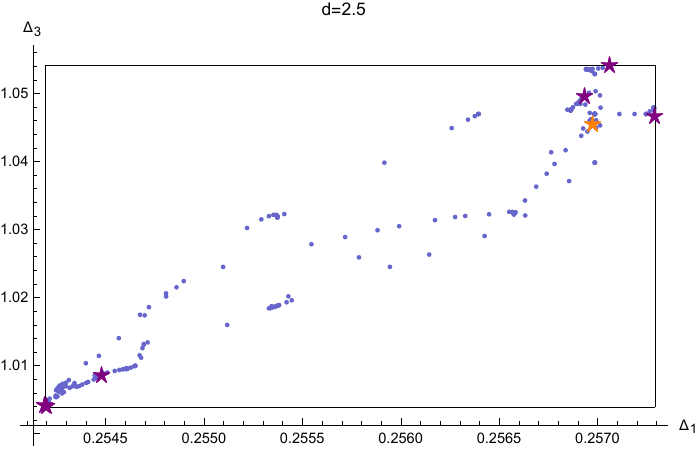}
\includegraphics[width=0.48\textwidth]{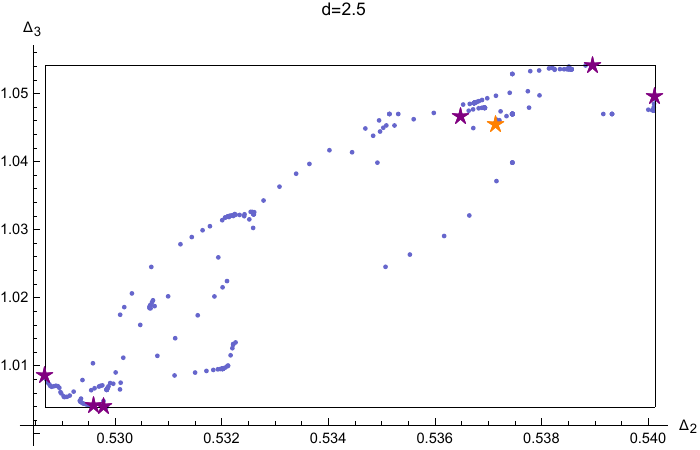}
\caption{All allowed navigator points in $d=2.5$. Same description as figure~\ref{fig:isl-275}.}\label{fig:isl-25}
\end{figure}

In figures~\ref{fig:isl-275} and \ref{fig:isl-25} we present the results from our navigator runs to maximise/minimise along the parameter directions, giving the size of the navigator islands in $d=2.75$ and $d=2.5$ dimensions. They result from runs with the following number of points:
\begin{align}
d&=2.75\!: & \text{max/min $\Delta_{\phi}$:}&\ 119/119,& \text{max/min $\Delta_{\phi^2}$:}&\ 49/113, & \text{max/min $\Delta_{\phi^3}$:}&\  133/67,
\\
d&=2.5\!: & \text{max/min $\Delta_{\phi}$:}&\ 40/94,& \text{max/min $\Delta_{\phi^2}$:}&\ 51/150, & \text{max/min $\Delta_{\phi^3}$:}&\  48/127.
\end{align}
In the figures, all these runs are superimposed alongside the run to minimise the navigator. The extremisation runs terminated at the extrema reported in table~\ref{tab:operatorsAndGaps} in the main text, and shown as purple stars in the figures. 

\subsection{Extremal spectra}
\label{app:extremalSpectra}

\begin{figure}
\centering
\includegraphics[height=0.48\textwidth]{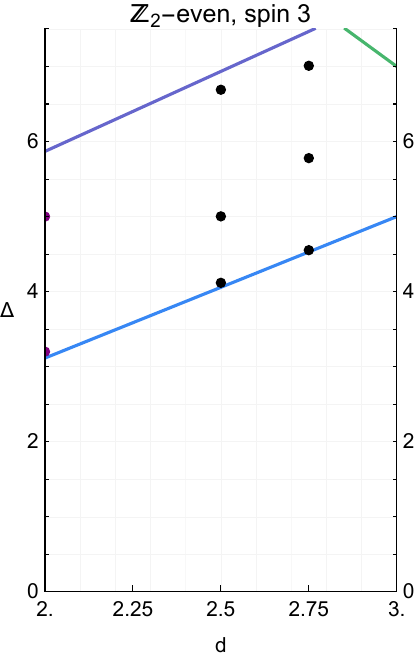}\qquad
\includegraphics[height=0.48\textwidth]{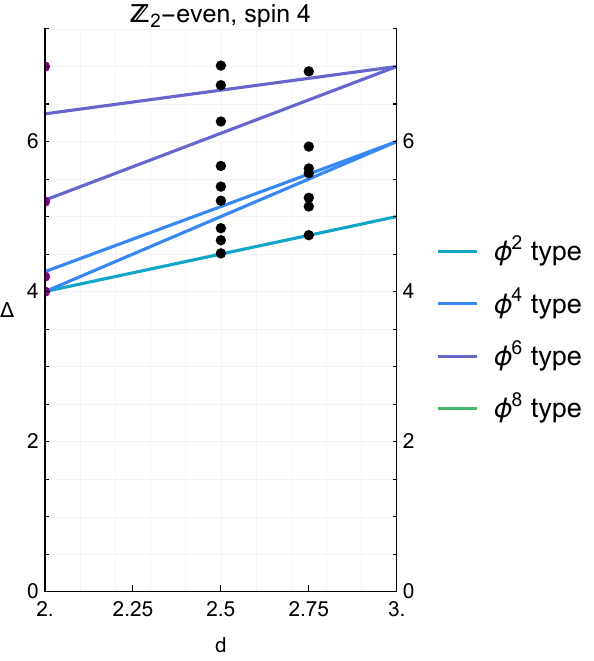}
\caption{Spectra for $\Z_2$-even operators of spin 3 and spin 4. The perturbative spin-$3$ spectrum looks rather sparse. This is because there is no spin-3 primary with six fields and $\Delta=6$, since the corresponding primary in the free theory is involved in a multiplet recombination effect with the spin-4 broken current of dimension $5$.}\label{fig:spec-E-3-4}
\end{figure}

\begin{figure}
\centering
\includegraphics[height=0.48\textwidth]{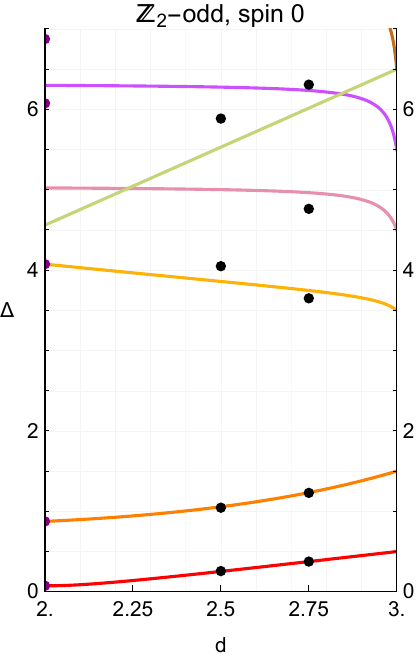}\qquad
\includegraphics[height=0.48\textwidth]{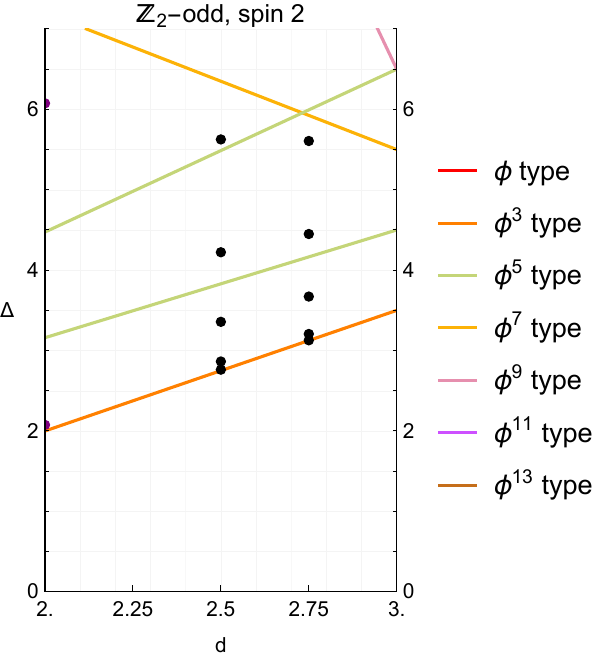}
\caption{Spectra for $\Z_2$-odd operators of spin 0 and spin 2.}\label{fig:spec-O-0-2}
\end{figure}

\begin{figure}
\centering
\includegraphics[height=0.48\textwidth]{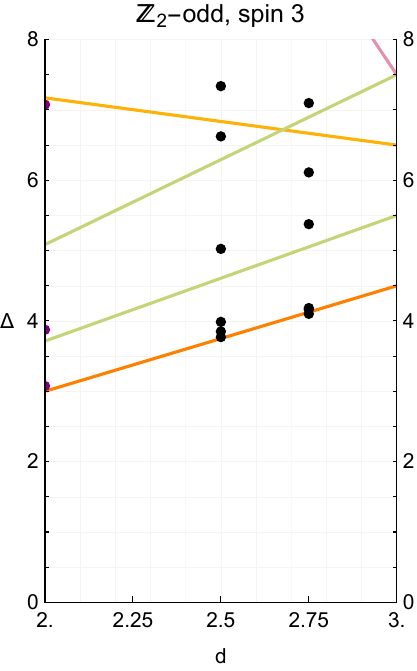}\qquad
\includegraphics[height=0.48\textwidth]{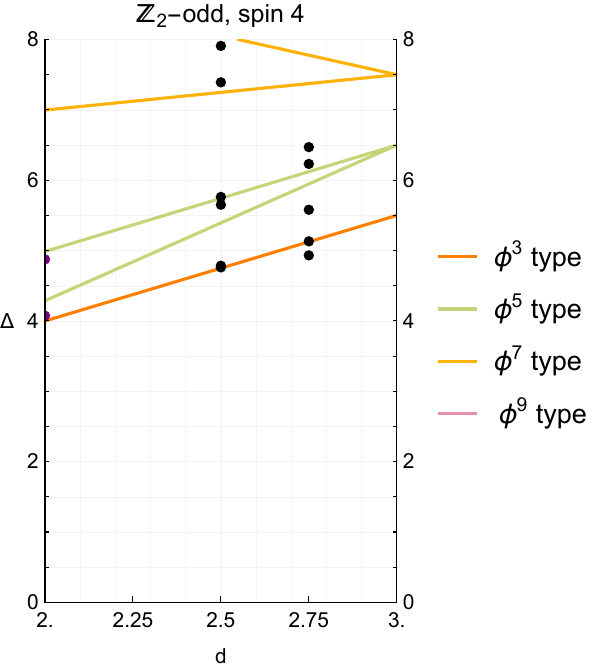}
\caption{Spectra for $\Z_2$-odd operators of spin 3 and spin 4.}\label{fig:spec-O-3-4}
\end{figure}

In figure~\ref{fig:spec-E-3-4} we show the extremal spectra for $\Z_2$-even operators of spin 3 and 4, complementing figure~\ref{fig:spec-E-0-2} in the main text. The $\Z_2$-odd extremal spectra are shown in figure~\ref{fig:spec-O-0-2} for spin 0 and 2, and in figure~\ref{fig:spec-O-3-4} for spin 3 and 4. As remarked in the main text, the extremal spectra show poor agreement with the perturbative estimates. 

\subsection{Running bootstrap in practice}
\label{app:inpractice}

Here we make a few remarks from the experience of running the bootstrap algorithms for this project. They represent some practical issues and are aspects to keep in mind for future similar studies.  
\begin{itemize}
\item The scanning algorithm with Delaunay triangularisation is effective for regions with rounded shapes, while some bootstrap islands from the literature have been rather elongated. Especially at pointy regions at the end of such elongated regions the algorithm performs badly. This can be mitigated by manually adding points during the run (possible in \texttt{Simpleboot}), or by a change of variables to make the size of the expected island roughly equal in all directions, as done e.g. in \cite{Chester:2019ifh,Chang:2024whx}.
\item In the navigator, one point is solved at the time before moving to the next iteration, meaning that a complete scan can extend over weeks. In this case, jobs might be killed by the cluster and have to be restarted. In \texttt{Simpleboot}, a job is restarted easily by moving the point to the evaluation queue, and if the job was terminated during an \texttt{SDPB} run, this leads to no problems. However, if the job was terminated at some other stage, for instance while the blocks were computed, it may corrupt some ancillary files which have to be deleted by hand.
\item Using hot-starting (c.f. section~\ref{sec:remarksOnBootstrap}), a check-point from a nearby iteration is loaded into \texttt{SDPB}, which can sometimes lead to stalling. Ultimately \texttt{Simpleboot} will detect this, delete the check-point and start a new \texttt{SDPB} run without check-point leading to significantly longer running time. This can be avoided by a manual intervention, in which we remove by hand the ``bad'' checkpoint and replace it with another (hopefully better) check-point from a nearby point. 
\item In this project, our runs to maximise/minimise along the directions in the scan space took significant number of iterations before even getting close to the relevant region, since our island was rather elongated. A simple proposal for reducing the total number of iterations when working with elongated islands is to first explore the extent of the island in both directions along just one axis, and then start the other scans from points near the two ends. 
\end{itemize}

\section{Review of results from perturbative renormalisation}
\label{app:renormalisation}

\textbf{Note added.} \textsl{Since the publication of this paper, some mistakes in the results in \cite{Hager:2002uq} have been found \cite{Adzhemyan:2026whu,Jack:2026npe}, see also \cite{Bednyakov:2025usv} and \cite{AdzhemyanFuture}. Equations~\eqref{eq:betafuncApp}, \eqref{eq:critCouplApp}, \eqref{eq:gamma4expl}, and \eqref{eq:gamma6expl} have been updated with correct expressions. No equations in the main text have been updated.}

For convenience, we reproduce here the renormalisation constants and beta functions from \cite{Hager1999,Hager:2002uq}, specialising to the case $n=1$.
These papers constitute a six-loop renormalisation of the action \eqref{eq:actionIntro}, giving $O(\eps^3)$ data for the operators $\phi$, $\phi^2$, $\phi^4$ and $\phi^6$. The beta function for $\lambda$ is
\begin{align}
\label{eq:betafuncApp}
\beta(\bar \lambda)&=-\mu \frac{\de \bar\lambda}{\de\mu}=
-2 \epsilon  \bar{\lambda }+\frac{10 \bar{\lambda }^2}{3}+\left(-\frac{45 \zeta_2 }{4}-\frac{281}{15}\right) \bar{\lambda }^3
\\
&\quad +  
\left(
90 (\beta_2\zeta_2+4 \beta_4)+\frac{3095  }{24}\zeta_2-\frac{525 }{4}\zeta_3+\frac{3375   }{16}\zeta_4+\frac{225}{4} \zeta_2 \ln2+\frac{3643}{18}
\right)\bar{\lambda }^4
+O(\bar\lambda^5),
\nonumber
\end{align}
and the critical coupling is
\begin{align}
\nonumber
\bar{\lambda}_*&=\frac{3 \eps }{5}+\frac{9 (675 \zeta_2 +1124) \eps^2}{5000}-\bigg(
\frac{729}{125} (\beta_2\zeta_2+4\beta_4)-\frac{401571}{50000}\zeta_2-\frac{1701}{200}\zeta_3
\\
&\qquad \qquad\qquad \qquad\qquad\qquad\qquad
+\frac{2187}{1600}\zeta_4+\frac{729}{200} \zeta_2 \ln 2-\frac{165519}{312500}
\bigg)\eps^3+O(\eps^4),
\label{eq:critCouplApp}
\end{align}
where the bar denotes the absorption of factors of $32\pi^2$.
In these expressions, $\zeta_k$ denote the usual Riemann zeta values, and $\beta_k$ denote the Dirichlet beta function $\sum_{n=0}^{\infty}(-1)^n/(2n+1)^k$. 

Computing the field renormalisation constants and beta functions, and evaluating at the critical coupling $\bar\lambda_*$, gives
\begin{align}
\label{eq:gamma1expl}
\gamma_\phi &=\frac{\eps^2}{1000}+\left(\frac{2279}{375000}+\frac{27 \pi ^2}{40000}\right)\eps^3+O(\eps^4),
\\
\gamma_{\phi^2}&=\frac{4 \eps^2}{125}+\left(\frac{513 \zeta_2}{5000}+\frac{3866}{46875}\right)\eps^3
+O(\eps^4),
\label{eq:gamma2expl}
\\
\nonumber
\gamma_{\phi^4} &=\frac{4 \eps }{5}+
\frac{60750 \zeta_2+72120}{75000} \eps^2
-\bigg(\frac{2592 \beta_4}{125}-\frac{1701 \zeta_3}{250}+\frac{2187 \zeta_4}{800}
\\\label{eq:gamma4expl}
&
\quad\qquad+3\frac{ 43200 \beta_2-17701+18900 \ln2}{25000}\zeta_2+\frac{77821}{234375}\bigg)\eps^3+O(\eps^4),
\\
\nonumber
\gamma_{\phi^6}&=4 \eps-\frac{20250 \zeta_2+33720}{5000}\eps^2+\bigg( \frac{3888 \beta_4}{25}-\frac{567 \zeta_3}{10}+\frac{8019 \zeta_4}{160}+\frac{655476}{15625}\\\label{eq:gamma6expl}
&
\quad\qquad+{9}\frac{10800 \beta_2+301+6750 \ln2}{2500}\zeta_2\bigg)+O(\eps^4).
\end{align}
Expressions for general $n$ are given in \cite{Hager:2002uq}.\footnote{In studying \cite{Hager:2002uq}, we found two typos, which can be corrected by resolving internal inconsistency: 1) in $\gamma_u$, line four, $-3805$ should read $-3508$, 2) in $\varphi(\bar w_R^*)$, second line should have the term $\pi^2(n^3+8n^2-496n-2888)$ (two extra minus signs compared to the paper).
\textbf{Note added.} \textsl{Correct expressions are given in \cite{Jack:2026npe}.}
}

\subsection{Other large expressions}
\label{app:large}

Here we write explicitly a set of corrections to the OPE coefficients from \cite{Henriksson:2020jwk}, where they were only given implicitly. Denoting $\mathcal J_\ell=\phi\de^\ell\phi$, the OPE coefficients with $\phi$ are
\begin{equation}
\label{eq:OPEJell}
\lambda_{\phi\phi\mathcal J_\ell}^2=\left(\lambda_{\phi\phi\mathcal J_\ell}^{\mathrm{free},3-\eps}\right)^2\big[1+\eps^2\alpha_\ell+O(\eps^3)\big],
\end{equation}
where
\begin{equation}
\alpha_\ell=\frac{15}{4(\ell-\frac{1}{2}) (\ell+\frac{1}{2})^2}+\frac{\ell^2-4}{\ell^2-\frac14}\big[S_1(\ell-1)-S_1(2\ell-1)\big]+S_1\left(\ell-\tfrac12\right)+2\ln 2\,,
\end{equation}
and 
\begin{equation}
\label{eq:FreeTheoryOPE}
\left(\lambda_{\phi\phi\mathcal J_\ell}^{\mathrm{free},d}\right)^2=\frac{2 \Gamma (\frac{d}{2}+\ell-1)^2 \Gamma (d+\ell-3)}{\Gamma \left(\frac{d}{2}-1\right)^2 \Gamma (\ell+1) \Gamma (d+2 \ell-3)}
\end{equation}
are the free OPE coefficients in $d$ dimensions. $S_1$ denotes the analytic continuation of the harmonic numbers. Despite the apparent appearance of $\ln 2$, the corrections $\alpha_\ell$ are rational numbers, $\alpha_2=\frac{23}{3750}$, $\alpha_4=\frac{9349 }{1653750}$, etc.

\section{Character decompositions}
\label{eq:characterDecomp}

In $d=3$ dimensions, we use the Hilbert series methods from \cite{Dolan:2005wy,Henning:2017fpj} to write the character decomposition of the spectrum of the free scalar field in terms of conformal primary operators.
In table~\ref{tab:3dspec}, we give the spectrum of parity-even operators. In this table we emphasised the contribution of certain operators: The multiplet of the stress tensor $T_{\mu\nu}$ is short in both the free and the interacting theory. The multiplets of $\phi$ and the currents $\mathcal J_{\ell}$ are short in the free theory, but long in the interacting theory. This means that certain primaries higher in the spectrum, denoted $E$ (equation-of-motion) are missing from the tricritical CFT. In table~\ref{tab:3dspecOddP}, we give the parity-odd spectrum. Notice in particular that the first parity-odd scalar has $\Delta=11$.

\begin{table}\caption{Spectrum of parity-even operators in the theory of a free 3d scalar field. The horizontal axis denotes $\Delta$, the vertical axis $\ell$. $E$ denotes operators missing in the interacting theory due to equation-of-motion effects. }
\label{tab:3dspec}
\centering
\footnotesize
\def\arraystretch{1.5}
\begin{tabular}{|c|cccccccccccccccccccccc|}
\hline
$8  $  &  &&&&&&&&&&&&&&&& &  $\!\!\! \mathcal{J}_8 \!\!\! $  &  $ 2  $  &  $ 4  $  &  $ 5  $  &  $ 6  $  \\  $
 7  $  &&&&&&&&&&&&&&&&    $ 0  $  &  $ 1  $  &  $ 2  $  &  $ 3  $  &  $\!\!\!\!\!E+2 \!\!\!\!\! $  &  $ 4  $  &  $ 6  $  \\  $
 6  $  &&&&&&&&&&&&&  &  $ \!\!\!\mathcal{J}_6 \!\!\!$  &  $ 2  $  &  $ 3  $  &  $ 3  $  &  $ 4  $  &  $ 4  $  &  $ 7  $  &  $ 9  $  &  $ 11  $  \\  $
 5  $  &  &&&&&&&&&&  &  $ 0  $  &  $ 1  $  &  $ 1  $  &  $ 2  $  &  $\!\!\!\!\!E+1\!\!\!\!\!$  &  $ 2  $  &  $ 3  $  &  $ 5  $  &  $ 5  $  &  $ 6  $  &  $ 7  $  \\  $
 4  $  &&&&&&&&&  &  $\!\!\! \mathcal{J}_4  \!\!\!$  &  $ 1  $  &  $ 2  $  &  $ 2  $  &  $ 2  $  &  $ 2  $  &  $ 4  $  &  $ 4  $  &  $ 5  $  &  $ 5  $  &  $ 8  $  &  $ 11  $  &  $ 14  $  \\  $
 3  $  &  &&&&&&  &  $ 0  $  &  $ 1  $  &  $ 1  $  &  $ 1  $  &  $ E  $  &  $ 1  $  &  $ 1  $  &  $ 2  $  &  $ 2  $  &  $ 2  $  &  $ 3  $  &  $ 5  $  &  $ 6  $  &  $ 7  $  &  $ 8  $  \\  $
 2  $  & &&&&  &  $ T  $  &  $ 1  $  &  $ 1  $  &  $ 1  $  &  $ 1  $  &  $ 1  $  &  $ 2  $  &  $ 2  $  &  $ 2  $  &  $ 2  $  &  $ 3  $  &  $ 4  $  &  $ 5  $  &  $ 5  $  &  $ 7  $  &  $ 9  $  &  $ 12  $  \\  $
 1  $  &&&&  $ 0  $  &  $ 0  $  &  $ 0  $  &  $ 0  $  &  $ 0  $  &  $ 0  $  &  $ 0  $  &  $ 0  $  &  $ 0  $  &  $ 0  $  &  $ 0  $  &  $ 1  $  &  $ 1  $  &  $ 1  $  &  $ 1  $  &  $ 3  $  &  $ 3  $  &  $ 4  $  &  $ 4  $  \\  $
 0  $  &  $ \phi   $  &  $ 1  $  &  $ 1  $  &  $ 1  $  &  $ \text{E}  $  &  $ 1  $  &  $ 1  $  &  $ 1  $  &  $ 1  $  &  $ 1  $  &  $ 1  $  &  $ 2  $  &  $ 2  $  &  $ 2  $  &  $ 2  $  &  $ 3  $  &  $ 3  $  &  $ 4  $  &  $ 4  $  &  $ 5  $  &  $ 5  $  &  $ 7  $  \\\hline  $
 $  &  $ \frac{1}{2}  $  &  $ 1  $  &  $ \frac{3}{2}  $  &  $ 2  $  &  $ \frac{5}{2}  $  &  $ 3  $  &  $ \frac{7}{2}  $  &  $ 4  $  &  $ \frac{9}{2}  $  &  $ 5  $  &  $ \frac{11}{2}  $  &  $ 6  $  &  $ \frac{13}{2}  $  &  $ 7  $  &  $ \frac{15}{2}  $  &  $ 8  $  &  $ \frac{17}{2}  $  &  $ 9  $  &  $ \frac{19}{2}  $  &  $ 10  $  &  $ \frac{21}{2}  $  &  $ 11  $  
\\\hline
\end{tabular}
\end{table}

\begin{table}\caption{Spectrum of parity-odd operators in the theory of a free 3d scalar field. The horizontal axis denotes $\Delta$, the vertical axis $\ell$.}
\label{tab:3dspecOddP}
\centering
\footnotesize
\def\arraystretch{1.5}
\begin{tabular}{|c|cccccccccccccccccccccc|}
\hline
$8 $  &  $   $  &  $   $  &  $   $  &  $   $  &  $   $  &  $   $  &  $   $  &  $   $  &  $   $  &  $   $  &  $   $  &  $   $  &  $   $  &  $   $  &  $   $  &  $   $  &  $   $  &  $ 0  $  &  $ 0  $  &  $ 0  $  &  $ 1  $  &  $ 3  $  \\  $
 7  $  &  $   $  &  $   $  &  $   $  &  $   $  &  $   $  &  $   $  &  $   $  &  $   $  &  $   $  &  $   $  &  $   $  &  $   $  &  $   $  &  $   $  &  $   $  &  $ 0  $  &  $ 0  $  &  $ 0  $  &  $ 1  $  &  $ 2  $  &  $ 3  $  &  $ 4  $  \\  $
 6  $  &  $   $  &  $   $  &  $   $  &  $   $  &  $   $  &  $   $  &  $   $  &  $   $  &  $   $  &  $   $  &  $   $  &  $   $  &  $   $  &  $ 0  $  &  $ 0  $  &  $ 0  $  &  $ 1  $  &  $ 2  $  &  $ 3  $  &  $ 3  $  &  $ 3  $  &  $ 6  $  \\  $
 5  $  &  $   $  &  $   $  &  $   $  &  $   $  &  $   $  &  $   $  &  $   $  &  $   $  &  $   $  &  $   $  &  $   $  &  $ 0  $  &  $ 0  $  &  $ 0  $  &  $ 0  $  &  $ 1  $  &  $ 1  $  &  $ 1  $  &  $ 1  $  &  $ 3  $  &  $ 5  $  &  $ 7  $  \\  $
 4  $  &  $   $  &  $   $  &  $   $  &  $   $  &  $   $  &  $   $  &  $   $  &  $   $  &  $   $  &  $ 0  $  &  $ 0  $  &  $ 0  $  &  $ 1  $  &  $ 1  $  &  $ 1  $  &  $ 1  $  &  $ 1  $  &  $ 2  $  &  $ 4  $  &  $ 5  $  &  $ 5  $  &  $ 7  $  \\  $
 3  $  &  $   $  &  $   $  &  $   $  &  $   $  &  $   $  &  $   $  &  $   $  &  $ 0  $  &  $ 0  $  &  $ 0  $  &  $ 0  $  &  $ 0  $  &  $ 0  $  &  $ 0  $  &  $ 0  $  &  $ 1  $  &  $ 2  $  &  $ 2  $  &  $ 2  $  &  $ 2  $  &  $ 4  $  &  $ 6  $  \\  $
 2  $  &  $   $  &  $   $  &  $   $  &  $   $  &  $   $  &  $ 0  $  &  $ 0  $  &  $ 0  $  &  $ 0  $  &  $ 0  $  &  $ 0  $  &  $ 0  $  &  $ 0  $  &  $ 1  $  &  $ 1  $  &  $ 1  $  &  $ 1  $  &  $ 2  $  &  $ 3  $  &  $ 4  $  &  $ 4  $  &  $ 5  $  \\  $
 1  $  &  $   $  &  $  $  &  $  $  &  $ 0  $  &  $ 0  $  &  $ 0  $  &  $ 0  $  &  $ 0  $  &  $ 0  $  &  $ 0  $  &  $ 0  $  &  $ 0  $  &  $ 0  $  &  $ 0  $  &  $ 0  $  &  $ 0  $  &  $ 0  $  &  $ 0  $  &  $ 0  $  &  $ 0  $  &  $ 1  $  &  $ 2  $  \\  $
 0  $  &  $ 0  $  &  $ 0  $  &  $ 0  $  &  $ 0  $  &  $ 0  $  &  $ 0  $  &  $ 0  $  &  $ 0  $  &  $ 0  $  &  $ 0  $  &  $ 0  $  &  $ 0  $  &  $ 0  $  &  $ 0  $  &  $ 0  $  &  $ 0  $  &  $ 0  $  &  $ 0  $  &  $ 0  $  &  $ 0  $  &  $ 0  $  &  $ 1  $  \\\hline  $
  $  &  $ \frac{1}{2}  $  &  $ 1  $  &  $ \frac{3}{2}  $  &  $ 2  $  &  $ \frac{5}{2}  $  &  $ 3  $  &  $ \frac{7}{2}  $  &  $ 4  $  &  $ \frac{9}{2}  $  &  $ 5  $  &  $ \frac{11}{2}  $  &  $ 6  $  &  $ \frac{13}{2}  $  &  $ 7  $  &  $ \frac{15}{2}  $  &  $ 8  $  &  $ \frac{17}{2}  $  &  $ 9  $  &  $ \frac{19}{2}  $  &  $ 10  $  &  $ \frac{21}{2}  $  &  $ 11  $ 
\\\hline
\end{tabular}
\end{table}

In $d=2$ we consider the torus partition function \eqref{eq:pfmn}. Using the expressions for the characters given in \cite{Castro:2011zq}, we expand each character up to level-10 descendants:
\begin{align}
\chi_{vac}&=\tilde\chi _{2,2}+\chi _{4,0}+\tilde \chi _{4,4}+\chi _{6,2}+2\tilde \chi _{6,6}+\chi _{8,0}+2 \chi _{8,4}+3 \tilde\chi _{8,8}+\chi _{9,9}+2 \chi _{10,2}+3 \chi _{10,6}
\nonumber
\\&\quad
+4\tilde \chi _{10,10}
\\
\chi_{\phi_{2,2}}&=\chi _{0,0}+\chi _{2,2}+\chi _{3,3}+\chi _{4,0}+\chi _{4,4}+\chi _{5,1}+2 \chi _{5,5}+\chi _{6,0}+\chi _{6,2}+2 \chi _{6,6}+\chi _{7,1}+2 \chi _{7,3}\nonumber\\&\quad+3 \chi _{7,7}+\chi _{8,0}+2 \chi
   _{8,2}+2 \chi _{8,4}+4 \chi _{8,8}+2 \chi _{9,1}
+2 \chi _{9,3}+3 \chi _{9,5}+5 \chi _{9,9}+4 \chi _{10,0}\nonumber\\&\quad+2 \chi _{10,2}+3 \chi _{10,4}+4 \chi _{10,6}+6 \chi _{10,10}\,,
\\
\chi_{\phi_{1,2}}&=\chi _{0,0}+\chi _{3,3}+\chi _{4,4}+\chi _{5,5}+\chi _{6,0}+2 \chi _{6,6}+\chi _{7,1}+2 \chi _{7,7}+\chi _{8,0}+\chi _{8,2}+3 \chi _{8,8}\nonumber\\&\quad+\chi _{9,1}+2 \chi _{9,3}+3 \chi _{9,9}+\chi
   _{10,0}+2 \chi _{10,2}+2 \chi _{10,4}+5 \chi _{10,10}\,,
\\
\chi_{\phi_{2,1}}&=\chi _{0,0}+\chi _{3,3}+\chi _{4,4}+\chi _{5,5}+\chi _{6,0}+2 \chi _{6,6}+\chi _{7,1}+2 \chi _{7,7}+\chi _{8,0}+\chi _{8,2}+2 \chi _{8,8}\nonumber\\&\quad+\chi _{9,1}+2 \chi _{9,3}+4 \chi _{9,9}+\chi
   _{10,0}+2 \chi _{10,2}+2 \chi _{10,4}+4 \chi _{10,10}\,,
\\
\chi_{\phi_{3,2}}&=\chi _{0,0}+\chi _{2,2}+\chi _{4,0}+2 \chi _{4,4}+\chi _{5,5}+2 \chi _{6,2}+2 \chi _{6,6}+\chi _{7,3}+2 \chi _{7,7}+4 \chi _{8,0}+2 \chi _{8,4}\nonumber\\&\quad+4 \chi _{8,8}+2 \chi _{9,1}+2 \chi
   _{9,5}+3 \chi _{9,9}+\chi _{10,0}+4 \chi _{10,2}+4 \chi _{10,6}+6 \chi _{10,10}\,,
\\
\chi_{\phi_{3,1}}&=\chi _{0,0}+\chi _{2,2}+\chi _{4,0}+\chi _{4,4}+\chi _{5,5}+\chi _{6,2}+2 \chi _{6,6}+\chi _{7,3}+\chi _{7,7}+\chi _{8,0}+2 \chi _{8,4}+3 \chi _{8,8}\nonumber\\&\quad+\chi _{9,1}+\chi _{9,5}+3 \chi
   _{9,9}+\chi _{10,0}+2 \chi _{10,2}+3 \chi _{10,6}+4 \chi _{10,10}\,,
\end{align}
Here $\chi_{n,\ell}$ is the character of a quasiprimary at level $n$ with dimension $n+\Delta_0$ (with $\Delta_0=\Delta_{\phi_{p,q}}$) and spin $\ell$:
\begin{equation}
\chi_{n,\ell}(q,\bar q)=\frac1{(1-q)(1-\bar q)}q^{\frac c{24}+h}\bar q^{\frac c{24}+\bar h}+(h\leftrightarrow \bar h), \qquad h+\bar h=\Delta_0+n,\ |h-\bar h|=\ell\,,
\end{equation}
where there is only one term if $h=\bar h$ ($\ell=0$). 
The characters $\tilde\chi_{n,n}$ in the decomposition of $\chi_{\mathrm{vac}}$ are short characters representing conserved currents, 
\begin{equation}
\tilde \chi_{n,n}(q,\bar q)=\chi_{n,n}(q,\bar q)-\chi_{n+1,n-1}(q,\bar q).
\end{equation}

\section{General multicritical theories}
\label{app:generalMulti}

We give some general results for $\phi^{2m}$ theory. The case of $\phi^{p}$ theory with odd $p$ is also interesting, although we do not consider it here, see e.g. \cite{Codello:2017epp,Gracey:2017okb,Zinati:2019gct}. 
Apart from diagrammatic studies, $\phi^{2m}$ theories have been studied with the non-perturbative RG in
\cite{Codello:2012sc,Hellwig:2015woa,Codello:2014yfa} and analytic bootstrap in \cite{Henriksson:2020jwk,Guo:2023qtt}. 

\subsection{Multicritical models in near their upper critical dimension}

The anomalous dimension for $\phi$ is given by \cite{Wegner1975} (see also \cite{Lewis:1978zz}, and section~5.A of \cite{Itzykson:1989sx})
 \begin{equation}
\Delta_\phi=\frac{d-2}2+\frac{2( m -1)^2\Gamma( m +1)^6}{\Gamma(2 m +1)^3}\eps^2+O(\eps^3).
\end{equation}
Scaling dimensions of spinning operators $\phi\de^\ell\phi$ (broken currents) were given in \cite{Gliozzi:2017gzh}. Putting $\ell=0$ generates agreement with the known formula for the anomalous dimension of $\phi^2$: 
\begin{equation}\label{eq:multirelationphiphi2}
\gamma^{(2)}_{\phi^2}=d-2+\frac{8 ( m  -1)^3 ( m  +1) \Gamma ( m  +1)^6}{( m  -2) \Gamma (2  m  +1)^3}\eps^2+O(\eps^3).
\end{equation}
In \cite{Henriksson:2020jwk} the correction to the central charge was computed:
\begin{equation}\label{eq:CTmulti}
\frac{C_T}{C_{T,\mathrm{free}}}=1-\frac{4( m -1)^3(3 m -1)\Gamma( m +1)^6}{ m (2 m -1)\Gamma(2 m +1)^3}\frac{}{}\eps^2+O(\eps^3),
\end{equation}
where $C_{T,\mathrm{free}}=\frac d{d-1}$ in the normalisation used here.\footnote{This corresponds to the conformal Ward identity written as $\lambda_{\O\O T}=\frac{-d\Delta_\O}{2(d-1)\sqrt{C_T}}$ \cite{Osborn:1993cr}, see appendix~A of \cite{Henriksson:2022rnm} for a discussion on conventions.} This was extracted for general results for the OPE coefficients $\lambda_{\phi,\phi,\phi\de^\ell\phi}$.

The leading anomalous dimension of operators $\phi^k$ in the $ m $-critical theory were given in \cite{Wegner1975}:
\begin{equation}
\Delta_{\phi^k}=k\frac{d-2}2+2( m -1)\frac{k! m !}{(2 m )!(k- m )!}\eps+O(\eps^2),
\end{equation}
where the $O(\eps)$ term vanishes for $k< m $. 
Reference \cite{ODwyer:2007brp} explains how to compute the next order, and evaluates it for $ m =2,3$. 

For the multicritical theory with 
$O(N)$ symmetry, an expression for the anomalous dimension of $\phi$ was given in \cite{Hofmann:1991ge}
\begin{equation}
\gamma_\phi^{(2)}=\frac{( m -1)^2\Gamma( m )^2\Gamma( m +1)\Gamma( m +\frac N2)}{4\Gamma(2 m )^2\Gamma(1+\frac N2)}{_3F_2}
\left( {1-\frac N2- m ,\frac12-\frac m 2,-\frac m 2}~\atop~{ 1,\frac12- m }\middle|1\right)^{-2}\!.
\end{equation}

\subsection{Multicritical models in 2d}

In $d=2$ dimensions, the multicritical $\phi^{2m}$ theory is described by the diagonal ($A$ series) modular invariants of the minimal models $\mathcal M_{m+2,m+1}$ with central charge $1-\frac6{(m+1)(m+2)}$ ($m=2$ is Ising, $m=3$ is tri-Ising, etc). This theory has a set of
$\frac{m(m-1)}2$ Virasoro primaries, where the weights of the Kac table are given by
\begin{equation}
h_{p,q}=\frac{((m+2)p-(m+1)q)^2-1}{2(m+1)(m+2)}, \qquad \Delta_{p,q}=2h_{p,q}\,.
\end{equation}
The first $2m-2$ operators of type $\phi^k$ can be identified with Kac table states according to the following \cite{Zamolodchikov:1986db}:\footnote{Our conventions have swapped $p\leftrightarrow q$ with respect to \cite{Zamolodchikov:1986db}.}
\begin{equation}
\label{eq:Identification}
\phi^k \leftrightarrow \begin{cases}
\phi_{k+1,k+1}\,, & k \leqslant m-1\,,
\\
\phi_{k-m+3,k-m+2} \,,& m\leqslant k\leqslant 2m-2\,.
\end{cases}
\end{equation}
This fills the diagonal entries of the Kac table, and also the entries directly above the diagonal. 
The identification of the remaining states is more complicated. For instance, \cite{Zamolodchikov:1986db} gives that $\de_\mu\phi\de^\mu\phi \leftrightarrow \phi_{3,1}$ with dimension $\Delta_{3,1}=2+\frac4{m+1}$, but it is not clear how to interpret this in a continuation above two dimensions since $\de_\mu\phi\de^\mu\phi$ is not a primary in the $d>2$ free theory. For tri-critical Ising $m=3$, we find it natural to identify $\phi_{3,1} \leftrightarrow\phi^6$.

\subsection[Four-point function of $\phi_{2,2}$ in 2d minimal models]{Four-point function of $\boldsymbol{\phi_{2,2}}$ in 2d minimal models}
\label{app:fourpointSigma}
Here we reproduce the expression for the four-point function of $\phi$ in general 2d minimal models $\mathcal M_{m+2,m+1}$ under the identification $\phi=\sigma=\phi_{2,2}$. It takes the form 
\begin{align}
\label{eq:Gsigma2d}
\G_{\sigma\sigma\sigma\sigma}^{2d}(z,\bar z)&\!=\! \left|\mathscr F_{\1}(p|z)\right|^2\!+\!\lambda_{\sigma\sigma\phi_{3,3}}^2\left|\mathscr F_{\phi_{3,3}}(p|z)\right|^2\!+\!
\lambda_{\sigma\sigma\phi_{1,3}}^2\left|\mathscr F_{\phi_{1,3}}(p|z)\right|^2\!+\!\lambda_{\sigma\sigma\phi_{3,1}}^2\left|\mathscr F_{\phi_{3,1}}(p|z)\right|^2\!,
\end{align}
 where the conformal blocks were given in \cite{Zamolodchikov:1986db} for general $m=p-1$:\footnote{We had to correct the expression for $\mathscr F_{\phi_ {1,3}}(p|z)$ compared to that reference, to match with output from the code \texttt{ConformalBlock} by Ying-Hsuan Lin (unpublished).}
\begin{align}
\mathscr F_{\1}(p|z)&=\frac1{(1-z)^{\Delta_\phi}}\bigg[(1-z){_2F_1}\left(\tfrac1{p+1},\tfrac3{p+1};\tfrac2{p+1};z\right){_2F_1}\left(\tfrac{p-3}p,\tfrac{p-1}p;\tfrac{p-2}p;z\right)
\nonumber\\
&\qquad\qquad\qquad+ \frac{z}{2(p-2)}{_2F_1}\left(\tfrac1{p+1},\tfrac3{p+1};\tfrac{p+3}{p+1};z\right){_2F_1}\left(\tfrac{p-3}p,\tfrac{p-1}p;\tfrac{2p-2}p;z\right)\bigg]
,
\label{eq:VirBlock1}
\\
\mathscr F_{\phi_{1,3}}(p|z)&=\frac{z^{h_{1,3}}}{(1-z)^{\Delta_\phi}}\bigg[(z-1)(p-2){_2F_1}\left(\tfrac{p+2}{p+1},\tfrac p{p+1};\tfrac{2p}{p+1};z\right){_2F_1}\left(\tfrac{p-3}p,\tfrac{p-1}p;\tfrac{p-2}p;z\right)
\nonumber\\
&\qquad\qquad\qquad +(p-1){_2F_1}\left(\tfrac1{p+1},-\tfrac1{p+1};\tfrac{p-1}{p+1};z\right){_2F_1}\left(\tfrac{p-3}p,\tfrac{p-1}p;\tfrac{2p-2}p;z\right)\bigg]
,
\label{eq:VirBlock2}\\
\mathscr F_{\phi_{3,1}}(p|z)&=\frac{2(p+2)(p+3)z^{h_{3,1}-1}}{3(1-z)^{\Delta_\phi}}\bigg[(z-1){_2F_1}\left(\tfrac{1}{p+1},\tfrac3{p+1};\tfrac2{p+1};z\right){_2F_1}\left(\tfrac{p+1}p,\tfrac{p-1}p;\tfrac{p+2}p;z\right)
\nonumber\\
&\qquad\qquad\qquad +{_2F_1}\left(\tfrac1{p+1},\tfrac3{p+1};\tfrac{p+3}{p+1};z\right){_2F_1}\left(\tfrac1p,-\tfrac1p;\tfrac2p;z\right)\bigg]
,
\label{eq:VirBlock3}\\
\mathscr F_{\phi_{3,3}}(p|z)&=\frac{z^{h_{3,3}}}{(1-z)^{\Delta_\phi}}\bigg[\frac{z(1-z)}{2(p-1)}{_2F_1}\left(\tfrac{p+2}{p+1},\tfrac{p}{p+1};\tfrac{2p}{p+1};z\right){_2F_1}\left(\tfrac{p+1}p,\tfrac{p-1}p;\tfrac{p+2}p;z\right)
\nonumber\\
&\qquad\qquad\qquad +{_2F_1}\left(\tfrac1{p+1},-\tfrac1{p+1};\tfrac{p-1}{p+1};z\right){_2F_1}\left(\tfrac1p,-\tfrac1p;\tfrac2p;z\right)\bigg]
,
\label{eq:VirBlock4}
\end{align}
Here $\Delta_\phi=\frac3{2p(p+1)}$. Moreover, the squared OPE coefficients that enter \eqref{eq:Gsigma2d} are \cite{Zamolodchikov:1986db}
\begin{align}\lambda^2_{\sigma,\sigma,\phi_{3,3}}&=
\frac{\Gamma \big(1-\frac{2}{p}\big)^2 \Gamma \big(1+\frac{1}{p}\big) \Gamma
   \big(1+\frac{3}{p}\big) \Gamma \big(1-\frac{3}{p+1}\big) \Gamma \big(1-\frac{1}{p+1}\big)
   \Gamma \big(1+\frac{2}{p+1}\big)^2}{\Gamma \big(1-\frac{3}{p}\big) \Gamma
   \big(1-\frac{1}{p}\big) \Gamma \big(1+\frac{2}{p}\big)^2 \Gamma \big(1-\frac{2}{p+1}\big)^2
   \Gamma \big(1+\frac{1}{p+1}\big) \Gamma \big(1+\frac{3}{p+1}\big)},
\\
\lambda^2_{\sigma,\sigma,\phi_{1,3}}&=
\frac{3 \Gamma \big(1-\frac{3}{p+1}\big) \Gamma \big(1-\frac{1}{p+1}\big) \Gamma
   \big(1+\frac{2}{p+1}\big)^2}{4 (p-2)^2 (p-1)^2 \Gamma \big(1-\frac{2}{p+1}\big)^2 \Gamma
   \big(1+\frac{1}{p+1}\big) \Gamma \big(1+\frac{3}{p+1}\big)},
\\
\lambda^2_{\sigma,\sigma,\phi_{3,1}}&=
\frac{3 \Gamma \big(1-\frac{2}{p}\big)^2 \Gamma \big(1+\frac{1}{p}\big) \Gamma
   \big(1+\frac{3}{p}\big)}{4 (p+2)^2 (p+3)^2 \Gamma \big(1-\frac{3}{p}\big) \Gamma
   \big(1-\frac{1}{p}\big) \Gamma \big(1+\frac{2}{p}\big)^2}.
\end{align}
As a check, the well-known Ising correlator is recovered for $p=3$ ($m=2$)
\begin{equation}
\left|\mathscr F_{\1}(3|z)\right|^2+\frac14\left|\mathscr F_{\phi_{1,3}}(3|z)\right|^2
=
\frac{\sqrt{1+\sqrt{1-z}}\sqrt{1+\sqrt{1-\zb}}+\sqrt{1-\sqrt{1-z}}\sqrt{1-\sqrt{1-\zb}}}{2(1-z)^{1/8}(1-\zb)^{1/8}},
\end{equation}
where only two of the four operators contribute. 

For completeness, we also give the leading expansion of the Virasoro conformal blocks \eqref{eq:VirBlock1}--\eqref{eq:VirBlock4} (for $m=4$) in terms of global conformal blocks. 
Up to level-6 descendants, the expansions are\footnote{In $\left|\mathscr F_{\phi_{3,3}}\right|^2$, we find that there are no descendants of twist $\tau=\frac15+4$ (for $m=4$).}
\begin{align}
\left|\mathscr F_{\1}\right|^2&=g_{0,0}+\frac{9 g_{0,2}}{2240}+\frac{1083 g_{0,4}}{24371200}+\frac{209819 g_{0,6}}{163774464000}+\frac{81 g_{4,0}}{5017600}+\frac{9747 g_{4,2}}{54591488000},
\\
\left|\mathscr F_{\phi_{3,3}}\right|^2&=g_{1/5,0}+\frac{93 g_{1/5,4}}{4259840}+\frac{89 g_{1/5,6}}{97976320},
\\
\left|\mathscr F_{\phi_{1,3}}\right|^2&=g_{6/5,0}+\frac{9 g_{6/5,2}}{9152}+\frac{137091 g_{6/5,4}}{2657607680}+\frac{37429777 g_{6/5,6}}{16935794769920}+\frac{81
   g_{26/5,0}}{83759104}\nonumber\\&\quad +\frac{1233819 g_{26/5,2}}{24322425487360},
\\
\left|\mathscr F_{\phi_{3,1}}\right|^2&=g_{3,0}+\frac{63 g_{3,2}}{10880}+\frac{121 g_{3,4}}{819200}+\frac{1764437 g_{3,6}}{329777152000}+\frac{3969 g_{7,0}}{118374400}+\frac{7623 g_{7,2}}{8912896000},
\end{align}
where
\begin{equation}
g_{\tau,\ell}=\frac{k_{h}(z)k_{\bar h}(\bar z)+k_{\bar h}(z)k_{h}(\bar z)}{1+\delta_{\ell,0}} \quad h,\bar h=\frac{\Delta\pm \ell}2, \quad k_\beta(x)=x^\beta{_2F_1}(\beta,\beta;2\beta;x)
\end{equation}
and 
$\tau=\Delta-\ell$.

\let\oldbibliography\thebibliography
\renewcommand{\thebibliography}[1]{%
  \oldbibliography{#1}%
  \setlength{\itemsep}{0pt}%
}

\bibliographystyle{JHEP}
\bibliography{bibl}

\end{document}